%% file: ms.tex
\documentclass{aa}  

\usepackage{natbib}
\usepackage{array}
\pdfoutput=1 

\usepackage[varg]{txfonts}
\bibpunct{(}{)}{;}{a}{}{,} 

\usepackage{subcaption}
\usepackage{graphicx}

\usepackage{txfonts}

\mathchardef\mhyphen="2D

\date{}

\begin{document} 
	
	\title{Optimising and comparing source extraction tools using objective segmentation quality criteria}

	\author{Caroline Haigh\inst{1}
		\and
		Nushkia Chamba\inst{2,3}
		\and
		Aku Venhola\inst{4,5}
		\and
		Reynier Peletier\inst{4}
		\and
		Lars Doorenbos\inst{1}
		\and
		Matthew Watkins\inst{1}
		\and
		Michael H. F. Wilkinson\inst{1}
		\fnmsep
	}
	
	\institute{Bernoulli Institute for Mathematics, Computer Science
		and Artificial Intelligence,\\ University of Groningen,
		P.O. Box 407, 9700 AK Groningen, The Netherlands\\
		\email{(c.haigh,m.h.f.wilkinson)@rug.nl}
		\and
		Instituto de Astrof\'\i sica de Canarias, Calle V\'\i a L\'actea,\\ s/n, 38205 San Crist\'obal de La Laguna, Santa Cruz de Tenerife, Spain\\
		\email{chamba@iac.es}
		\and
		Departamento de Astrofísica, Universidad de La Laguna,\\
		38205, La Laguna, Tenerife, Spain
		\and
		Kapteyn Astronomical Institute, University of Groningen,\\
		P.O. Box 800, 9700 AV Groningen, The Netherlands\\
		\email{peletier@astro.rug.nl}    
		\and
		Space Physics and Astronomy Research Unit, University of Oulu,\\ Pentti Kaiteran katu 1, 90014 Oulu, Finland\\
		\email{aku.venhola@oulu.fi}
	}
	
	\date{June 2020}
	
	\abstract
	{With the growth of the scale, depth, and resolution of astronomical imaging surveys, there is an increased need for highly accurate automated detection and extraction of astronomical sources from images. This also means there is a need for objective quality criteria, and automated methods to optimise parameter settings for these software tools.}
	{We present a comparison of several tools which have been developed to perform this task: namely SExtractor, ProFound, NoiseChisel, and MTObjects. In particular, we focus on evaluating performance in situations which present challenges for detection -- for example, faint and diffuse galaxies; extended structures, such as streams; and objects close to bright sources. Furthermore, we develop an automated method to optimise the parameters for the above tools.}
	{We present four different objective segmentation quality measures, based on precision, recall, and a new measure for the correctly identified area of sources. Bayesian optimisation is used to find optimal parameter settings for each of the four tools on simulated data, for which a ground truth is known. After training, the tools are tested on similar simulated data, to provide a performance baseline. We then qualitatively assess tool performance on real astronomical images from two different surveys.}
	{We determine that when area is disregarded, all four tools are capable of broadly similar levels of detection completeness, while only NoiseChisel and MTObjects are capable of locating the faint outskirts of objects. MTObjects produces the highest scores on all tests on all four quality measures, whilst SExtractor obtains the highest speeds. No tool has sufficient speed and accuracy to be well-suited to large-scale automated segmentation in its current form.}
	{}
	
	
	\maketitle

	\section{Introduction}
	Segmentation maps -- images showing which pixels in an image belong to which source -- are used extensively to preprocess observational data for analysis. They are used for masking sources, estimating sky backgrounds, and creating catalogues, amongst other applications. It is therefore essential that the tools used to create these maps are accurate and reliable. Otherwise, the later scientific process may be invalidated by errors in the measurements of sources.
	
	Unfortunately, astronomical images have many properties which cause problems for traditional image segmentation algorithms. Images may be highly noisy, and have an extremely large dynamic range. Objects generally have no clear boundaries, and their outer regions may extend below the level of background noise. As many generic segmentation algorithms are edge-based \citep{pal1993review,wilkinson98:_autom}, they are unable to accurately process these images.
	
	In addition, with the growth of the scale of astronomical surveys, there is an increased need for a fast and accurate tool for segmentation. This is illustrated by current projects such as the Legacy Survey of Space and Time (LSST), which aims to produce around 15TB of raw data per night \citep{ivezic2008lsst}. With surveys of this scale, human intervention will no longer be feasible, meaning that the tools should ideally be robust to variations in images without manual tuning.
	
	Because of these unique challenges, a number of tools have been developed for the sole purpose of accurately detecting sources in astronomical images. The most well-known of these for optical data is SExtractor \citep{bertin1996sextractor}. In recent years, however, a number of alternatives have been proposed, including ProFound \citep{robotham2018profound}, NoiseChisel \citep{akhlaghi2015noise}, and MTObjects \citep{teeninga2013bi,teeninga16:_statis}.
	
	In this paper, we evaluate and compare these segmentation tools, in order to study their strengths and weaknesses. A thorough comparison provides a means for astronomers to choose which algorithm is best suited for their scientific goals. In addition, several of these tools are still under active development, and such an analysis can help to direct future advancements.
	
	For this comparison, we develop numerical measures for segmentation quality (Sect. \ref{sec:metrics}), and propose a method for automatic configuration of tool parameters (Sect. \ref{sec:opt}). This approach to evaluating segmentation maps aims to provide an objective measure of quality. To do this, we use simulated images, with a known ground truth (Sect. \ref{sec:data}), to provide evaluations which are not dependent on visual biases and preconceptions. We supplement our results by demonstrating the performance of our automatically configured parameters on real survey images (Sect. \ref{sec:real}).
	
	\begin{figure}
		\centering
		\includegraphics[width=\columnwidth]{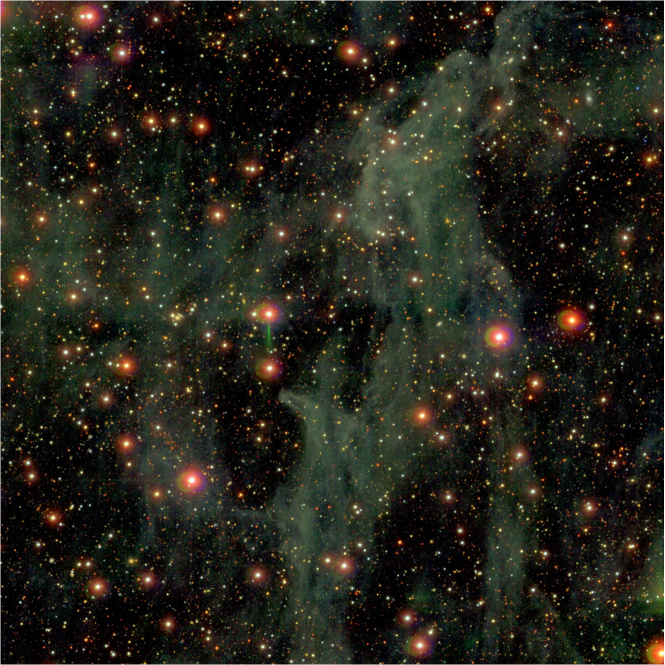}
		\caption{A \textit{gri}-composite image of IAC Stripe82 field \texttt{f0363\_g.rec.fits} showing a large structure of galactic cirri. Such complex, overlapping structures are challenging for source detection tools.}  
	\end{figure}
	
	Throughout this paper we use the terms `segmentation', `source detection', and `source extraction' interchangeably, to refer to the process of identifying unique sources in astronomical images and marking the pixels of the image in which each source is the dominant contributor.
	
	\section{Source extraction methods}
	\subsection{Previous methods}
	For as long astronomical images have been produced, it has been necessary for their contents to be catalogued and measured, in order that they may be used for scientific applications. As manually locating and outlining objects is a slow and subjective process, particularly when considering the faint outskirts of objects, many attempts have been made at automating this process.
	
	Early automatic tools directly scanned photographic plates to locate sources and produce measurements. Notable is COSMOS \citep{pratt1977cosmos}, which used a process of repeated thresholding to produce `coarse measurements' of images -- essentially quantising the image over an estimated local background level. It then used `fine measurements' to produce more accurate measurements of objects' profiles. Later additions included improved deblending of adjacent sources \citep{beard1990cosmos}.
	
	Whilst modern tools no longer use digitised photographic plates, instead working directly with data captured by CCDs, the overall process used in recent tools are fundamentally very similar to those used in their predecessors. Almost all tools follow the same four main steps:
	
	\begin{enumerate}
		\item Identify and measure the background level.
		\item Threshold the image relative to the background.
		\item Locate (and deblend) sources appearing above the threshold.
		\item Produce a catalogue of sources and their measured properties.
	\end{enumerate}
	
	SExtractor, described in more detail below, uses a very similar method of repeated thresholding to COSMOS. In contrast, several other tools make use of dendrograms -- hierarchical representations of images, in which nodes representing local maxima are connected at the highest brightness level where thresholding would show a single, unbroken object. Users may subsequently `prune' the dendrogram by removing nodes connecting very small or faint regions, and may automatically or manually mark objects meeting some criteria. Dendrograms have been used to visualise and analyse hierarchical structure in both infrared images \citep{houlahan1992recognition} and radio data cubes \citep{rosolowsky2008dendrograms, goodman2009role}.
	
	Other tools have deviated from a thresholding-based approach. Many of these tools and their methods are described in \citet{masias2012review}.
	
	\subsection{Deblending}
	Deblending -- the process of separating overlapping or nested sources -- is closely linked to source extraction; all of the tools we discuss in this paper make some attempt at deblending. However, for some scientific purposes, the tools do not produce sufficiently accurate separation of sources, leading to problems such as poor photometry \citep{abbott2018dark, huang2018characterization}, and systematic measurements of physical properties such as redshift \citep{boucaud2020photometry} and cluster mass \citep{simet2015background}. Consequently, several tools also exist to perform deblending as a separate process.
	
	As these tools are predominantly either designed to use the results of another source extraction tool (such as SCARLET\citep{melchior2018scarlet}, which uses SExtractor for initial source detection), or are predominantly designed for smaller images with only a few galaxies (such as the machine learning-based methods proposed in \citet{reiman2019deblending}), we chose to not include them in the comparisons in this paper. However, the evaluation process we define in section \ref{sec:method} could equally be applied to compare deblending-specific tools in future work.
	
	\subsection{Compared tools}
	We chose to focus our comparison on four tools -- SExtractor \citep{bertin1996sextractor}, which is in common use, and three recent alternatives: ProFound \citep{robotham2018profound}, NoiseChisel + Segment \citep{akhlaghi2015noise}, and MTObjects \citep{teeninga2013bi,teeninga16:_statis}. We chose to exclude several other source extraction tools from this comparison for various reasons -- notably DeepScan \citep{prole2018automated}, which is dependent on the use of another tool (such as SExtractor) to produce an initial mask; and AstroDendro \citep{astrodendro}, which was prohibitively slow to run on large images.
	
	\subsubsection{SExtractor}
	SExtractor \citep{bertin1996sextractor} is a widely-used tool for the creation of segmentation maps. It was developed with the goal of producing catalogues of astronomical sources from large scale sky surveys.
	
	The first step in the SExtractor pipeline is the estimation and subtraction of the background. The image is divided into tiles, and a histogram is produced for each. Values more than 3 standard deviations from the median are removed. Tiles are then classified into crowded and uncrowded fields based on the change in histogram distribution, and a background value is estimated based on each tile's median and mode.
	
	The image is then thresholded at an fixed number of exponentially spaced levels above a user-defined threshold. This converts the light in the image into trees, with branches representing bright areas within larger, fainter objects. Pixels in branches which contain at least a given proportion of the light of their parent objects are marked as individual objects, whilst branches containing a lower amount of light are regarded as part of the parent object. Pixels in the outskirts of objects are allocated labels based on the probability that a pixel of that value is present at that point, using profiles fitted to the detected sources. 
	
	In practice, SExtractor may be used in multiple passes, particularly when detecting extended sources. For example, a hot/cold method may be used, wherein a sensitive pass captures the outskirts of objects, and a less sensitive pass identifies which objects are not false positive detections \citep{rix2004gems}.  It may also be used to identify candidate objects, which are then manually verified. 
	
	SExtractor version 2.19.5 was used for this comparison, using the default filter -- a convolution with a $3\times3$ pyramidal function, which approximates Gaussian smoothing. We found in subsequent testing, described in Appendix \ref{filterapp}, that using a 9x9 Gaussian PSF with a full width at half maximum of 5 pixels produced marginally better results, although this difference was not significant, and did not affect the general conclusions of this paper.
	
	\subsubsection{ProFound}
	ProFound \citep{robotham2018profound}, like SExtractor, was designed as a general purpose package for detecting and extracting astronomical sources; however, it aims to produce a more accurate segmentation, which may be used for galaxy profiling.
	
	Instead of using multiple thresholds, ProFound uses a single threshold after the background estimation stage in order to demarcate pixels containing sources. These pixels are then processed in descending order of brightness, with a watershed process being used to allocate less bright neighbouring pixels (within some tolerance) to the object of the brightest pixel in a region, until all pixels bordering the object are either allocated to other objects, are marked as background, or have higher flux than neighbouring pixels within the object. 
	
	Following this process, the background is re-estimated, and an iterative process of calculating photometric properties of the segments and repeatedly dilating them is performed, to produce a final segmentation map. ProFound version 1.1.0 was used for this comparison.
	
	\subsubsection{NoiseChisel + Segment}
	NoiseChisel \citep{akhlaghi2015noise} was designed with the goal of finding `nebulous objects', such as irregular or faint galaxies, accurately. NoiseChisel is intended to be hand-tuned for individual images -- the tutorial states that that configurations are `not generic' \citep{nctutorial}.
	
	NoiseChisel separates the image into areas containing light from objects, and areas containing only background. To do this, it uses a threshold below the estimated background level, and performs a series of binary morphological operations to create an initial detection map. Further morphological operations are then performed on the `objects' and `background' separately, and area and signal-to-noise thresholds are used to remove false detections. Segment then produces a map of  `clumps' by locating connected regions around local maxima in the image with a watershed-like process. It then discards those that do not meet a signal-to-noise threshold, and grows the remaining clumps to create a final segmentation map.
	
	Since the publication of the original paper, the program has been split into two separate tools within the GNU Astronomy Utilities package: NoiseChisel, and Segment. For the purposes of this comparison, the tools are treated as a single pipeline, and evaluated together -- we examine only this final `objects' output. The latest version as of the start of the comparison was used -- version 0.7.42a. Several new versions have since been released, which may contain different parameters and produce different results.
	
	\subsubsection{MTObjects}
	MTObjects \citep{teeninga2013bi,teeninga16:_statis} takes a similar approach to SExtractor; both operate on the principle that after a background subtraction step, objects can be detected by a thresholding process. However, where SExtractor uses a small number of fixed thresholds, MTObjects uses tree-based morphological operators. 
	
	A max-tree \citep{salembier1998antiextensive} is constructed from the smoothed and background-subtracted image. The max-tree is a tree of the image: the leaves represent local maximum pixels, nodes represent increasingly large connected areas of the image, with decreasing minimum pixel values, and the root represents the entire image. This tree is then filtered, by using tests to determine which nodes of the tree -- or areas of the image -- contain an amount of flux, given their area, that is statistically significant relative to their background. If a node has no significant parent, or its parent has another child with greater flux, it is marked as an object. Despite representing all connected components at all grey levels in the image, building the max-tree is very efficient ($O(N \log N)$ typically for floating point images \citep{carlinet14:_compar_review_compon_tree_comput_algor}). 
	
	The max-tree structure used in MTObjects is very similar to the dendrogram, used in several astronomical applications as described above \citep{houlahan1992recognition, rosolowsky2008dendrograms}. There are two main differences. Firstly, the dendrogram only contains nodes where areas connect, whereas the max-tree contains a node for every difference in brightness value. Secondly, MTObjects uses a single statistical significance test to detect objects, combining multiple attributes of the node, whilst the dendrogram methods frequently filter small and faint objects at fixed thresholds.
	
	There have been no official software releases of MTObjects -- we used a Python and C implementation,\footnote{https://github.com/CarolineHaigh/mtobjects} which we adapted from the software used in the original paper. We used significance test 4 as recommended by \citet{teeninga16:_statis}.
	
	\section{Methodology}
	\label{sec:method}
	\subsection{Data}
	In this section we describe the data upon which we tested the tools. Simulated data (Sects. \ref{sec:simdata}, \ref{sec:simgt}) allows us to accurately quantify performance on a simplified version of the problem, whilst survey images (Sect. \ref{sec:realdata}) allow us to qualitatively explore behaviour in a range of different real situations.
	
	\label{sec:data}
	\subsubsection{Simulated data}
	Testing source detection algorithms on real observational data has several limitations. Firstly, the ground truth is not known -- even if objects have been manually labelled, it is possible that objects have been missed, or incorrectly measured. In particular, it is difficult to establish the true extent of objects at a low brightness level, as their outer regions may not be clearly visually distinguishable from background.
	
	Secondly, many features of interest -- such as ultra-diffuse galaxies -- are not thought to be very common. This means that it is difficult to establish a statistical measure of how accurately they can be detected. As these objects are also more difficult for algorithms to detect, a larger sample is required to determine the accuracy of the algorithms. 
	
	By using simulated data, we gain the ability to test the algorithms on large datasets with a known ground truth. This means that we can make accurate measures of precision and accuracy for faint features, as well as taking into account the true extent of objects. We can also measure accuracy of algorithms in different controlled conditions, such as with high noise, background variation, and overlapping sources.
	
	We created ten frames of data emulating images in the r'-band of data in the Fornax Deep Survey (FDS). This is a deep, medium-sized ground based survey of the nearby Fornax cluster, which is located at the distance of 20 Mpc \citep{iodice2016, venhola2018fornax}. Each simulated image contained approximately 1500 `stars', 4000 `cluster galaxies' and 50 `background galaxies'. Stars were simulated as point sources, and galaxies as S\'ersic models \citep{sersic1968atlas}. The number and structural parameters of the stars and galaxies were drawn from distributions similar to those found in the FDS. In the simulated images, stars have magnitudes between 10 and 23 mag, and galaxies have mean effective surface brightnesses between 21 mag arcsec$^{-2}$ and 31 mag arcsec$^{-2}$. Background galaxies have effective radii between 0.5 and 3.5 arcsec and S\'ersic indices between 2 and 4. Cluster galaxies have effective radii between 2.5 and 40 arcsec, and S\'ersic indices between 0.5 and 2. Axis ratios varied from 0.3 to 1.0. To replicate observation conditions, images were convolved with the $r$-band point spread function of the OmegaCAM, and Poissonian and Gaussian noise were added \citep{venhola2018fornax}. For further details of the process, see \citet[Chapter 5]{venhola19:PhDthesis}.

	\begin{figure}
		\centering
		
		\begin{subfigure}[b]{0.48\columnwidth}
			\includegraphics[width=\columnwidth]{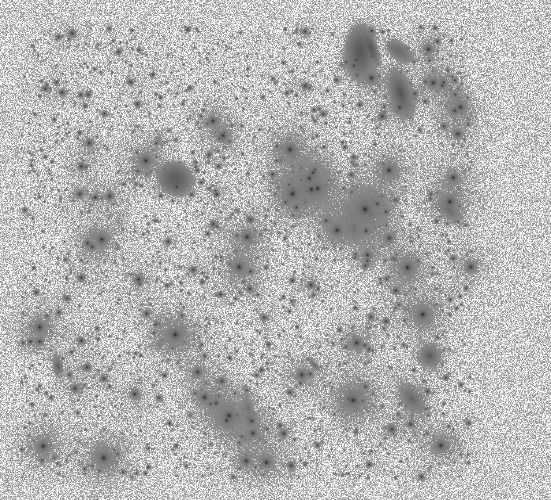}
		\end{subfigure}
		\begin{subfigure}[b]{0.48\columnwidth}
			\includegraphics[width=\columnwidth]{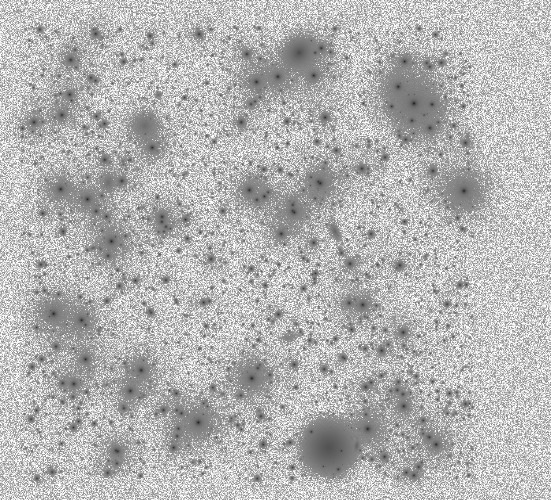}
		\end{subfigure}
		
		\begin{subfigure}[b]{0.48\columnwidth}
			\includegraphics[width=\columnwidth]{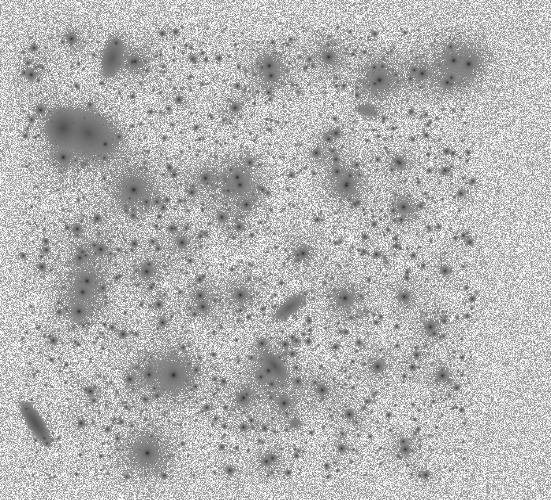}
		\end{subfigure}
		\begin{subfigure}[b]{0.48\columnwidth}
			\includegraphics[width=\columnwidth]{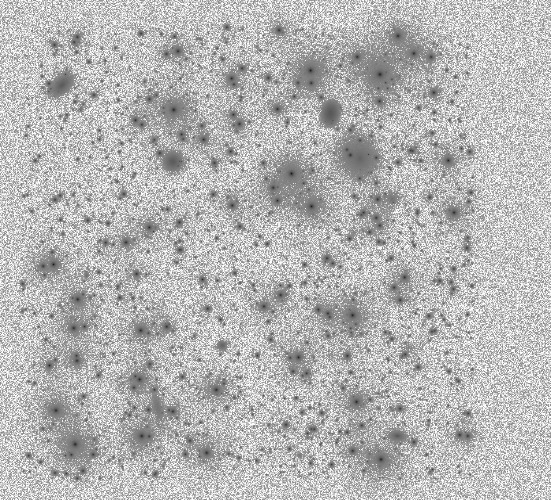}
		\end{subfigure}
		
		\caption{Simulated survey images}
		\label{simages}
		
	\end{figure}
	
	\label{sec:simdata}
	
	\subsubsection{Choosing a ground truth}
	Astronomical sources have no clear boundary; their light merely becomes insignificant in relation to noise and background light at some point in their outskirts. This means that when we create a ground truth for simulated images -- a `correct' segmentation map -- we need to choose a threshold, $t$, below which we judge light from sources to be undetectable. Assuming a flat background, this threshold can be expressed as a sum of the background level, $bg$, and some multiple, $n$, of the standard deviation of the noise, $\sigma$:
	
	$$t = bg + (n * \sigma)$$
	
	Sources may also overlap, meaning that each pixel contains light from multiple sources. In segmentation maps, each pixel is allocated to a single source; therefore, it is necessary to determine which source has the strongest relationship with a given pixel. It should be noted that whilst segmentation maps are the traditional method of demarcating sources within an image, they are limited by their inability to represent the reality that pixels contain light belonging to multiple sources \footnote{A new data format would be required to clearly represent this nested data. This could prove to be a challenging problem, due to the complexity of allocating multiple labels and proportional brightnesses to each pixel}. Consequently, tree-based methods, which inherently model nested objects, are unable to capture this structure within segmentation maps. As such, information contained in the models is lost, and not measured in the evaluation.
	
	We initially considered allocating each pixel to the source which contributed the most flux to it. However, this meant that fainter sources in the vicinity of bright sources were entirely erased, as they had a lower raw flux contribution.
	
	Instead, we chose to allocate labels based on a combination of the importance of the pixel to the source and the importance of the source to the pixel. For a source with total flux $F_{s}$, contributing a flux $f_{s,p}$ to a pixel with total flux $F_{p}$, the pixel contains $\frac{f_{s,p}}{F_{s}}$ of the source's flux. Conversely, the source contributes $\frac{f_{s,p}}{F_{p}}$ of the light contained within the pixel.
	
	These measures may be combined to give a single measure:
	
	$$\frac{f_{s,p}}{F_{s}} \times \frac{f_{s,p}}{F_{p}}$$
	
	When allocating a pixel to a source, $F_{p}$ will be constant for all sources contributing to the pixel. Therefore, the pixel may be allocated to the source with the highest value for
	
	$$\frac{(f_{s,p})^{2}}{F_{s}}$$ 
	
	and $f_{s,p} \geq t$
	
\begin{figure*}
		\centering
		\begin{subfigure}[b]{0.35\textwidth}
			\includegraphics[width=\columnwidth]{figs/sim/cluster3_0.jpg}
			\caption{Simulated image}
		\end{subfigure}
		\begin{subfigure}[b]{0.35\textwidth}
			\includegraphics[width=\columnwidth]{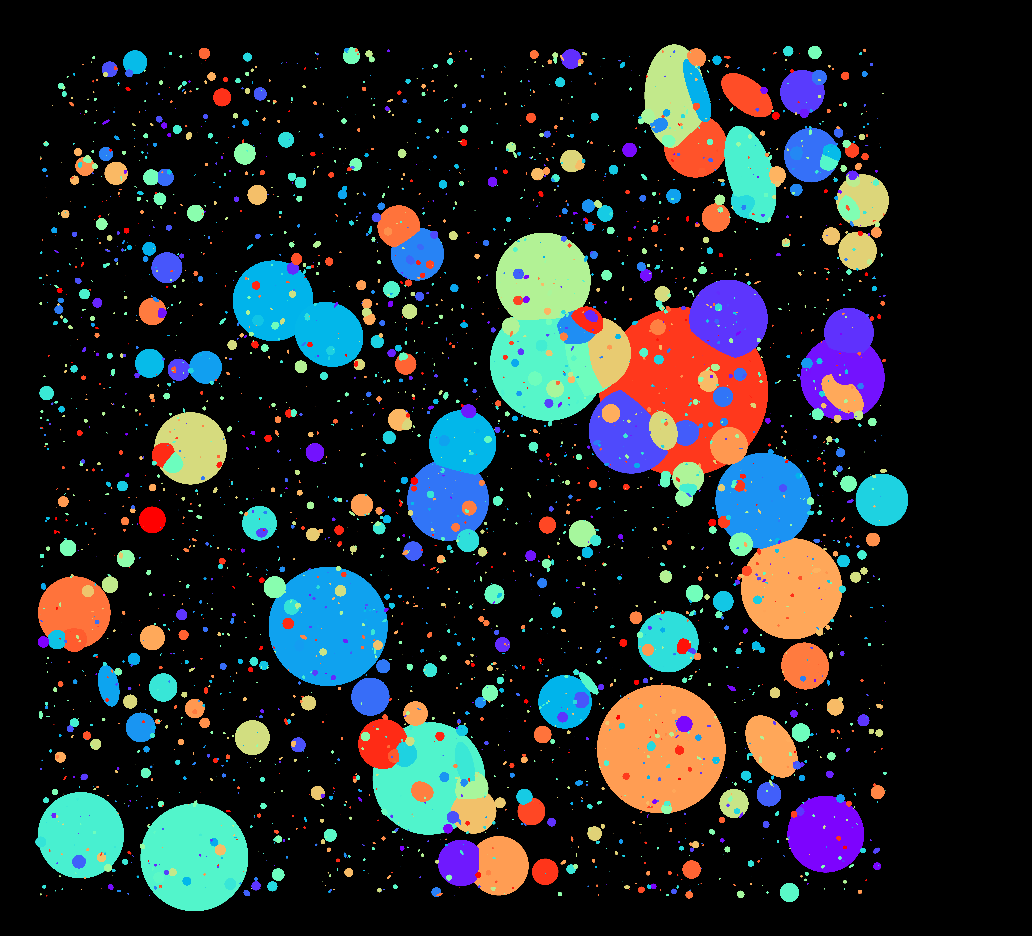}
			\caption{Ground truth (1.0 $\sigma$)}
		\end{subfigure}
		
		\begin{subfigure}[b]{0.35\textwidth}
			\includegraphics[width=\columnwidth]{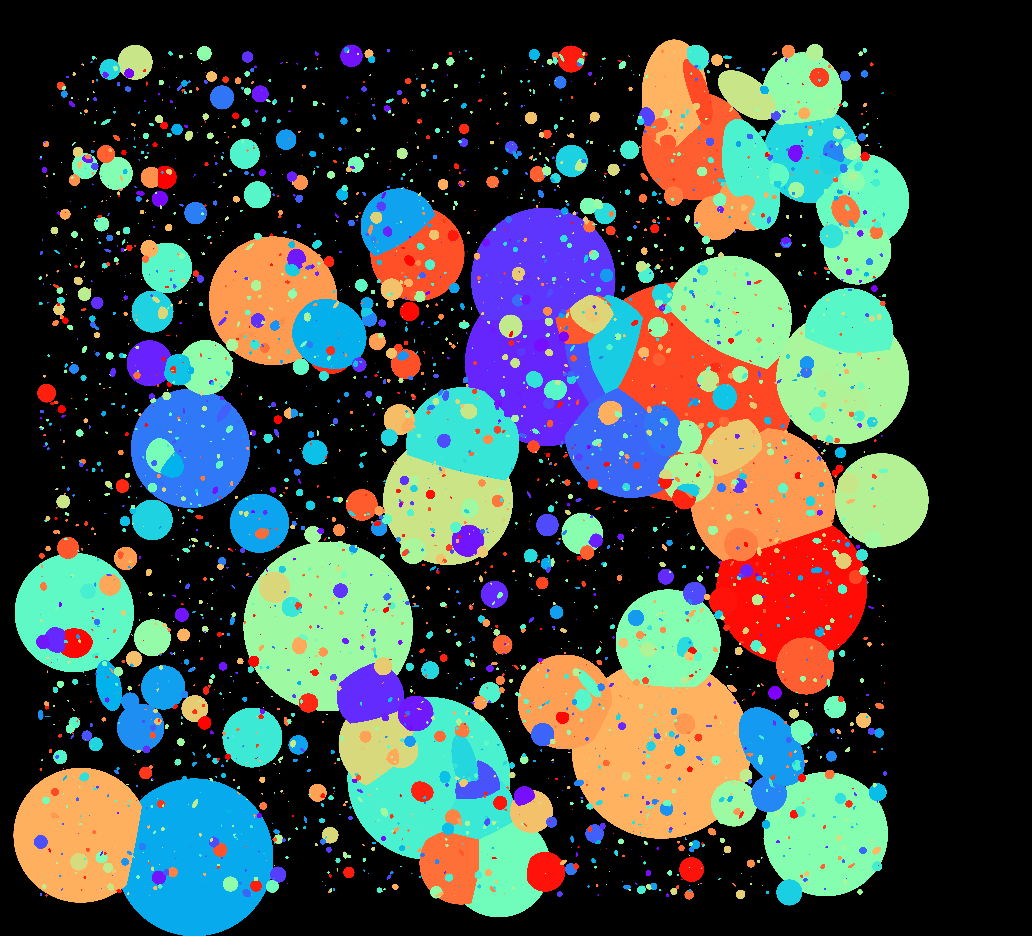}
			\caption{Ground truth (0.5 $\sigma$)}
		\end{subfigure}
		\begin{subfigure}[b]{0.35\textwidth}
			\includegraphics[width=\columnwidth]{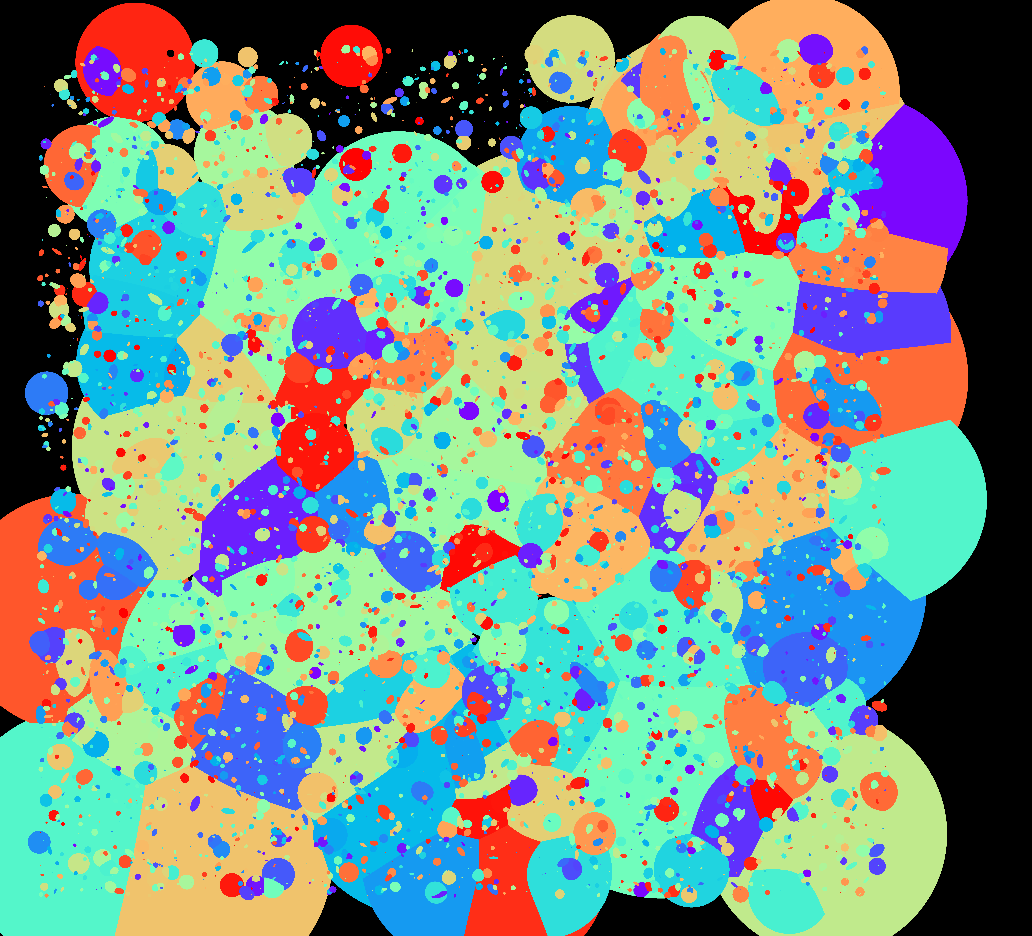}
			\caption{Ground truth (0.1 $\sigma$)}
		\end{subfigure}
		
		\caption{Ground truth segmentations of a simulated image, with a varying threshold ($n * \sigma$). The coloured regions label distinct objects, and the black regions make up the background.}
		\label{groundtruths}
	\end{figure*}

	The value of $n$ has a substantial effect on the areas allocated to objects, as shown in Fig. \ref{groundtruths}. Consequently, it has an large impact on the evaluation of the segmentation maps produced by the tools.
	
	As we wished to evaluate the performance of the tools at levels of low surface brightness, we chose to use a value of $n=0.1$ for our ground truths. This pushes the tools to optimise their parameters to capture and correctly allocate as much of the light in the images as possible.
	
	\label{sec:simgt}
	
	\subsubsection{Real-world data}
	Whilst testing algorithms on real-world data has several limitations, as discussed above, it is nevertheless essential. It allows us to subjectively evaluate performance on structures and conditions which cannot be easily simulated, such as streams, spiral galaxies, and unusual artefacts.

	With this in mind, we selected a number of images in the optical which contained examples of these features. We chose images from the Fornax Deep Survey \citep[FDS;][]{iodice2016, venhola2018fornax}, IAC Stripe 82 Legacy Project\footnote{\protect\url{http://research.iac.es/proyecto/stripe82/}} \citep[hereafter IAC Stripe82;][]{fliri2016s82, roman2018s82} which is a 2.5 degree stripe ($-50^{\circ} <$ R.A. $< 60^{\circ}, -1.25^{\circ} <$ Dec $< 1.25^{\circ}$) with a total area of 275 square degrees in all the five Sloan Digital Sky Survey (SDSS) bands and the Hubble Ultra Deep Field \citep[HUDF;][]{beckwith2006hubble}, a 11\,arcmin$^2$ region in the Southern Sky.  As the simulated images were designed to mimic the FDS, using real images from this survey allowed us to test the optimised parameters with similar imaging conditions, where they would be expected to perform well. The additional use of IAC Stripe82 and HUDF images allows us to examine the consistency of parameters on images with very different imaging conditions.\par
	
	While the FDS and IAC Stripe82 are deep surveys using ground-based telescopes, the VLT Survey Telescope (VST) and the SDSS Telescope respectively, the well-studied HUDF extends our analysis to the higher resolution, space-based data from the \textit{Hubble Space Telescope}. In terms of depth, the HUDF is the deepest with a point source depth of $\sim 29$\,mag which corresponds to a surface brightness limit of $\mu_{V_{606}} \sim 32.5$\,mag/arcsec$^2$\ in the $V_{606}$-band, computed as a $3\sigma$ fluctuation with respect to the background of the image in $10\times10$\,arcsec$^2$ boxes ($3\sigma;10\times10$\,arcsec$^2$). The FDS images in the SDSS $r$-band have a limiting depth of $\mu_r \sim 29.8$\,mag/arcsec$^2$\,($3\sigma;10\times10$\,arcsec$^2$) and the IAC Stripe82 survey is $\sim 1$\,mag shallower than FDS with a limiting surface brightness depth of $\mu_r \sim 28.6$\,mag/arcsec$^2$\,($3\sigma;10\times10$\,arcsec$^2$) and $\mu_g \sim 29.1$\,mag/arcsec$^2$\,($3\sigma;10\times10$\,arcsec$^2$). In order to select the deepest imaging from all these surveys in the optical regime, we use the $V_{606}$-band images in the HUDF and the SDSS $r$- and $g$-band images from FDS and IAC Stripe82 respectively. \par 
	
    Additionally, the FDS, IAC Stripe 82 and HUDF datasets collectively represent deep data with different surface brightness depths and spatial resolutions: FDS is $>1$\,mag deeper and two times higher in spatial resolution than IAC Stripe82 (0.2\,arcsec/pixel resolution in FDS (rebinned from the 0.21 arsec/pixel of the VST) compared to 0.396\,arcsec/pixel in SDSS) and the HUDF is $>2$\,mag deeper than FDS, with the best resolution currently possible from space $\sim 0.05$\,arcsec/pixel. Therefore, the optimised parameters of each algorithm are tested on real images with varying depth and resolution. However, in this work we specifically chose images in the optical wavelengths to test the limits of current detection algorithms for upcoming deeper and wider surveys such as LSST. In future work, a similar analysis to that presented herein can readily be extended to other wavelengths.
    
    \label{sec:realdata}
    
	\subsection{Parameter optimisation}
	\label{sec:opt}
	To produce a fair comparison of the algorithms' capabilities, they should be tested with parameters that are as close to optimal as possible. Due to the extremely large parameter spaces of some of the tools, it was not feasible to manually optimise the tools, or to test every possible combination of parameters.
	
	We therefore chose to use an automatic method to select good parameters for each tool. We initially considered use of a genetic algorithm for this purpose; however, this proved to be prohibitively slow, as a high number of time-consuming runs of each tool was required. Instead, we used Bayesian optimisation.
	
	\subsubsection{Bayesian optimisation}
	Bayesian optimisation is a method of black-box optimisation which is well-suited for functions which take a long time to evaluate \citep{jones1998efficient}. It operates by creating a model of how the function behaves, identifying the regions in parameter space where it may perform well or where it may not be well-fitted, and choosing points in these regions to evaluate, in order to improve the model.
	
	In the context of source extraction tools, the input takes the form of a set of relevant parameters, as dictated by each tool's documentation. The parameters are evaluated by running the tool on a training image, comparing the output to a known ground truth, and choosing a metric (as detailed in Sect. \ref{sec:metrics}) as the output score to optimise.
	
	We used the GPyOpt optimisation library \citep{gpyopt2016} to perform the optimisations. For each metric, each tool was optimised on every image individually, and the found parameters were then applied to all of the remaining images, to assess their performance. The tools' default parameters were used as a starting point. 120 evaluations were performed on each image in batches of four, using the local penalisation method, and the best set of parameters was chosen.
	
	\subsection{Metrics}
	\label{sec:metrics}
	The quality of a segmentation can be measured both in terms of the presence and absence of ground truth objects, and the similarity between the true objects and segmented shapes.
	
	\subsubsection{Matching detections}
	When measuring detection rates, it is necessary to match detected objects with ground truth objects. It may be the case that a detected object covers the area of multiple true objects, or conversely that multiple detected objects are found within the area of a single true object. Therefore, a one-to-one mapping is required, in order to prevent algorithms from being rewarded for failing to correctly distinguish between sources.
	
	We chose to use the brightest pixel in each object as an identifier -- the detected object containing the brightest pixel in a ground truth object was matched to it. In the event that a detected object contained the brightest pixel of multiple ground truth objects, the object containing the pixel with the highest flux was chosen as a unique match.
	
	Three measures made use of this matching procedure:
	
	\begin{itemize}
		\item Detection recall (completeness) -- the proportion of objects which are detected.
		\item Detection precision (purity) -- the proportion of segments which can be matched to real objects.
		\item F-score -- the harmonic mean of precision and recall: $$F\mhyphen score = 2  \times \frac{precision \times recall}{precision + recall}$$
	\end{itemize}
	
	\subsubsection{Evaluating areas}
	
	In order to quantify the accuracy of the areas of segmented objects, we used a modified version of over-merging and under-merging scores \citep{levine1981experimental}. The under-merging score measures the extent to which objects which should be a single segment are broken into multiple pieces by the segmentation tool. The over-merging score measures the opposite -- the extent to which multiple objects are incorrectly combined into a single segment by the tool. Combining these scores gives a measure of the overall quality of the segmentation.
	
	In the original method, the ground truth segmentation is divided into N segments, $R_1 ... R_N$, with areas $A_1 ... A_N$, and the test segmentation is divided into M segments, $T_1 ... T_M$, with areas $a_1 ... a_M$. The original metrics are calculated by finding $R_k$ to maximise $T_j \cap R_k$,for each test segment, $T_j$:
	
	\begin{itemize}
			\item Under-merging error (UM):
			$$UM = \sum_{j=1}^{M}\frac{(A_k - (T_j \cap R_k))(T_j \cap R_k)}{A_k}$$
			\item Over-merging error (OM):
			$$OM = \sum_{j=1}^{M} (a_j - (T_j \cap R_k))$$
	\end{itemize}
	
	In these original definitions, we found that the over-merging score did not penalise segmentations which divided large objects into many small pieces. This meant that tools could find enormous numbers of false positives, fragmenting the `background' segment, without penalty. Consequently, we chose to redefine the over-merging score to become symmetric to the under-merging score, which better takes into account the number and size of segments. We also defined an area score, which combined the two measures to give an overall score.
	
	\begin{itemize}
		\item Over-merging error (OM) -- for each reference segment, $R_k$, find $T_j$ to maximise $T_j \cap R_k$
		$$OM = \sum_{k=1}^{N}\frac{(a_j - (T_j \cap R_k))(T_j \cap R_k)}{a_j}$$
		\item Area score -- $$Area\:score = 1 - \sqrt{OM^2 + UM^2}$$
	\end{itemize}
	
	\noindent
	As the area score alone does not take into account precision and recall, we also defined two combined scores. These give us the ability to optimise for a balanced F-score and area score.
	
	\begin{itemize}
		\item Combined score A -- $$\sqrt{Area\:score^2 + F\mhyphen score^2}$$
		\item Combined score B -- $$\sqrt[3]{(1-OM) \times (1-UM) \times F\mhyphen score}$$
	\end{itemize}
	
	\noindent
	We additionally measure speed -- the rate at which images can be processed, measured in megapixels per second.
	
	\section{Results}
	Whilst the original intent was to compare all four programs on all metrics, ProFound proved to be very slow to optimise and run, making it impractical for use on large images and surveys. As such, it was optimised only on F-score and area score. Processing speeds are discussed in more detail in Sect. \ref{speedsect}.
	
	\subsection{Detection accuracy}
	
\begin{figure}
		\centering
		\begin{subfigure}[b]{\columnwidth}
			\includegraphics[width=\columnwidth]{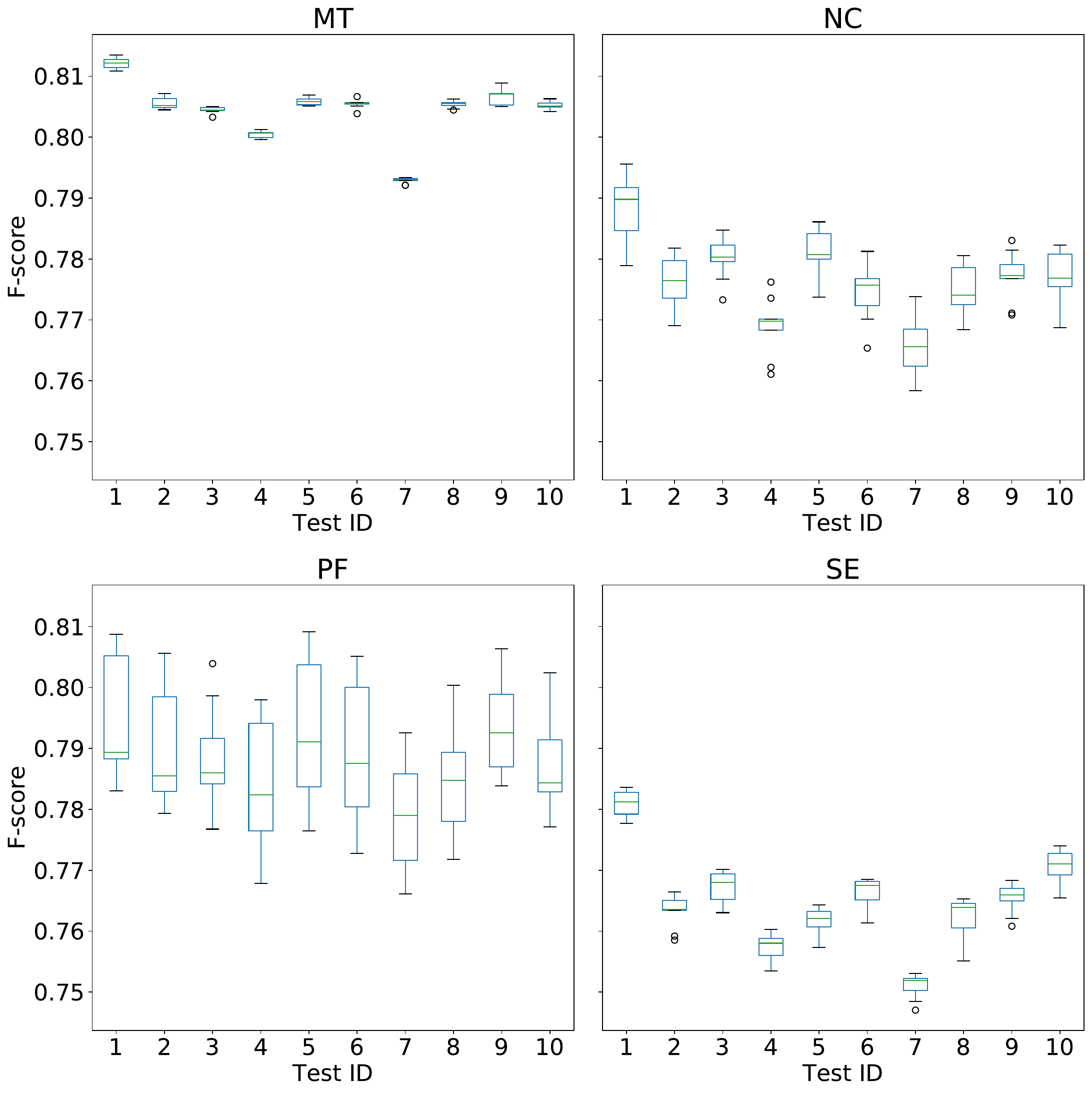}
			\caption{F-scores grouped by image evaluated}
		\end{subfigure}
		
		\begin{subfigure}[b]{\columnwidth}
			\includegraphics[width=\columnwidth]{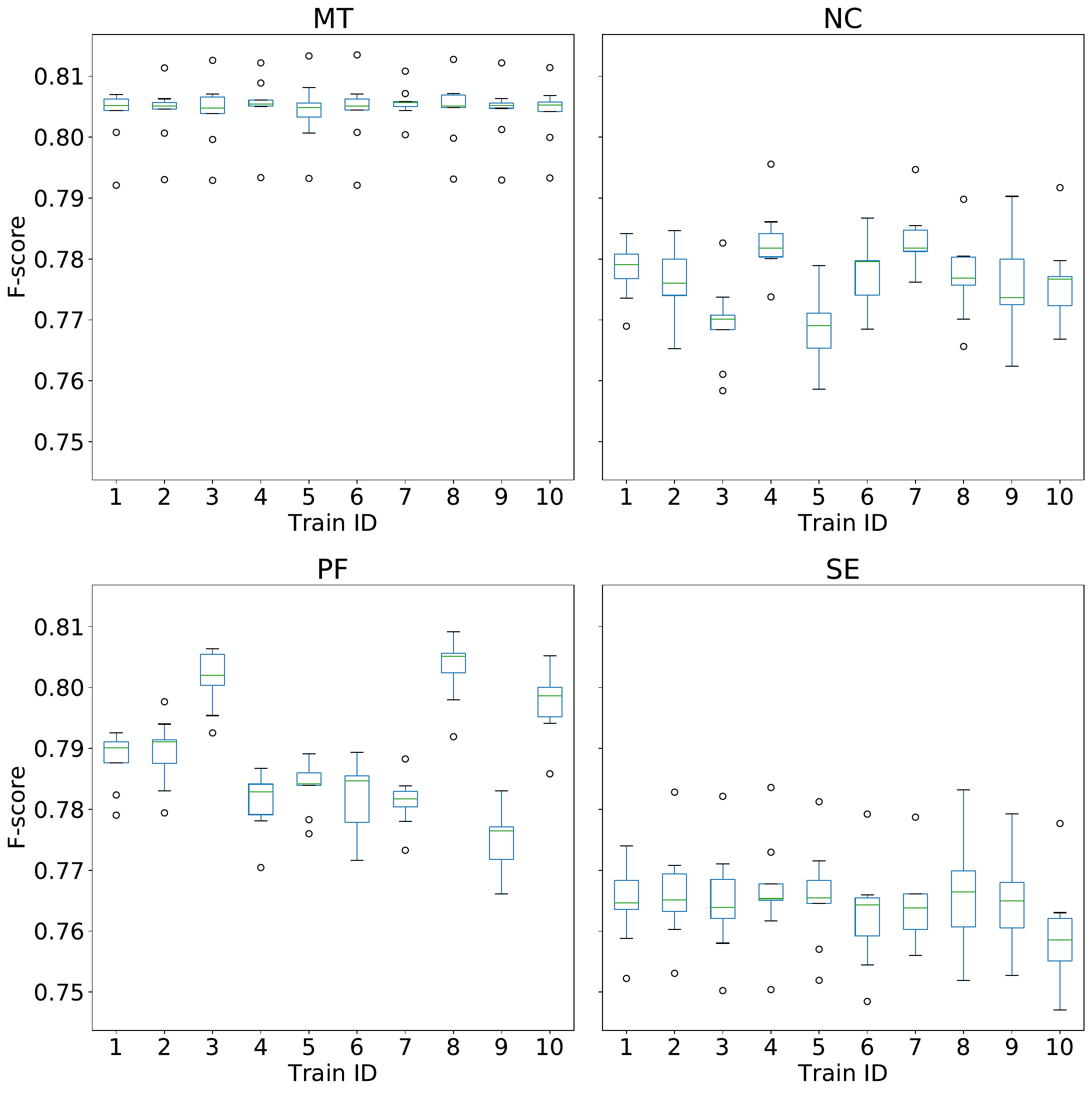}
			\caption{F-scores grouped by image used to optimise parameters}
		\end{subfigure}
		
		\caption{F-score test distributions. Each tool's parameters were optimised for F-score on each of the ten images, and evaluated on the remaining nine images. Boxes extend from first (Q1) to third (Q3) quartiles of the results, with median values marked; whiskers extend to the furthest F-score less than $1.5 * (Q3-Q1)$ from each end of the box.}
		\label{Fgraph}
		
	\end{figure}

	Figure \ref{Fgraph} shows the range of F-scores produced when each tool is optimised for F-score. Two plots are shown for each tool: one in which the scores are grouped by the image being evaluated, and one in which the scores are grouped by the training image used to optimise the parameters. The scores of the training image are excluded from both graphs.
	
	For both MTObjects and SExtractor, it is notable that the scores have smaller interquartile ranges and more varied medians when grouped by test image. This suggests that for these tools, the factor limiting the performance is the structure of each individual test image, rather than the particular parameter set chosen. In contrast, ProFound has a smaller interquartile range when scores were grouped by optimisation image, suggesting that in this case, performance is limited by the image used in the optimisation process.
	
	Overall, we see the strongest performance from MTObjects, with median scores of over 0.80 for the majority of images. The weakest performance was produced by SExtractor, with scores of under 0.78 in most cases.

	Examining the precision and recall scores that make up the F-scores shows that all programs are capable of broadly similar performance, with recall between 0.61 and 0.7 and precision greater than 0.93. Whilst the recall scores appear low, many of the faintest objects in the image are not even visible to the human eye, and may in fact be impossible to detect by any tool; these objects are included in order to fully explore the limits of the tools' capabilities. It is therefore useful to regard recall scores primarily as a relative measure, to compare the tools' performances.
	
\begin{figure}
		\centering
		\includegraphics[width=\columnwidth]{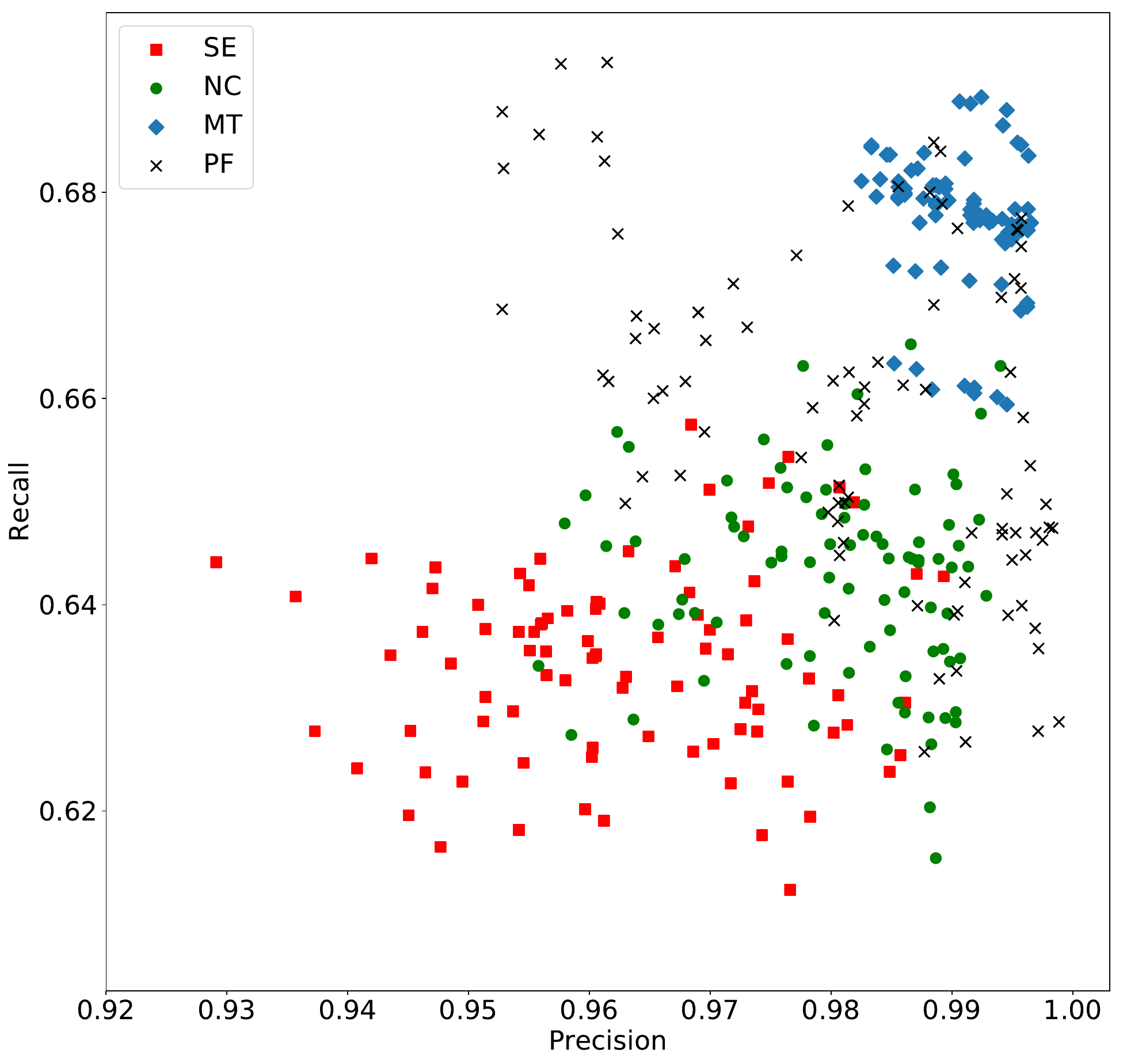}
		\caption{Precision vs recall -- each tool's parameters were optimised for F-score on each of the ten images, and evaluated on the remaining nine images.}
		\label{PRgraph}
	\end{figure}

	Differences between the programs become apparent when the scores are plotted against each other, as shown in Fig. \ref{PRgraph}. All the tools have a moderate spread of recall scores, which may be caused by differences in difficulty between the individual images.
	
	MTObjects and NoiseChisel both produce generally higher levels of precision than SExtractor; with MTObjects giving a slightly higher maximum value, and a lower spread. ProFound achieves the greatest values for both precision and recall, but has a very wide spread.
	
\begin{figure}
		\centering
		\includegraphics[width=\columnwidth]{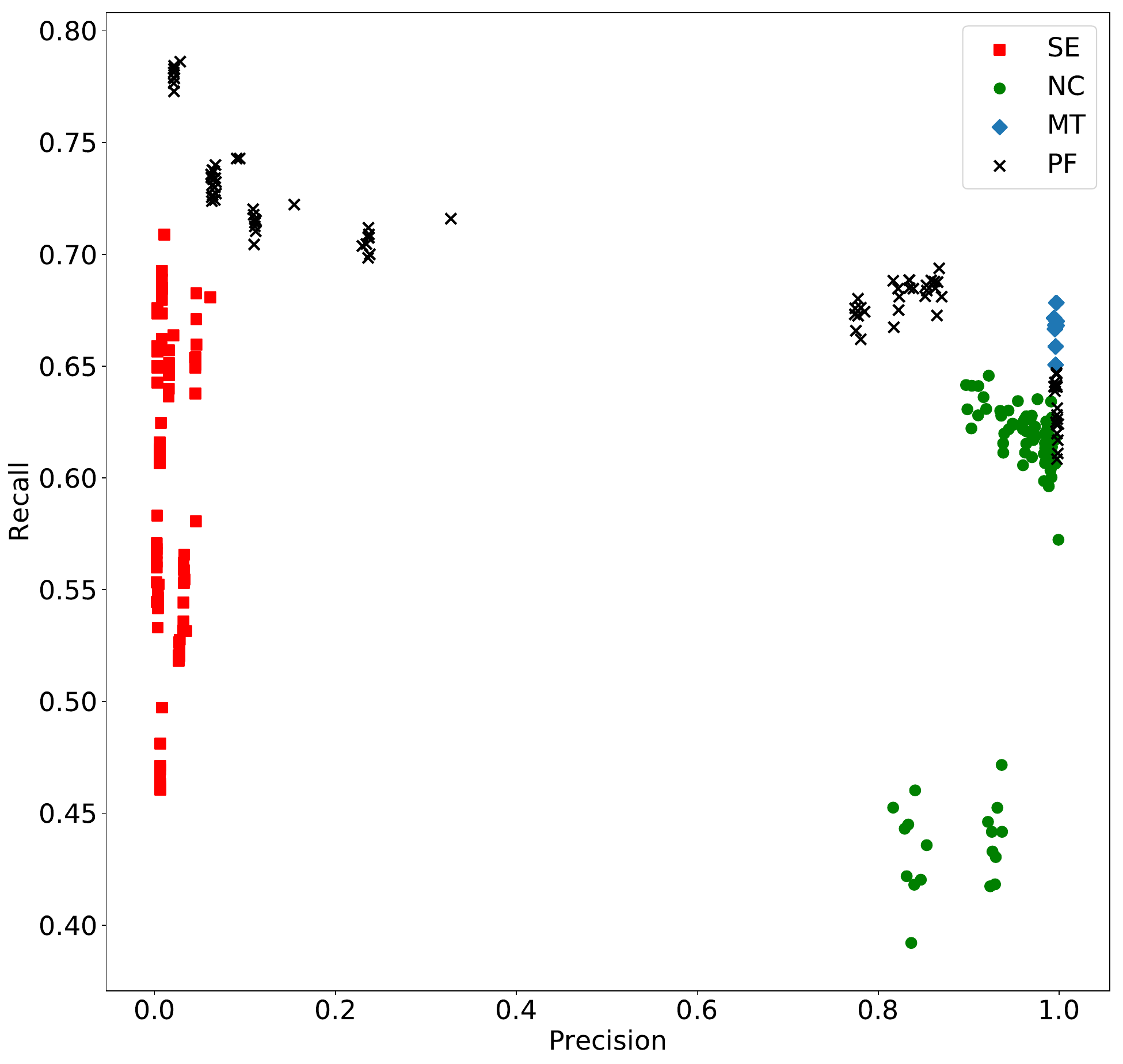}
		\caption{Precision vs recall -- each tool's parameters were optimised for area score on each of the ten images, and evaluated on the remaining nine images.}
		\label{PRareagraph}
	\end{figure}

	When optimised for area score, SExtractor performed with a substantially lower precision -- it found an enormous number of false positives, as shown in Fig. \ref{PRareagraph}. Here, we clearly see that optimising for area score is detrimental to the F-score results.  This appears to be a result of a very low threshold being selected in order to maximise the area of large shapes, meaning that a large number of small areas of noise are incorrectly marked as objects.
	
	In contrast, NoiseChisel and MTObjects were capable of increasing their area scores without compromising their F-scores substantially. ProFound performed inconsistently, covering the full range of precision scores across the ten optimisations. 
	
	\subsection{Area measures}
	Unsurprisingly, all tools were capable of reaching higher area scores when optimised for area score than F-score, as can be seen in Fig. \ref{summary}. 
	
\begin{figure}
		\centering
		\begin{subfigure}[b]{\columnwidth}
			\includegraphics[width=\columnwidth]{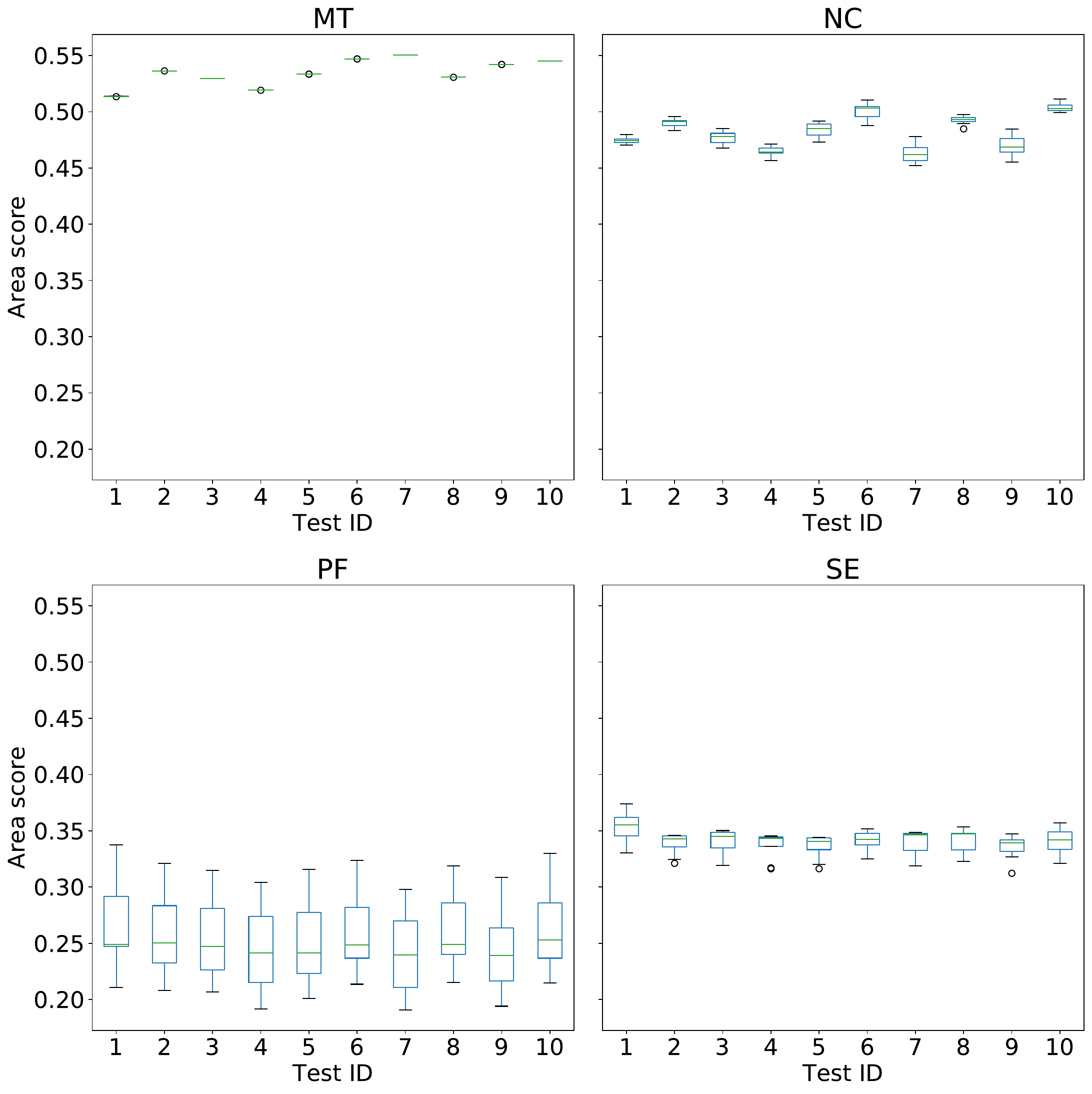}
			\caption{Area scores grouped by image evaluated}
		\end{subfigure}
		
		\begin{subfigure}[b]{\columnwidth}
			\includegraphics[width=\columnwidth]{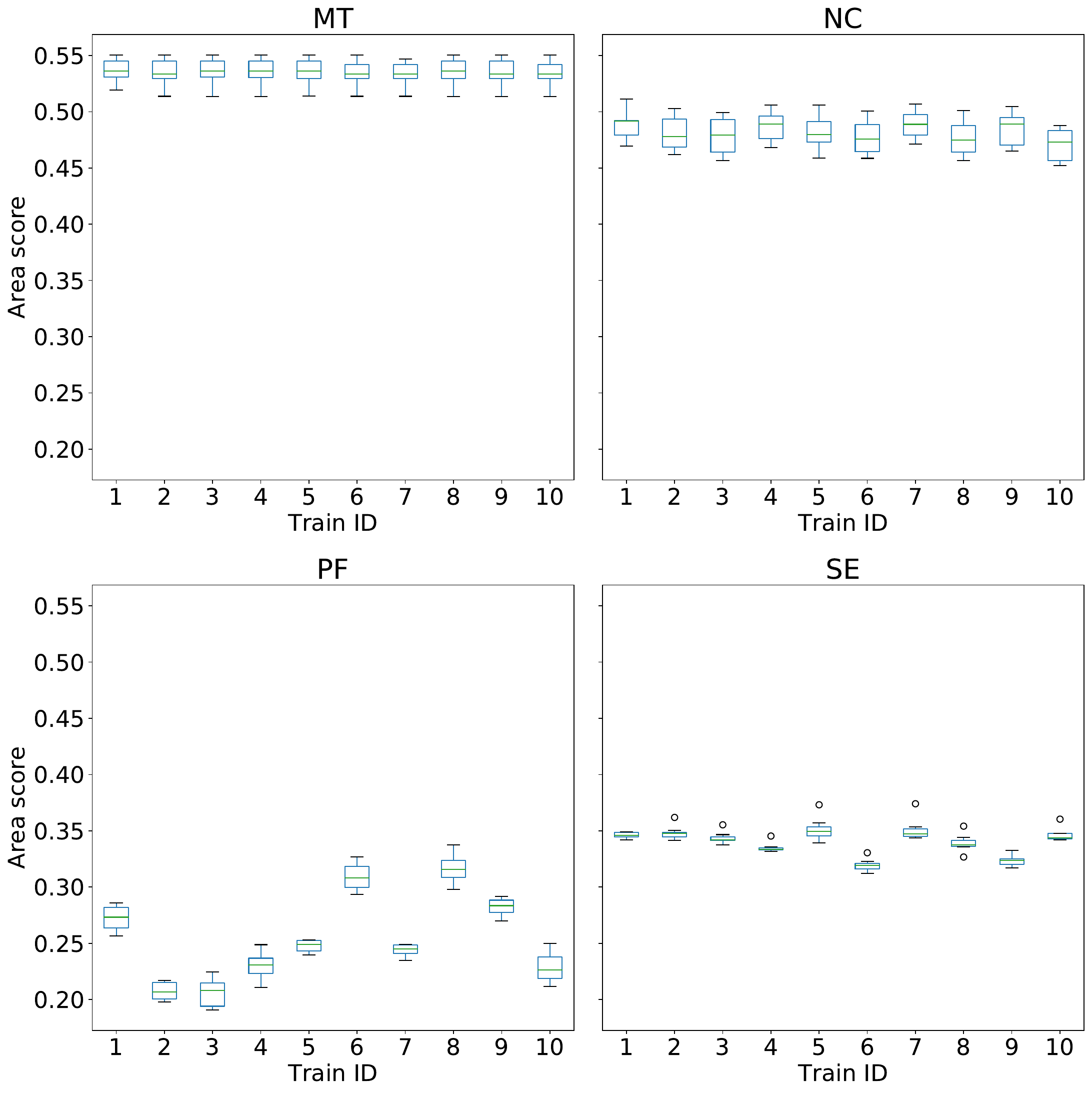}
			\caption{Area scores grouped by image used to optimise parameters}
		\end{subfigure}
		
		\caption{Area score test distributions. Each tool's parameters were optimised for area score on each of the ten images, and evaluated on the remaining nine images.}
		\label{Areagraph}
		
	\end{figure}

	When optimised for area score, NoiseChisel and MTObjects both performed well, with area scores substantially higher than the other two tools, and MTObjects having a slight edge over NoiseChisel. Both tools also showed lower variation when scores were grouped by test image, as shown in Fig. \ref{Areagraph}, suggesting that the tools' performance is being limited by the content of the test images, rather than the parameters found in the optimisation.
	
	In contrast, ProFound had much greater variability in area scores when grouped by test image, and indeed, substantial variation between the parameter sets. It also produced the weakest area scores overall. SExtractor was capable of producing higher area scores than ProFound, but at substantial cost to precision, as discussed previously.
	
	\subsection{Combined scores}
	The two combined metrics offered a way of optimising for both area and F-score, differing in the balance between the two measures. As such, optimising for these metrics gives an indication of the overall peak performance of the tools.
	
\begin{figure}
		\centering
		\begin{subfigure}[b]{\columnwidth}
			\includegraphics[width=\columnwidth]{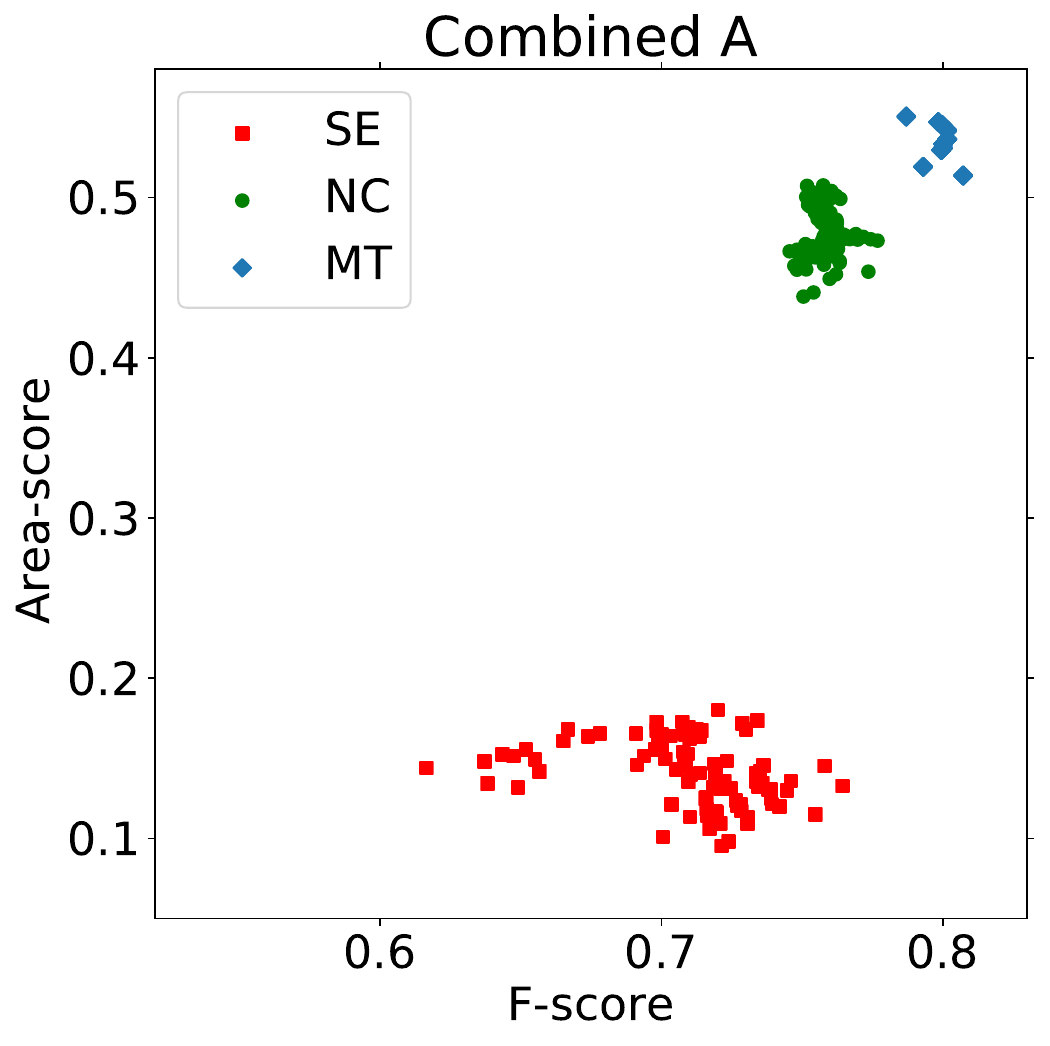}
			\caption{Optimised for Combined A}
		\end{subfigure}
		
		\begin{subfigure}[b]{\columnwidth}
			\includegraphics[width=\columnwidth]{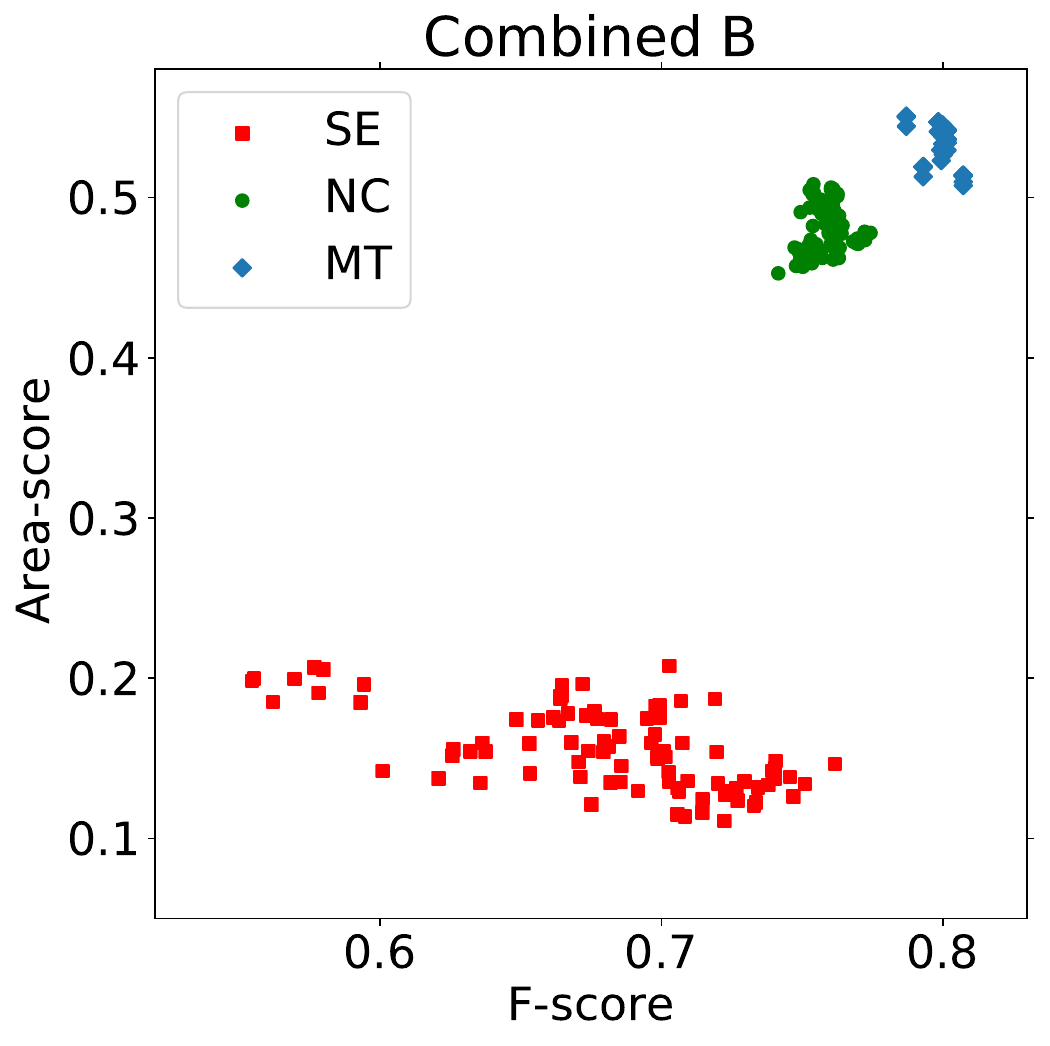}
			\caption{Optimised for Combined B}
		\end{subfigure}
		\caption{F-score vs area score -- each tool's parameters were optimised for the combined measures on each of the ten images, and evaluated on the remaining nine images.}
		\label{combo}
	\end{figure}

	In practice, both metrics produced broadly similar results in terms of both area and F-score, as shown in Fig. \ref{combo}. MTObjects produced the highest values for both F-score and area score, with NoiseChisel producing slightly lower values in both metrics. SExtractor produced lower F-scores, with a large degree of variability, and substantially lower area scores, as would be expected from its limited success when optimising purely for area. These results indicate that optimisation for combined scores prevents a large number of spurious detections being found by SExtractor, when compared to area score alone.
	
	\subsection{Overview of optimisation metrics}
\begin{figure*}
		\centering
		\includegraphics[width=\textwidth]{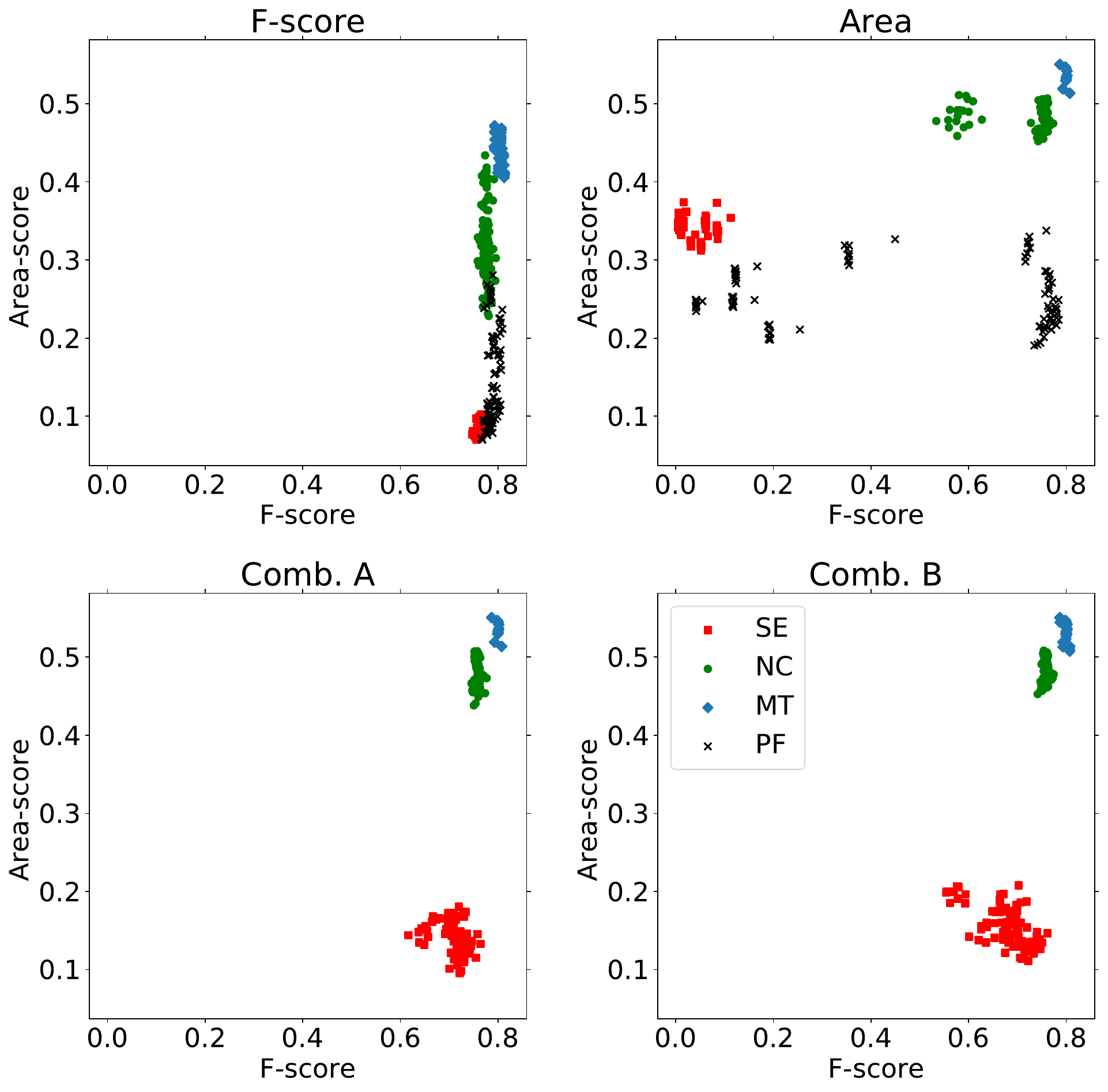}
		\caption{A summary of test scores for each program using each optimisation method. Each point represents the evaluation of the segmentation of one image using parameters found by optimising on a different image. Each plot shows results for a different optimisation metric. Note that ProFound was only optimised on F-score and area score.}
		\label{summary}
	\end{figure*}

	Figure \ref{summary} shows an overview of the results of the optimisation, in the form of scatter plots of F-score and area score. Points represent the result of evaluating the four tools on each image, using the parameters found by optimising for each metric on every other image individually. From this, we can make several observations about the tools' performance.
	
	Firstly, the tools designed specifically for locating low surface-brightness structures (NoiseChisel and MTObjects) are, unsurprisingly, capable of achieving higher area scores than the general-purpose tools. Secondly, all the tools must to some degree compromise F-score to obtain a higher area score, but this trade-off is much greater for the general-purpose tools. Thirdly, MTObjects has less spread than the other tools; indeed, it finds identical parameters and consequently produces identical results for nearly all optimisations over area or combined scores.
	
\begin{figure*}[p!]
		\centering
		\begin{subfigure}[b]{0.4\textwidth}
			\includegraphics[width=\textwidth]{figs/sim/cluster3_0.jpg}
			\caption{Original simulated image}
		\end{subfigure}
		\begin{subfigure}[b]{0.4\textwidth}
			\includegraphics[width=\textwidth]{figs/sim/sim_3_gt_01sig.png}
			\caption{Ground truth (0.1 $\sigma$)}
		\end{subfigure}

		\begin{subfigure}[b]{0.9\textwidth}
			\centering  
			\begin{tabular}{m{0.02\textwidth} m{0.2\textwidth} m{0.2\textwidth} m{0.2\textwidth} m{0.2\textwidth}}
				
				\multicolumn{1}{>{\centering\arraybackslash}m{0.02\textwidth}}{} 
				& \multicolumn{1}{>{\centering\arraybackslash}m{0.21\textwidth}}{\textbf{F-Score}}
				& \multicolumn{1}{>{\centering\arraybackslash}m{0.21\textwidth}}{\textbf{Area}}
				& \multicolumn{1}{>{\centering\arraybackslash}m{0.21\textwidth}}{\textbf{Combined A}}
				& \multicolumn{1}{>{\centering\arraybackslash}m{0.21\textwidth}}{\textbf{Combined B}}\\ 
				
				\textbf{SE} & \includegraphics[width=0.2\textwidth]{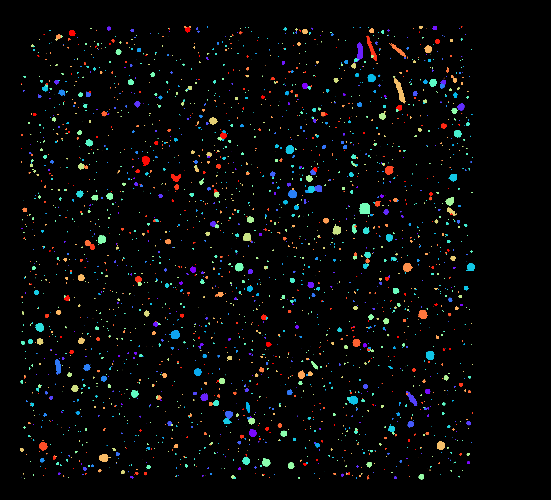} & \includegraphics[width=0.2\textwidth]{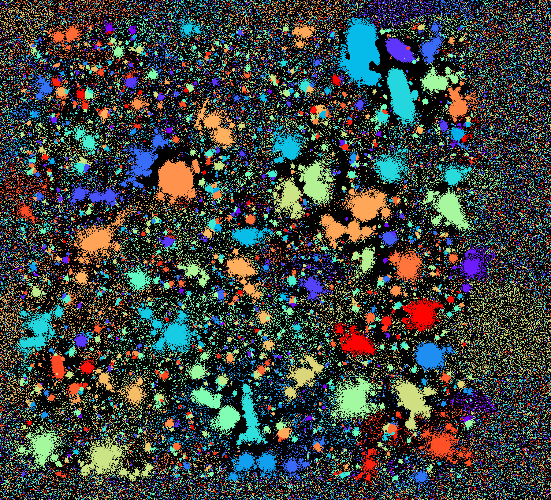} &
				\includegraphics[width=0.2\textwidth]{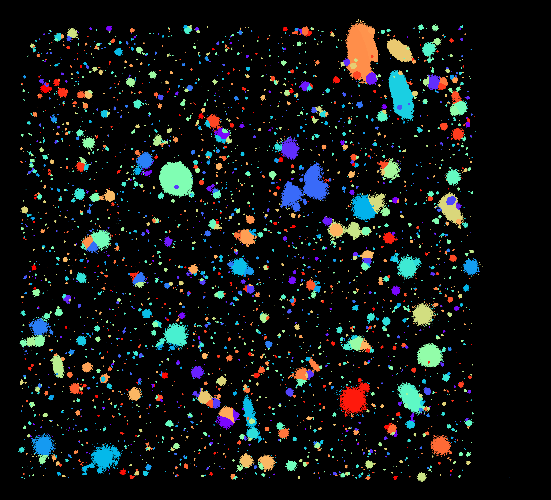}& 
				\includegraphics[width=0.2\textwidth]{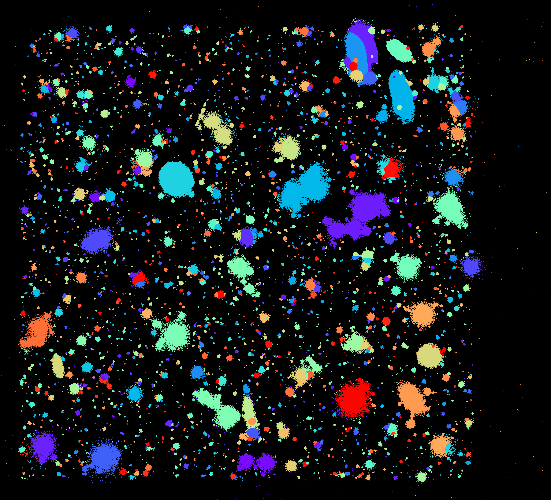}\\
				\textbf{NC} & \includegraphics[width=0.2\textwidth]{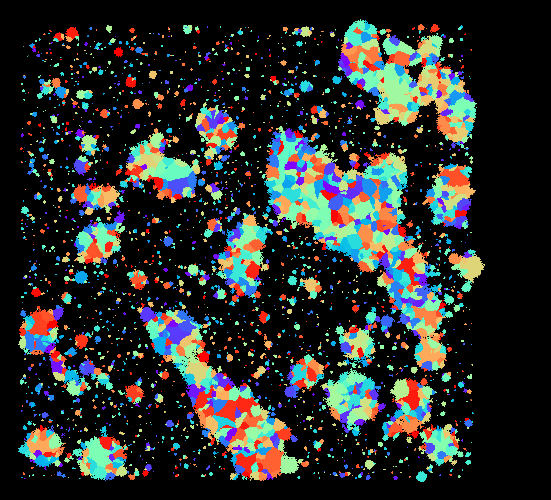} & \includegraphics[width=0.2\textwidth]{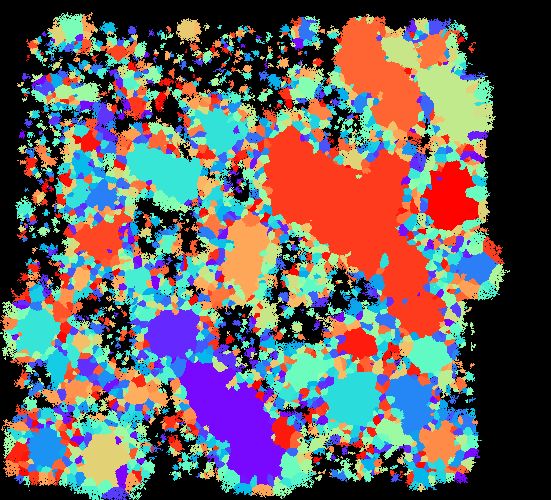} &
				\includegraphics[width=0.2\textwidth]{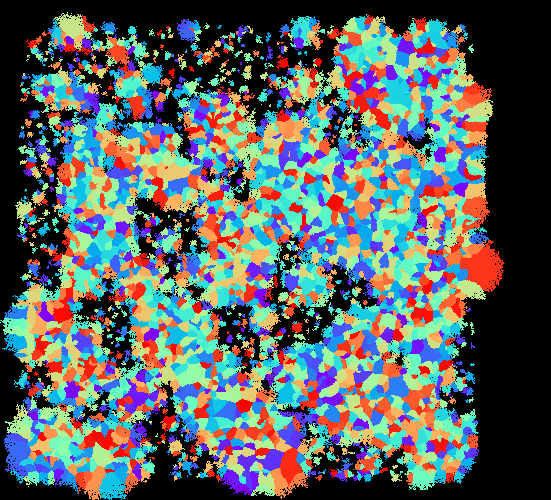}& 
				\includegraphics[width=0.2\textwidth]{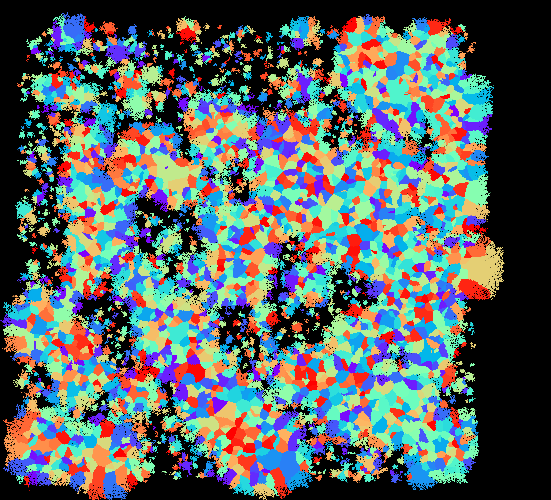}\\
				\textbf{MT}&
				\includegraphics[width=0.2\textwidth]{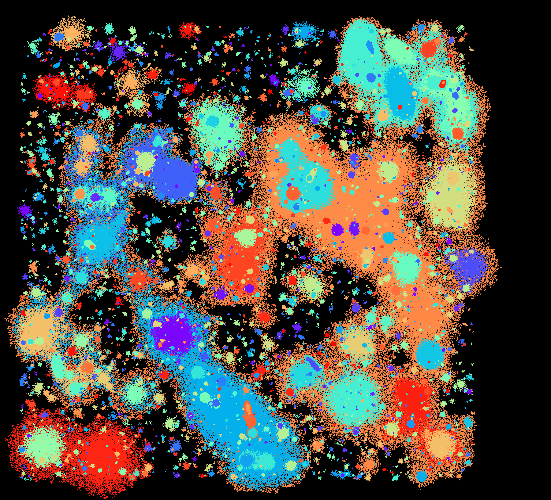} & \includegraphics[width=0.2\textwidth]{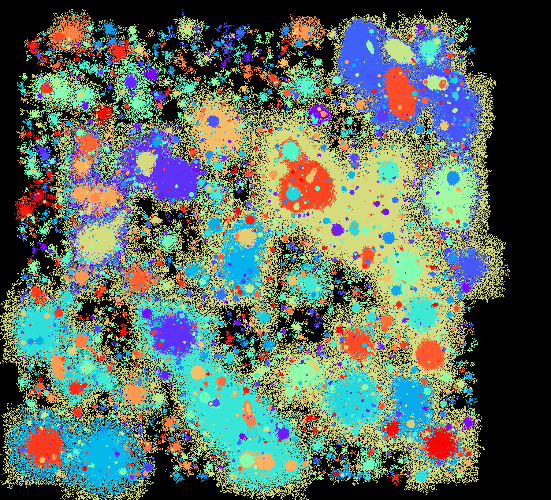} &
				\includegraphics[width=0.2\textwidth]{figs/sim/crop0/mto_1_0.png}& 
				\includegraphics[width=0.2\textwidth]{figs/sim/crop0/mto_1_0.png}\\
				\textbf{PF}&
				\includegraphics[width=0.2\textwidth]{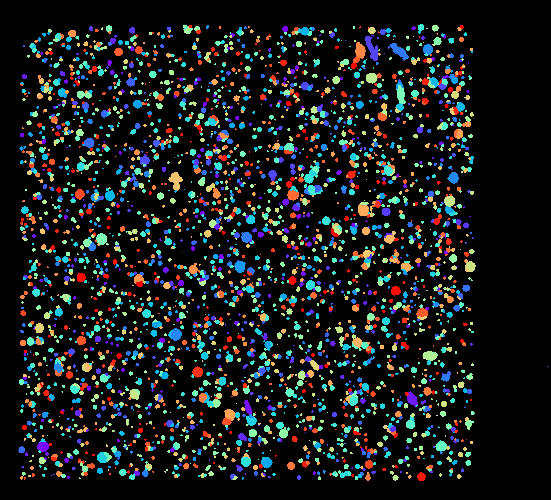} & \includegraphics[width=0.2\textwidth]{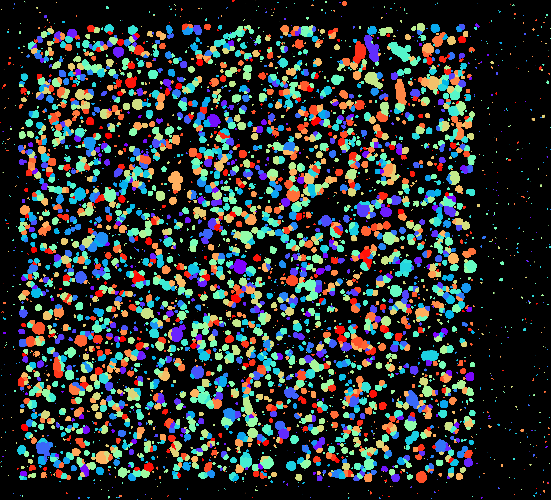} & & \\
			\end{tabular}
			\caption{Segmentation maps}
		\end{subfigure}
	
		\caption{Segmentations of a full simulated image, using the parameters which gave the highest median score for each combination of optimisation measure and tool on the simulated images: SExtractor (SE), NoiseChisel + Segment (NC), MTObjects (MT) and ProFound (PF). The coloured regions label distinct objects, and the black regions make up the background. Due to speed, PF was not optimised for Combined A and B.}
		\label{biglabels}
	\end{figure*}

\begin{figure*}
		\centering
		\begin{subfigure}[b]{0.4\textwidth}
			\includegraphics[width=\textwidth]{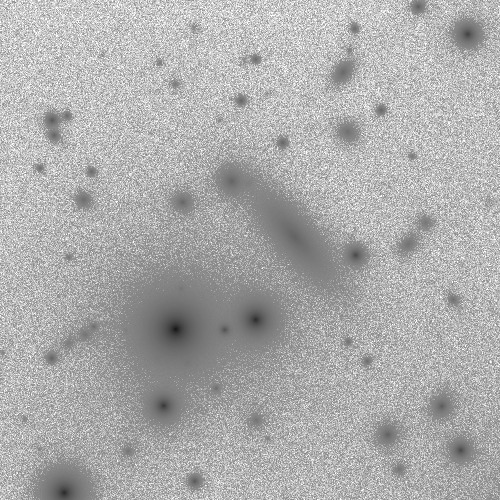}
			\caption{Original simulated image}
		\end{subfigure}
		\begin{subfigure}[b]{0.4\textwidth}
			\includegraphics[width=\textwidth]{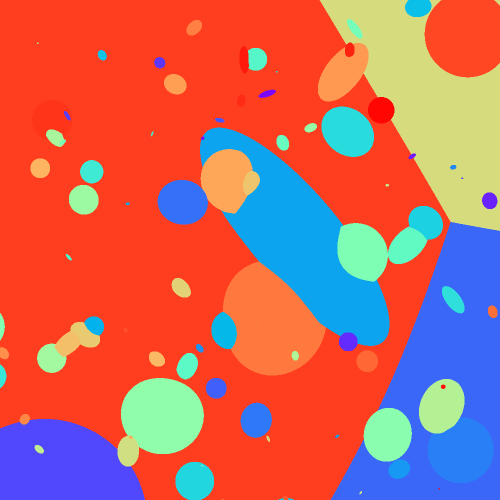}
			\caption{Ground truth (0.1 $\sigma$)}
		\end{subfigure}
		
		\begin{subfigure}[b]{0.8\textwidth}
			\centering  
			\begin{tabular}{m{0.02\textwidth} m{0.2\textwidth} m{0.2\textwidth} m{0.2\textwidth} m{0.2\textwidth}}
				
				\multicolumn{1}{>{\centering\arraybackslash}m{0.02\textwidth}}{} 
				& \multicolumn{1}{>{\centering\arraybackslash}m{0.2\textwidth}}{\textbf{F-Score}}
				& \multicolumn{1}{>{\centering\arraybackslash}m{0.2\textwidth}}{\textbf{Area}}
				& \multicolumn{1}{>{\centering\arraybackslash}m{0.2\textwidth}}{\textbf{Combined A}}
				& \multicolumn{1}{>{\centering\arraybackslash}m{0.2\textwidth}}{\textbf{Combined B}}\\ 
				
				\textbf{SE} & \includegraphics[width=0.2\textwidth]{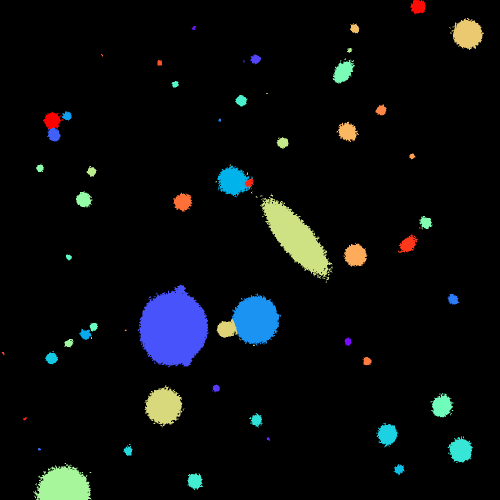} & \includegraphics[width=0.2\textwidth]{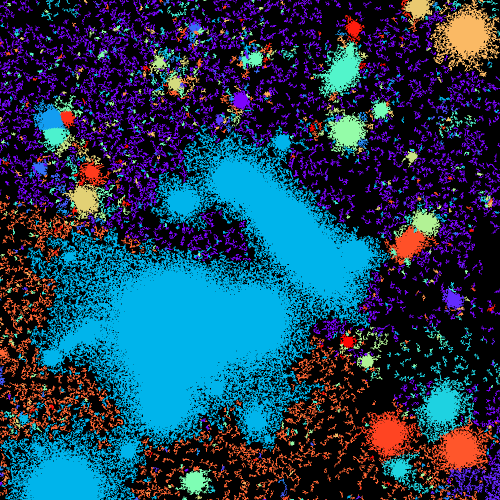} &
				\includegraphics[width=0.2\textwidth]{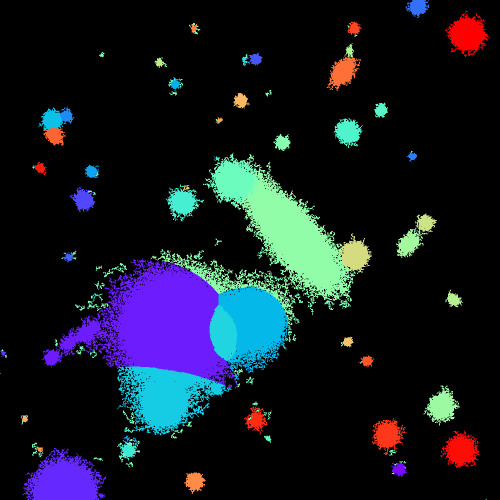}& 
				\includegraphics[width=0.2\textwidth]{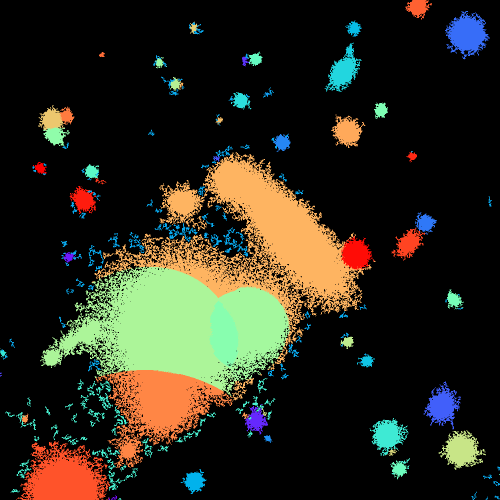}\\
				\textbf{NC} & \includegraphics[width=0.2\textwidth]{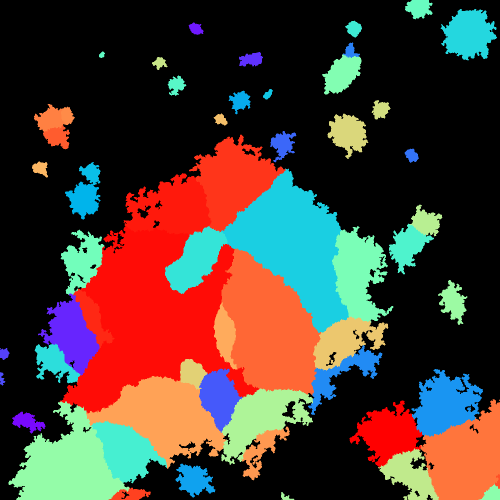} & \includegraphics[width=0.2\textwidth]{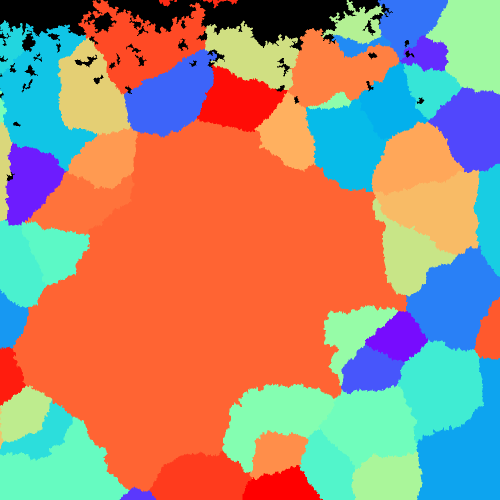} &
				\includegraphics[width=0.2\textwidth]{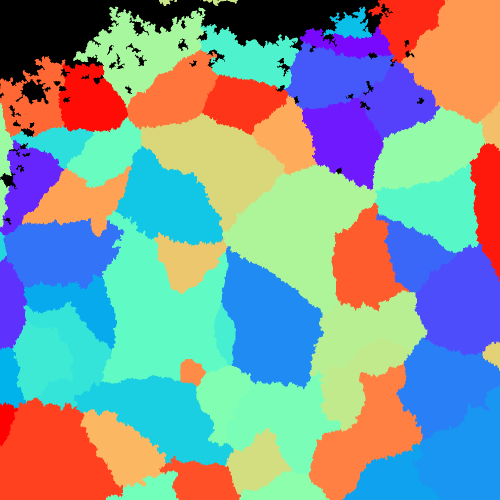}& 
				\includegraphics[width=0.2\textwidth]{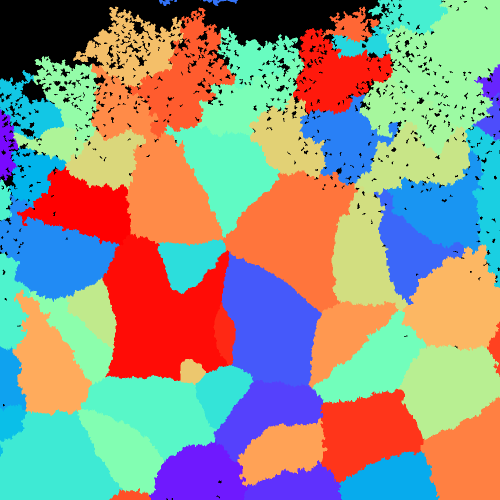}\\
				\textbf{MT}&
				\includegraphics[width=0.2\textwidth]{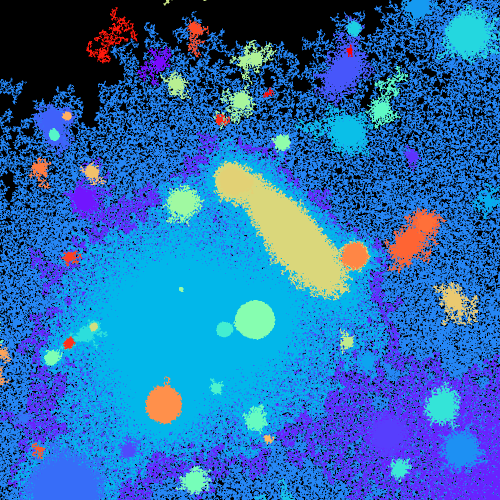} & \includegraphics[width=0.2\textwidth]{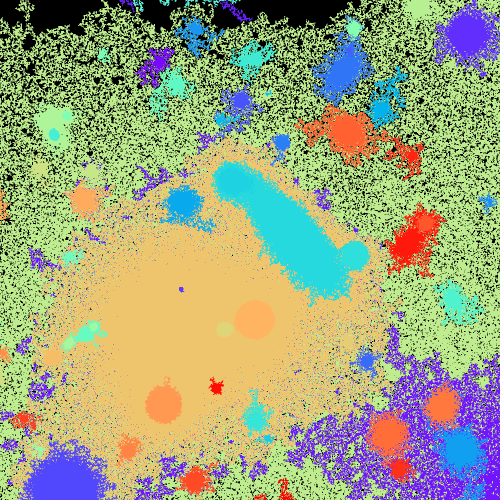} &
				\includegraphics[width=0.2\textwidth]{figs/sim/crop1/mto_1_1.png}& 
				\includegraphics[width=0.2\textwidth]{figs/sim/crop1/mto_1_1.png}\\
				\textbf{PF}&
				\includegraphics[width=0.2\textwidth]{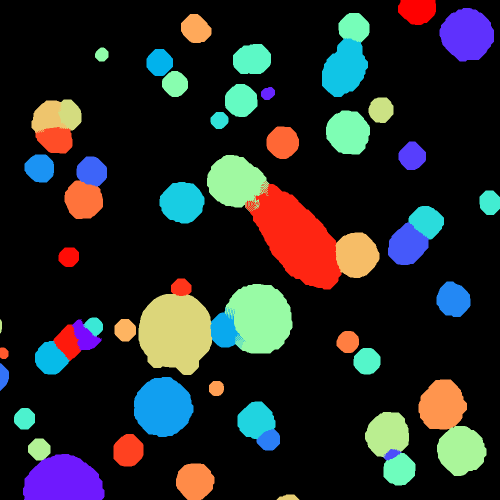} & \includegraphics[width=0.2\textwidth]{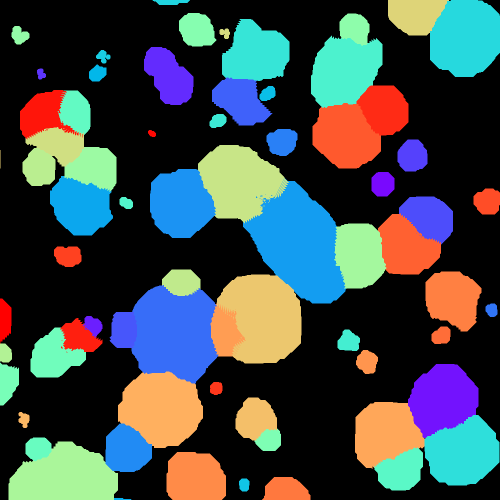} & & \\
			\end{tabular}
			\caption{Segmentation maps}
		\end{subfigure}  
		
		\caption{Segmentations of a section of a simulated image, using the parameters which gave the highest median score for each combination of optimisation measure and tool. For more information, see Fig. \ref{biglabels}.}
		\label{smalllabels}
	\end{figure*}

	Examining Figs. \ref{biglabels} and \ref{smalllabels} gives further insight into the behaviour behind these scores. We see that both NoiseChisel and MTObjects capture regions of light with visually similar boundaries, but that MTObjects marks many small, fractured sections in the outer regions as background. Meanwhile, NoiseChisel captures an area of light with fewer holes, but segments it into objects rather arbitrarily. In contrast, SExtractor and ProFound, which both have generally lower area scores, capture the compact centres of objects and only limited areas of the outskirts.
	
	\subsection{Background values}
	Each program makes internal estimations of background, which may be global or localised. We may also examine the pixels in the image which are not allocated to any segment in the final map. As the simulated images have a flat background with a mean of 0, we can use the mean value of these unallocated pixels as an indication of whether pixels containing no source light are being incorrectly allocated to sources, or conversely whether pixels are incorrectly regarded as belonging to sources.
	
	ProFound and SExtractor both consistently overestimated the background, giving values in the order of $10^{-1}\sigma$, where $\sigma$ is the standard deviation of the background noise ($1.1 \times 10 ^{-12}$ for the simulated images). This suggests that they are not detecting some parts of the sources, which visual inspection of Figs. \ref{biglabels} and \ref{smalllabels} confirm to be the case. There was one exception to this behaviour: SExtractor generally underestimated the background when optimised for area, with values in the order of $-10^{-2}\sigma$. This corresponds to the large number of small false positive detections made under this optimisation, thanks to the low background threshold used (see Table \ref{app:se}).
	
	MTObjects also underestimated the background, with values of around $-10^{-1}\sigma$ when optimised for metrics including area measures; it underestimated to a lesser degree ($-10^{-2}\sigma$) when optimised for F-score. This behaviour may be a consequence of the holes in the outskirts of objects causing the optimisation process to select parameters which overestimate the size of objects, thereby increasing the solid area within objects but also the number of incorrectly labelled background pixels.
	
	The strongest background estimation performance was produced by NoiseChisel. Whilst optimising for F-score gave an overestimation in a similar range to SExtractor and ProFound, it produced mean backgrounds in the order $\pm 10^{-3}\sigma$ when optimised for a metric including area measures. Not only were the values closer to the goal of 0, there was also no evidence of systematic over- or under-estimation.

	\subsection{Speed}
	\label{speedsect}
	The speed at which an image can be processed is very important when we consider the size and quantity of images produced by modern surveys. 
	
	\begin{figure}
		\centering
		\includegraphics[width=\columnwidth]{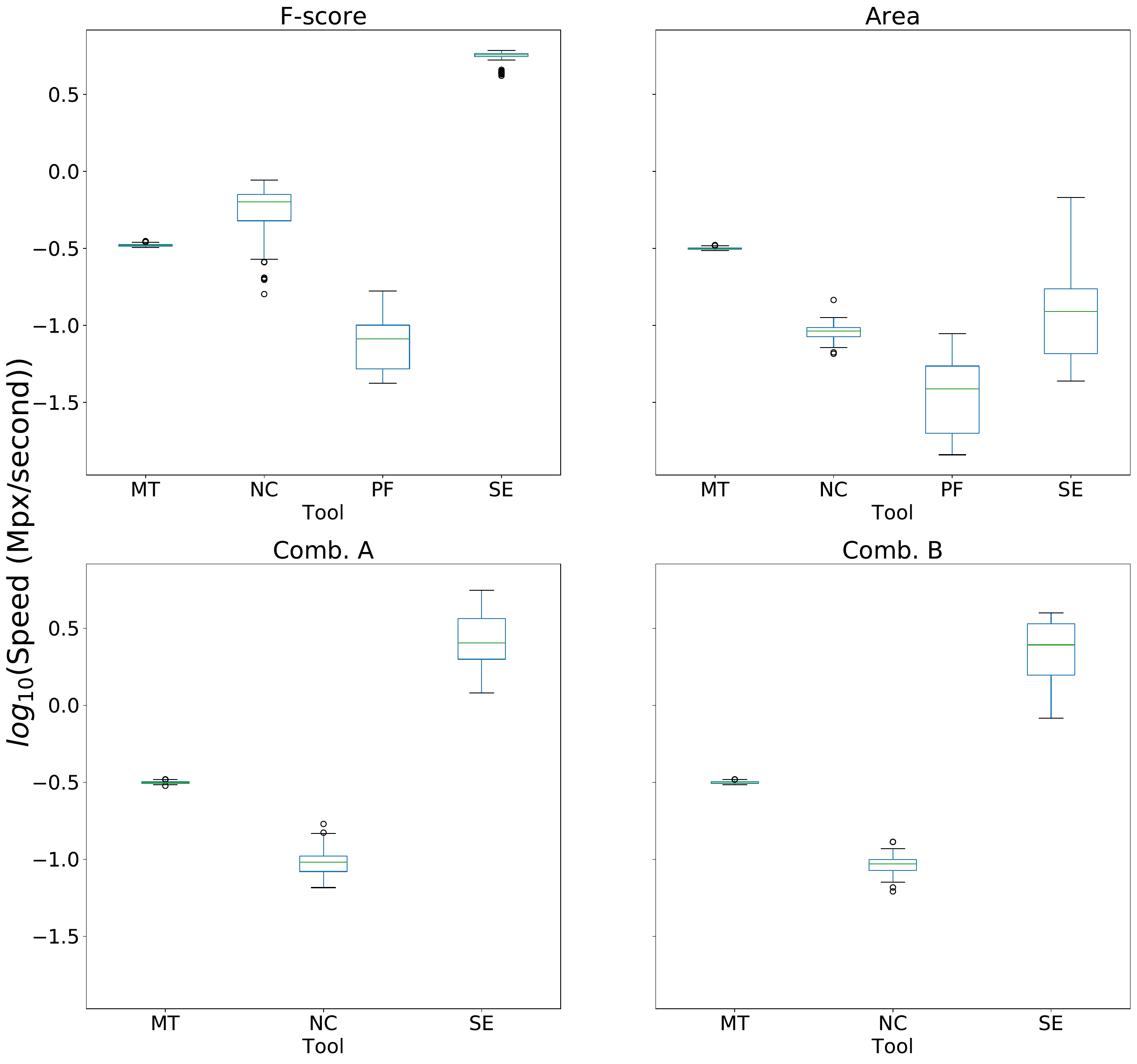}
		\caption{Distributions of processing speed across all combinations of images and optimised parameters for each tool and optimisation metric.}
		\label{Speedgraph}
	\end{figure}
	
	At its best, SExtractor was the fastest of all the tools by a considerable margin, as shown in Fig. \ref{Speedgraph}. When optimised for area, this advantage vanished completely -- potentially due to the vast increase in the number of false positives and large objects. When optimised for combined metrics, processing speed depended heavily on the individual set of parameters, producing a wide spread of speeds.
	
	MTObjects had the most consistent speed across optimisations. Neither SExtractor or MTObjects used parallel processing, which potentially reduced their speed. It should be noted that the original C implementation of MTObjects is faster than our current Python and C++ implementation, as \citet{teeninga16:_statis} reported SExtractor was only 2.5 times faster than the C version of MTObjects in terms of median performance, and only 1.3 times faster on average. Some code optimisation and using a parallel max-tree algorithm \cite{moschini17:_hybrid_shared_memor_paral_max} should be able to improve the performance in terms of speed.
	
	NoiseChisel showed fast performance when optimised for F-Score alone, but was much slower when area score was included in the optimisation criterion. This appears to be due to a combination of factors; predominantly a lower value for `detgrowquant', which affects how much objects are grown after detection.\footnote{We found that some non-optimal parameter combinations also caused substantial slowdown, due to the program requiring large amounts of memory and consequently writing some data structures to disk.}
	
	As mentioned previously, ProFound consistently had a very long processing time, which greatly reduced its viability as a tool for processing large images from surveys with many images. This is due in part to it writing temporary data to disk, which is discussed in the original ProFound publication \citep{robotham2018profound} -- ProFound offers a low-memory mode which reduces the amount of data stored, allowing the processing of larger images without a drastic slowdown; however, as noted, the method is fundamentally rather slow. The use of R as the implementation language may be further reducing the tool's potential speed. The authors of ProFound are rewriting parts of the code in C++, which should significantly improve its performance. 
	
	\subsection{Parameter consistency}
	MTObjects was by far the most consistent of the tools -- having only two relevant parameters, it had a much smaller parameter space to explore. While its optimised parameters varied slightly when optimising only over F-score, all other metrics gave the same optimal parameters for all cases but one, as shown in Table \ref{app:mt}.
	
	SExtractor and NoiseChisel, optimised over 6 and 20 parameters respectively, displayed far less consistency in the parameters that were found (Tables \ref{app:se}, \ref{app:nc}, \ref{app:ncs}). This could potentially have been reduced by increasing the number of iterations of the optimisation process. However, the similar scores produced using very different parameters suggest that there is no single best choice -- many combinations of settings perform equally well overall, but are better and worse in certain contexts.
	
	\subsection{Inserted galaxies}
	
	As a final step, we evaluated the performance of the tools on a sample of real galaxies, inserted into a frame of the Fornax Deep Survey (FDS), which the simulated data was designed to emulate. Testing the tools on real galaxies allows us to verify that the behaviour of the tools generalises to galaxies which are not perfect ellipticals.
	
	We selected a sample of 22 galaxies from the EFIGI catalogue \citep{baillard2011efigi}, which contains images from the fourth data release of the Sloan Digital Sky Survey (SDSS) \citep{adelman2006fourth}. Galaxies were selected with $D_{25}$ (diameter of the 25.0 mag arcsec$^{-2}$ isophote, in units of log 0.1 arcmin) between 1.7 and 1.999, a heliocentric velocity $<$ 2000 km/s, and a galactic latitude between 60$^\circ$ and 70$^\circ$. This is a representative sample of galaxies in the nearby Universe, with high-quality SDSS images and detailed morphological types. We isolated the galaxy at the centre of each image using k-flat filtering \citep{ouzounis2010kflat}, which removed areas of light not connected to the central pixel, whilst preserving the galaxy's internal detail. We then convolved each galaxy with the $r$-band point spread function of the OmegaCAM, and added Poissonian noise.
	
\begin{figure}
		\centering
		\begin{subfigure}[b]{0.4\columnwidth}
			\includegraphics[width=\columnwidth]{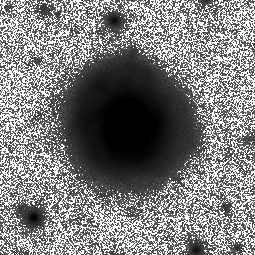}
			\caption{$10^{-10}$}
		\end{subfigure}
		\begin{subfigure}[b]{0.4\columnwidth}
			\includegraphics[width=\columnwidth]{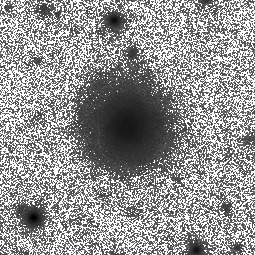}
			\caption{$10^{-11}$}
		\end{subfigure}
		
		\begin{subfigure}[b]{0.4\columnwidth}
			\includegraphics[width=\columnwidth]{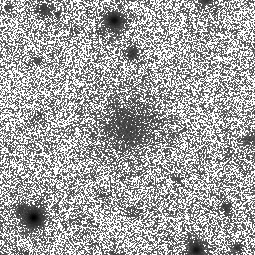}
			\caption{$10^{-12}$}
		\end{subfigure}
		\begin{subfigure}[b]{0.4\columnwidth}
			\includegraphics[width=\columnwidth]{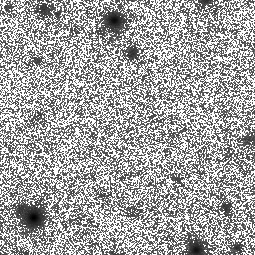}
			\caption{$10^{-13}$}
		\end{subfigure}
		
		\caption{A galaxy from the EFIGI sample, inserted into the FDS frame at the four given brightness scalings.}
		\label{scaled_galaxy}
	\end{figure}

	In order to examine the performance of the algorithms on galaxies of different brightnesses, we scaled the images to four different brightness levels, as shown in Fig. \ref{scaled_galaxy}. At the brightest level, the brightest pixel in each galaxy had a value in the same order as the brightest pixels in the image, (around $10^{-10}$, corresponding to a surface brightness of 21.5  mag arcsec$^{-2}$). At the faintest, the brightest pixels were barely visible to the human eye (around $10^{-13}$, corresponding to a surface brightness of 29  mag arcsec$^{-2}$)). We selected 22 locations in the FDS frame where there were very few objects present, in order to minimise interference with the inserted galaxies. We then created four images, with the 22 galaxies inserted into the same locations in the FDS frame, using a different brightness scaling for each image. We then ran all four tools on each image with the four sets of optimised parameters obtained on the simulated images.
	
	Whilst using inserted galaxies meant that we had a ground truth for those galaxies, there may still have been other objects present around them in the FDS frame, which would also be detected by the tools. This means that we are unable to rely on the previously defined metrics, as other detected objects would be marked as false positives and raise the under-merging error.
	
	Instead, we use a modified process to determine whether an inserted galaxy has been detected. If the brightest pixel in an object is contained within a non-background segment of the segmentation map, and it is also the brightest pixel in that segment, we determine that the object has been detected.
	
\begin{figure}
		\centering
		\begin{subfigure}[b]{0.49\columnwidth}
			\includegraphics[width=\columnwidth]{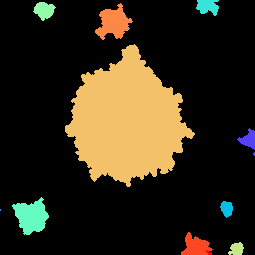}
			\caption{A `whole' detected galaxy.}
		\end{subfigure}
		\begin{subfigure}[b]{0.49\columnwidth}
			\includegraphics[width=\columnwidth]{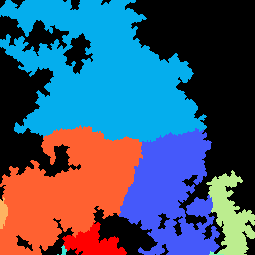}
			\caption{A fragmented galaxy.}
		\end{subfigure}
		\caption{Segmentation maps showing the two defined types of detection.}
		\label{fragmented_detect}
	\end{figure}

	Additionally, we classify detections into two types -- those where the galaxy has been mostly detected as a single object, and those where the algorithm has substantially fragmented the galaxy. To do this, we check for other detected segments whose brightest pixel is contained within the area of the inserted galaxy, suggesting that they are not primarily detecting some other background object. If there are multiple segments which meet this criteria, we check that the segment containing the most light from the inserted galaxy has at least $10\times$ the amount of light contained in the segment containing the second most light -- if it does, we mark the detection as `whole'; otherwise as fragmented. Whilst lacking the numerical accuracy of the previously defined area score, this provides an indication of the quality of detections. Examples of the two types of detection are shown in Fig. \ref{fragmented_detect}.

\begin{figure*}
		\centering
		\begin{subfigure}[b]{0.4\textwidth}
			\includegraphics[width=\columnwidth]{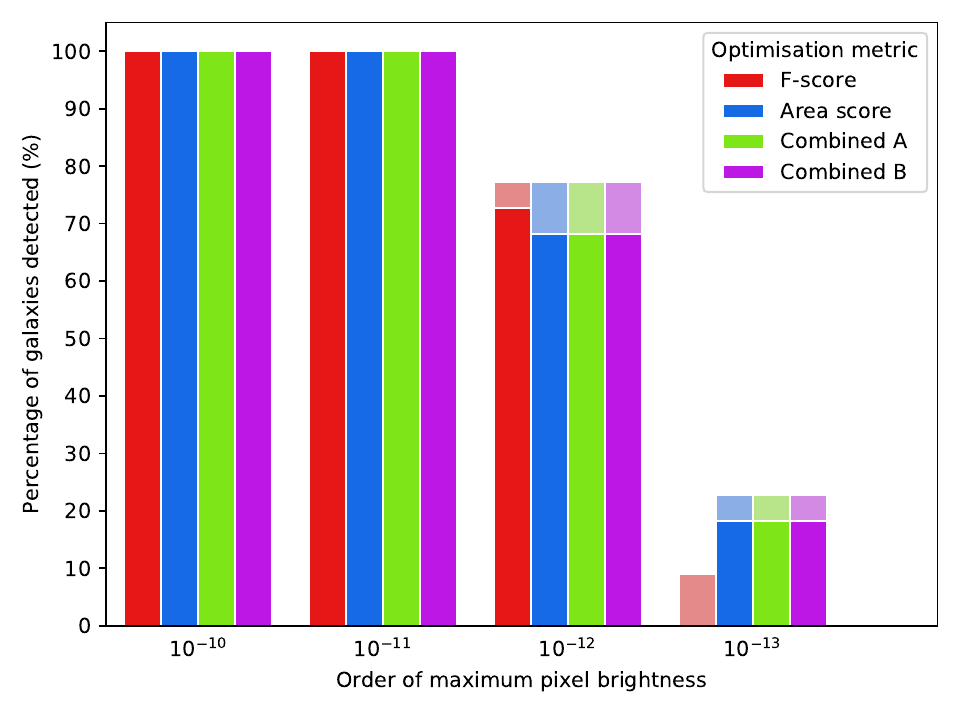}
			\caption{MTObjects}
		\end{subfigure}
		\begin{subfigure}[b]{0.4\textwidth}
			\includegraphics[width=\columnwidth]{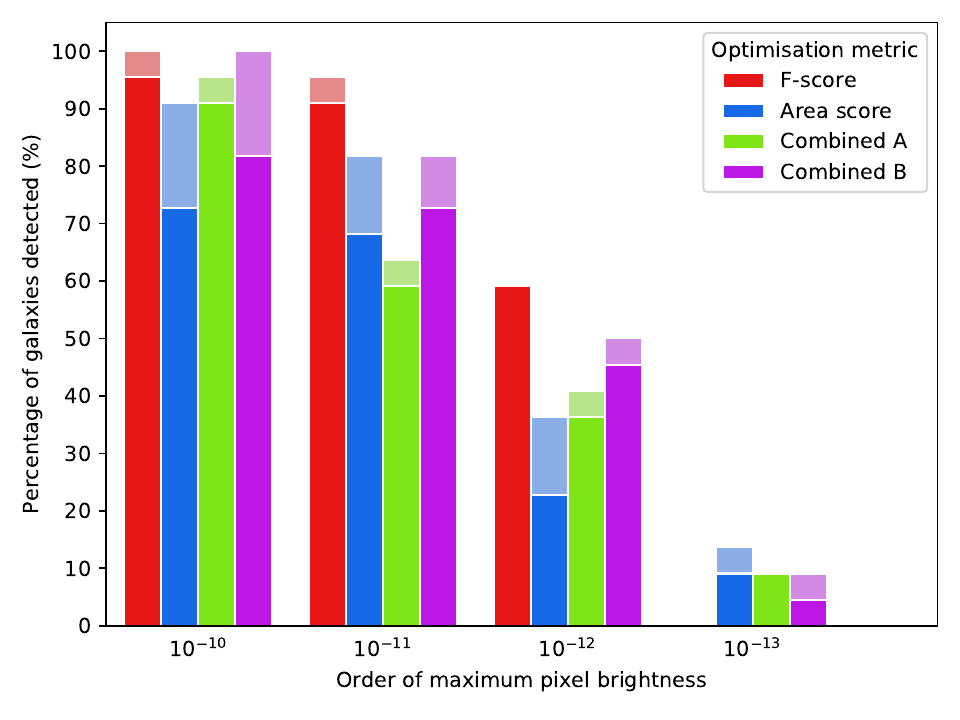}
			\caption{NoiseChisel}
		\end{subfigure}
		
		\begin{subfigure}[b]{0.4\textwidth}
			\includegraphics[width=\columnwidth]{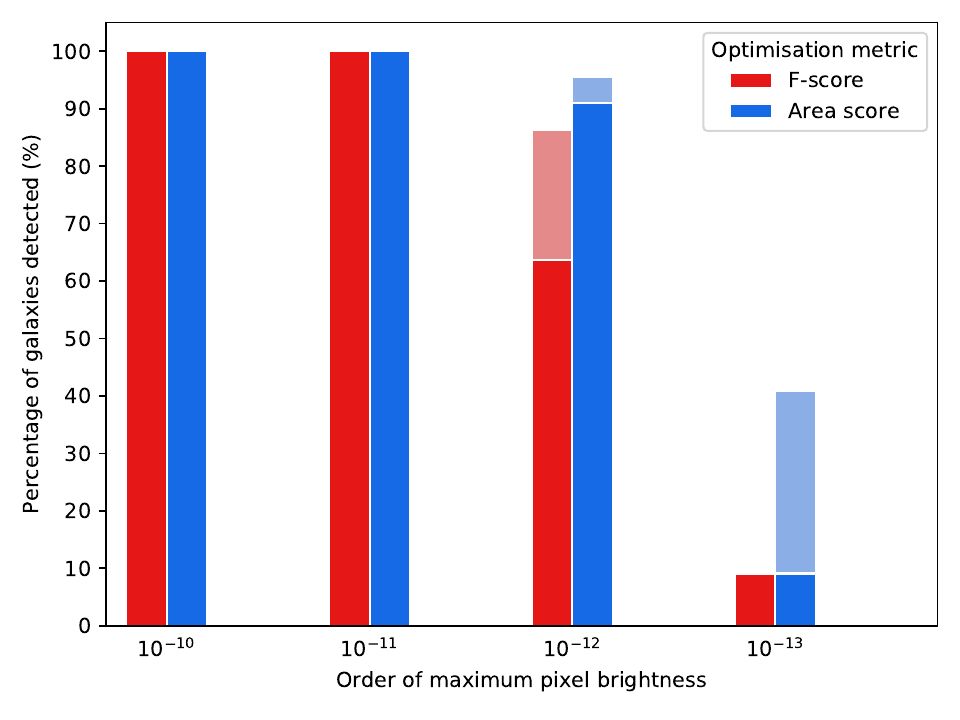}
			\caption{ProFound}
		\end{subfigure}
		\begin{subfigure}[b]{0.4\textwidth}
			\includegraphics[width=\columnwidth]{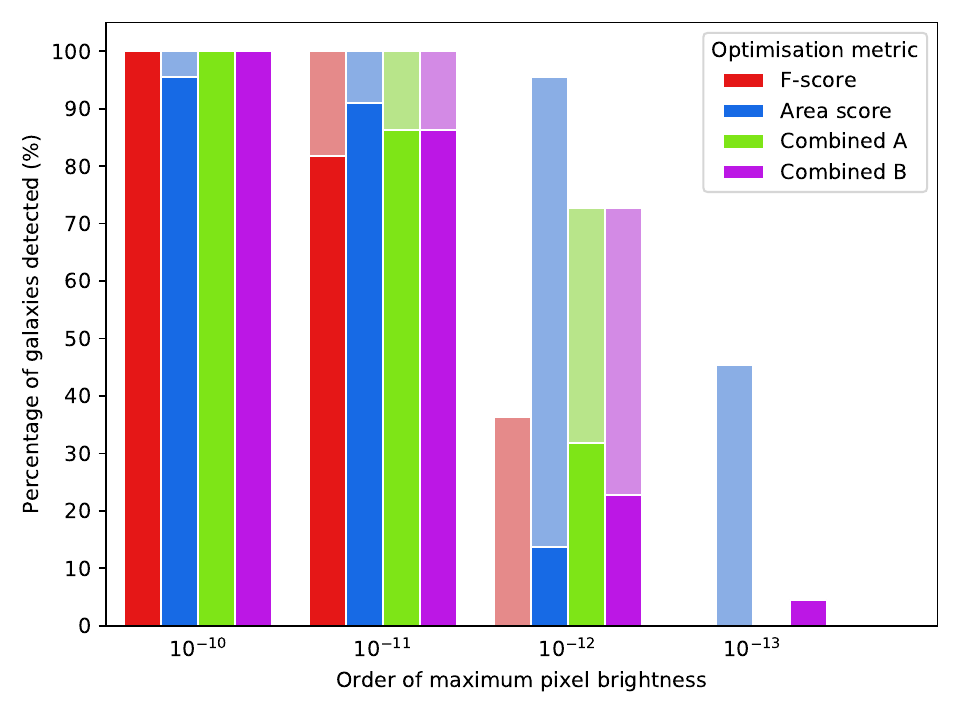}
			\caption{SExtractor}
		\end{subfigure}
		
		\caption{Percentage of inserted objects found, grouped by tool, brightness scaling, and optimisation metric, using the parameters which gave the highest median score for each combination of optimisation measure and tool. Lighter, stacked bars represent galaxies which are detected but substantially fragmented.}
		\label{inserted_graphs}
	\end{figure*}

	The results of this process are summarised in Fig. \ref{inserted_graphs}. At higher brightnesses, most tools perform well, with only NoiseChisel failing to detect any objects at the two highest brightness levels.
	
	At fainter levels, the tools show more variation. At the $10^{-12}$ brightness level, ProFound shows the strongest performance, fully detecting nearly over 90\% of the objects under an area score optimisation. SExtractor shows high levels of fragmentation at this level, consistent with its low area score found on the simulated data. NoiseChisel maintains a roughly consistent rate of fragmented detections, but with fewer detections overall, whilst MTObjects begins to show some fragmentation and a lower detection rate at this level.
	
	At the faintest brightness, very few of the inserted galaxies are visible to the human eye, and this is reflected in the results. Again, ProFound has a stronger performance than the other tools, with up to 40\% of galaxies detected, but a higher rate of fragmentation than at higher brightness levels. SExtractor reaches a similar detection rate under an area score optimisation, but only produces fragmented detections; visual inspection shows that this is due to the tool finding many tiny objects, as on the simulated images. Both NoiseChisel and MTObjects find very few objects at this low brightness level.
	
	These results are generally consistent with the results shown in the preceding sections -- all tools were capable of similar F-scores, and this is reflected in the similar detection rates found on the inserted galaxies. Similarly, variations in area score roughly correspond to the fraction of the inserted galaxies with substantial fragmentation for each tool, particularly at the $10^{-12}$ brightness level.
	
	It is notable that when the inserted galaxies are fainter, optimisations for F-score appear to be less effective than optimisations for area score. This may be due to the higher sensitivity to noise and lower thresholds generally found in area-based optimisations causing the fainter objects to be detected, whilst the F-score-based optimisations ignore these objects in order to minimise false detections.
	
	\section{Qualitative evaluation}
	\label{sec:real}
	
	In this section, we evaluate how the optimised parameters for each tool transfer to different surveys and instruments. We selected three surveys to apply the tools to, using the parameters with the highest median test score following the optimisation process: the Fornax Deep Survey (FDS) \citep{iodice2016,venhola2018fornax}; the IAC Stripe82 Legacy Project \citep{fliri2016s82, roman2018s82}; and the Hubble Ultra Deep Field (HUDF) \citep{beckwith2006hubble}. All of these datasets are deep surveys, with surface brightness limits fainter than $\mu \sim 28$ mag/arcsec$^2$, and have been used in several studies of galaxies of low surface brightness -- for example, \citet{akuFDS2017, venhola2019fornax, iodice2019} for FDS, \citet{roman2017a, roman2017b} for IAC Stripe82, and \citet{oesch2009structure, bouwens2008z} for HUDF. As far as the authors are aware, all these works have used SExtractor for masking sources of light and processing observational data. Therefore, evaluating the quality of segmentation for these deep datasets using the other available source extraction tools is an added value to ongoing research on faint structures of galaxies.
	
	Moreover, using the source extraction tools to derive segmentation maps of a completely new dataset with the `best' optimised parameters allows us to assess whether or not the parameters perform in a consistent manner across different datasets which have been acquired in very different conditions. It is also a test of the tool's practical applicability to large astronomical surveys of the future, such as those produced by Euclid \citep{amiaux2012euclid} and the LSST \citep{ivezic2008lsst}).
	
	The `best' parameters derived from our optimisation scheme for each test score that are used for the tools are highlighted with an asterisk in Appendix \ref{optables}.
	
	\begin{table*}
		\caption{Summary of Qualitative Evaluation}
		\label{table:qualsum}
		\centering  
		\begin{tabular}{c c c c c} 
			\hline\hline                       
			& MTObjects & NoiseChisel & ProFound & SExtractor \\  
			\hline
			Optimised parameters & 2 & 20 & 8 & 6 \\      
			Language & Python/C & C & R & C\\
			Clean edges of detected objects & - & \checkmark & \checkmark & Sometimes\\
			Detects elongated galaxy (FDS - Fig. \ref{r11_1}) & \checkmark & Fragmented & - & Fragmented \\
			Detects galaxy close to star (FDS - Fig. \ref{r11_2}) & \checkmark & Fragmented & - & Fragmented \\
			Detects cirrus (Stripe82 - Fig. \ref{s82_f0363}) & \checkmark & \checkmark & - & Sometimes\\
			Isolates spiral substructures (HUDF - Fig. \ref{hubble_2}) & \checkmark & - & - & -\\
			
			\hline  
		\end{tabular}
	\end{table*}

	\begin{figure*}
		\centering
		\begin{subfigure}[b]{0.35\textwidth}
			\includegraphics[width=\textwidth]{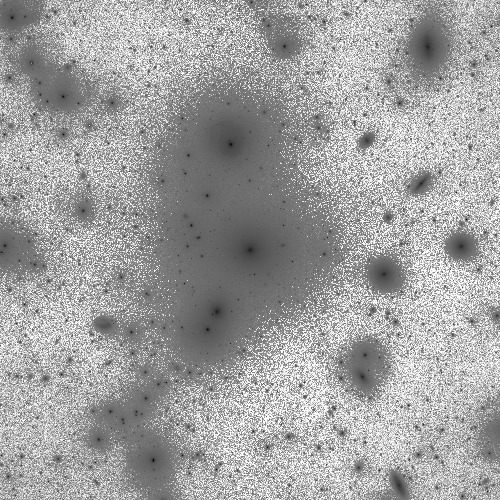}
			\caption{Input image -- the r-band of field 11 of the FDS.}
		\end{subfigure}
		\begin{subfigure}[b]{0.84\textwidth}
			\centering  
			\begin{tabular}{m{0.02\textwidth} m{0.2\textwidth} m{0.2\textwidth} m{0.2\textwidth} m{0.2\textwidth}}
				
				\multicolumn{1}{>{\centering\arraybackslash}m{0.02\textwidth}}{} 
				& \multicolumn{1}{>{\centering\arraybackslash}m{0.2\textwidth}}{\textbf{F-Score}}
				& \multicolumn{1}{>{\centering\arraybackslash}m{0.2\textwidth}}{\textbf{Area}}
				& \multicolumn{1}{>{\centering\arraybackslash}m{0.2\textwidth}}{\textbf{Combined A}}
				& \multicolumn{1}{>{\centering\arraybackslash}m{0.2\textwidth}}{\textbf{Combined B}}\\ 
				
				\textbf{SE} & \includegraphics[width=0.2\textwidth]{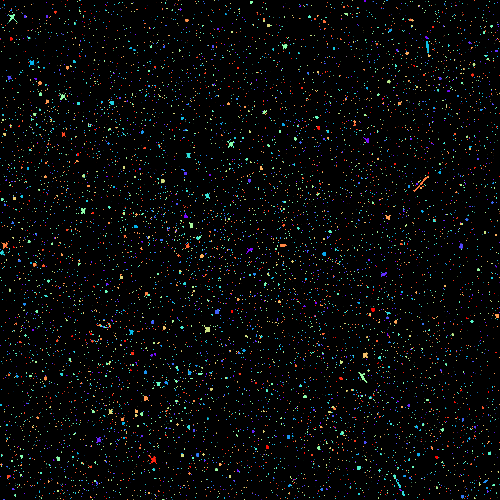} & \includegraphics[width=0.2\textwidth]{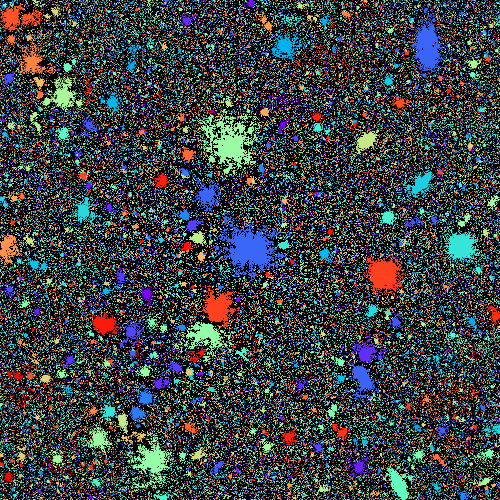} &
				\includegraphics[width=0.2\textwidth]{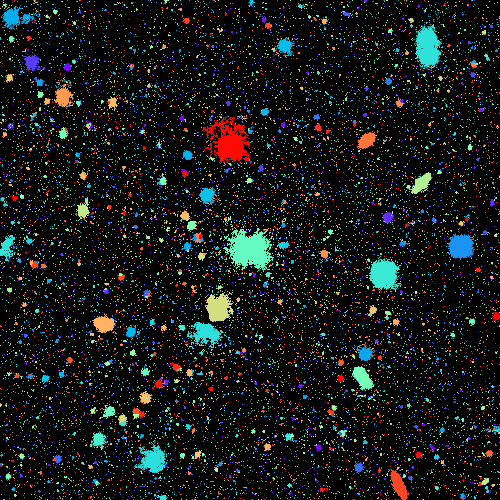}& 
				\includegraphics[width=0.2\textwidth]{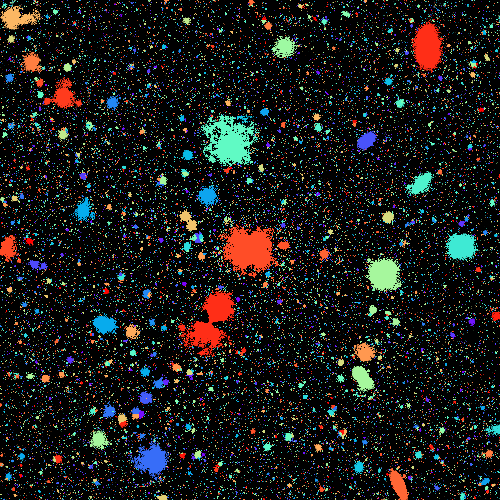}\\
				\textbf{NC} & \includegraphics[width=0.2\textwidth]{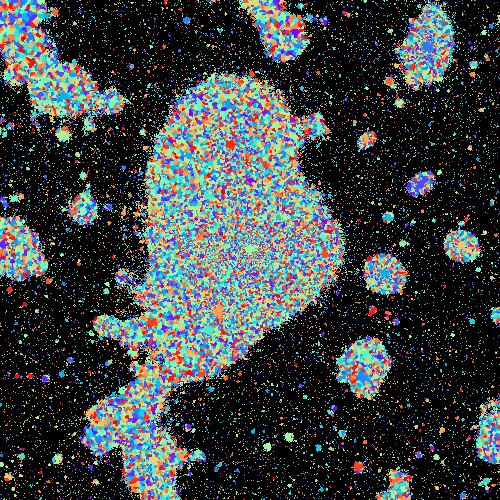} & \includegraphics[width=0.2\textwidth]{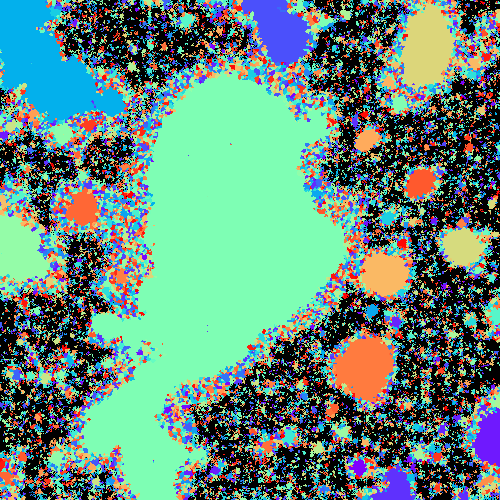} &
				\includegraphics[width=0.2\textwidth]{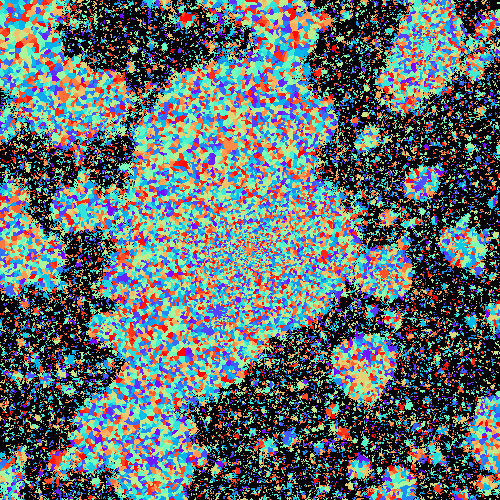}& 
				\includegraphics[width=0.2\textwidth]{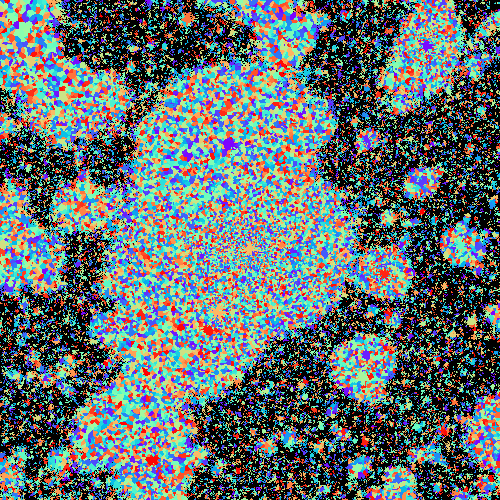}\\
				\textbf{MT}&
				\includegraphics[width=0.2\textwidth]{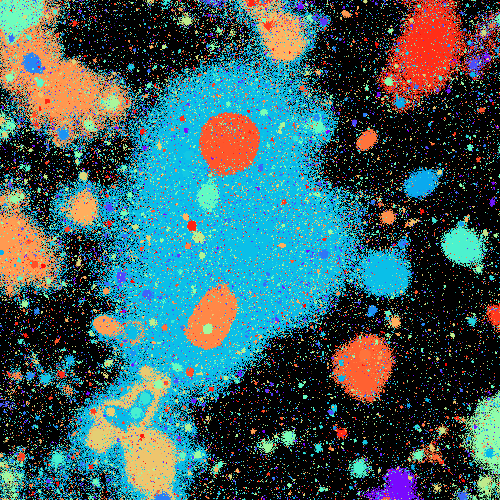} & \includegraphics[width=0.2\textwidth]{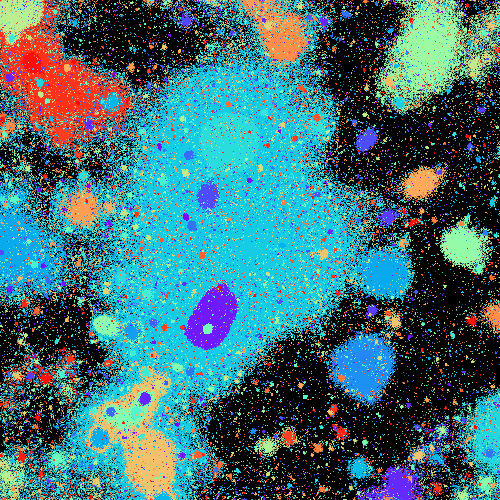} &
				\includegraphics[width=0.2\textwidth]{figs/fds/crop0/mto_1_0.png}& 
				\includegraphics[width=0.2\textwidth]{figs/fds/crop0/mto_1_0.png}\\
				\textbf{PF}&
				\includegraphics[width=0.2\textwidth]{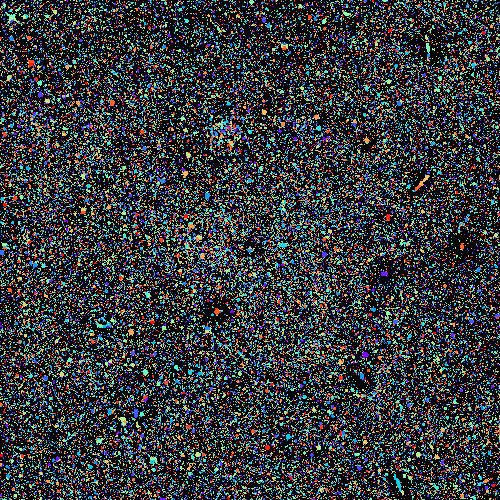} & \includegraphics[width=0.2\textwidth]{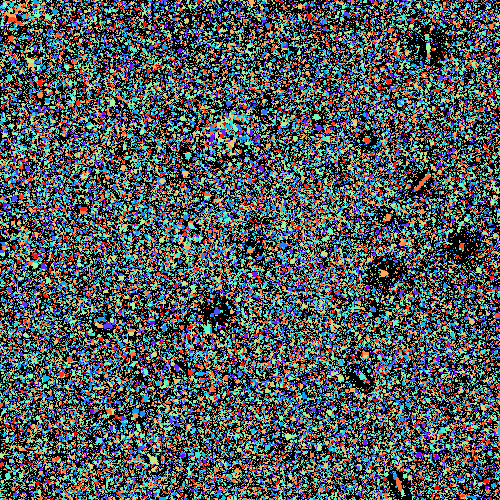} & & \\
			\end{tabular}
			\caption{Segmentations of the field, using the parameters which gave the highest median score for each combination of optimisation measure and tool. For more information, see Fig. \ref{biglabels}.}
		\end{subfigure} 
		
		\caption{Segmentations of a complete FDS field (field 11).}
		\label{r11_0}
	\end{figure*}

	\subsection{Fornax Deep Survey (FDS)}
	As the simulated images were created using the characteristics of the Fornax Deep Survey, using images from the real survey allows us to check that the parameters found on simulated data perform similarly on data which contains more unusual structures. The limiting surface brightness for $r$-band images of FDS is 29.8\,mag/arcsec$^2$ (3$\sigma$; $10\times10$\,arcsec$^2$) \citep{akuFDS2017}.
	
	\begin{figure*}
		\centering
		\begin{subfigure}[b]{0.25\textwidth}
			\includegraphics[width=\textwidth]{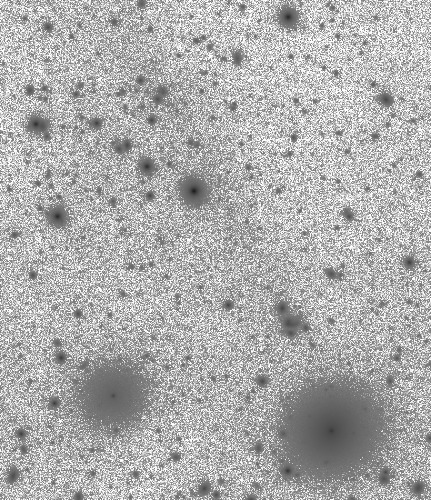}
			\caption{Input image -- the r-band of field 11 of the FDS.}
		\end{subfigure}
		
		\begin{subfigure}[b]{0.8\textwidth}
			\centering  
			\begin{tabular}{m{0.02\textwidth} m{0.2\textwidth} m{0.2\textwidth} m{0.2\textwidth} m{0.2\textwidth}}
				
				\multicolumn{1}{>{\centering\arraybackslash}m{0.02\textwidth}}{} 
				& \multicolumn{1}{>{\centering\arraybackslash}m{0.2\textwidth}}{\textbf{F-Score}}
				& \multicolumn{1}{>{\centering\arraybackslash}m{0.2\textwidth}}{\textbf{Area}}
				& \multicolumn{1}{>{\centering\arraybackslash}m{0.2\textwidth}}{\textbf{Combined A}}
				& \multicolumn{1}{>{\centering\arraybackslash}m{0.2\textwidth}}{\textbf{Combined B}}\\ 
				
				\textbf{SE} & \includegraphics[width=0.2\textwidth]{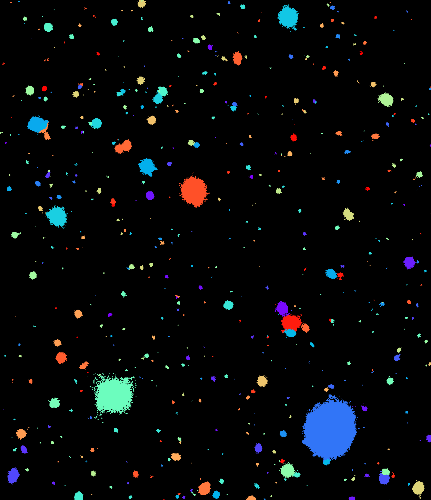} & \includegraphics[width=0.2\textwidth]{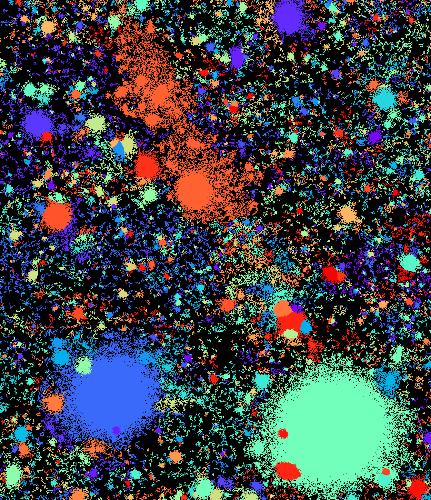} &
				\includegraphics[width=0.2\textwidth]{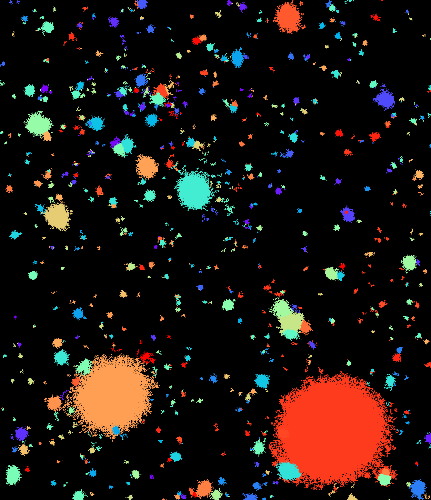}& 
				\includegraphics[width=0.2\textwidth]{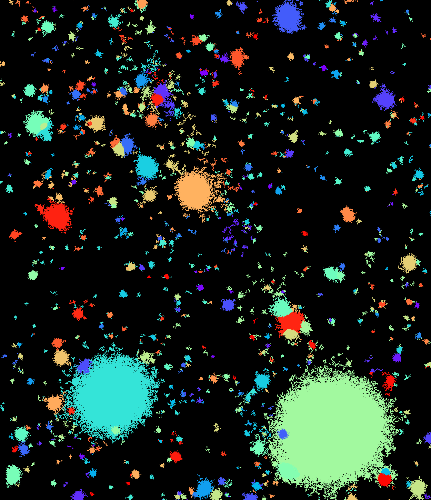}\\
				\textbf{NC} & \includegraphics[width=0.2\textwidth]{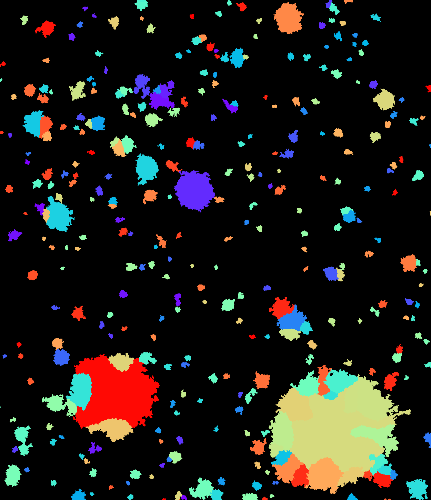} & \includegraphics[width=0.2\textwidth]{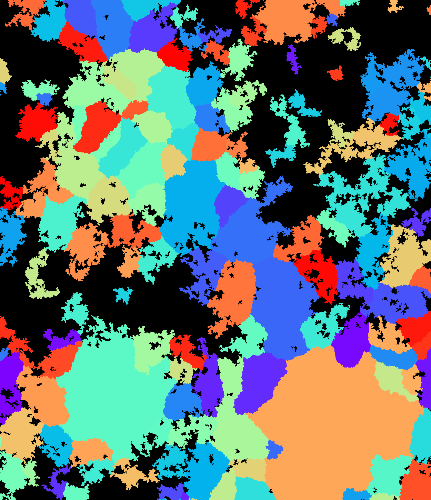} &
				\includegraphics[width=0.2\textwidth]{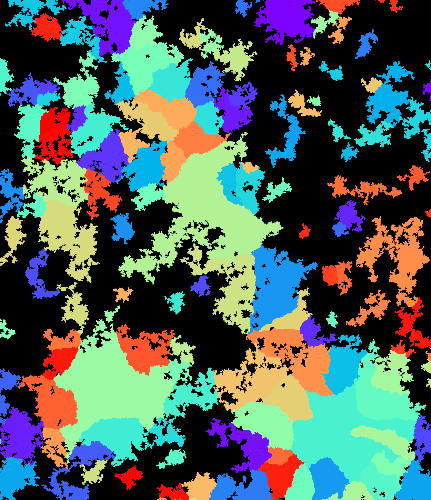}& 
				\includegraphics[width=0.2\textwidth]{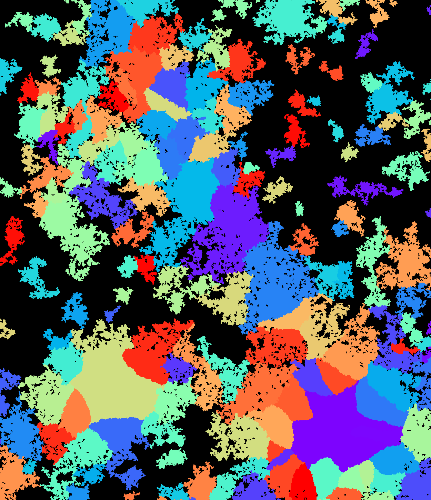}\\
				\textbf{MT}&
				\includegraphics[width=0.2\textwidth]{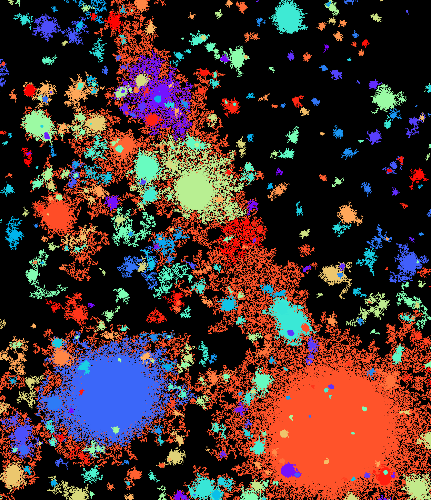} & \includegraphics[width=0.2\textwidth]{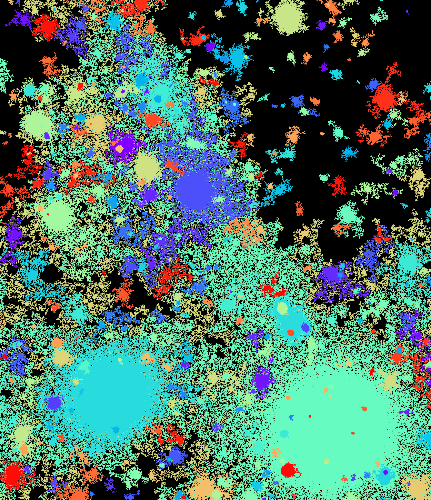} &
				\includegraphics[width=0.2\textwidth]{figs/fds/crop1/mto_1_1.png}& 
				\includegraphics[width=0.2\textwidth]{figs/fds/crop1/mto_1_1.png}\\
				\textbf{PF}&
				\includegraphics[width=0.2\textwidth]{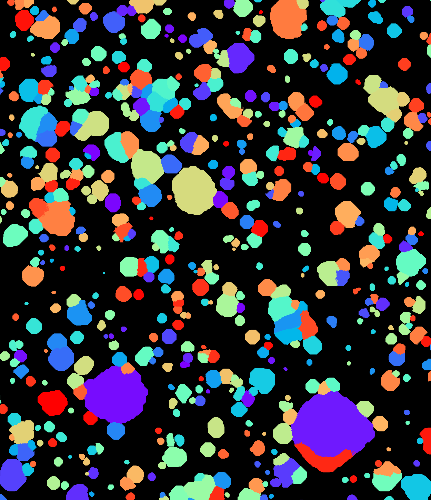} & \includegraphics[width=0.2\textwidth]{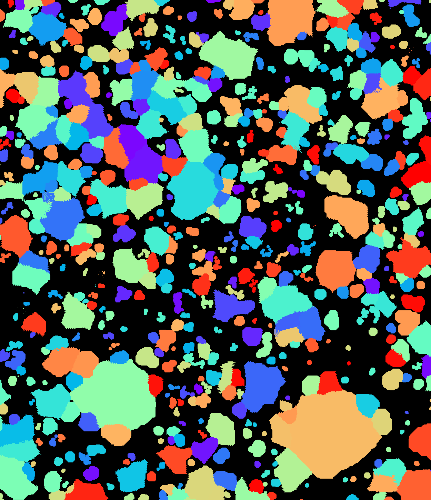} & & \\
			\end{tabular}
			\caption{Segmentations of the field section, using the parameters which gave the highest median score for each combination of optimisation measure and tool. For more information, see Fig. \ref{biglabels}.}
		\end{subfigure}  
		
		\caption{Segmentations of a section of an FDS field (field 11), showing a low-surface brightness galaxy.}
		\label{r11_1}
	\end{figure*}
	
	\begin{figure*}
		\centering
		\begin{subfigure}[b]{0.35\textwidth}
			\includegraphics[width=\textwidth]{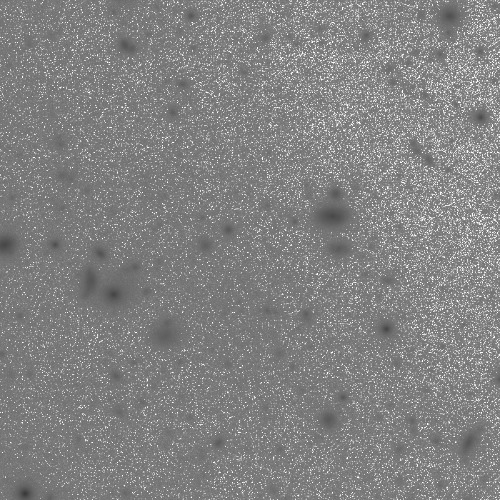}
			\caption{Input image -- the r-band of field 11 of the FDS.}
		\end{subfigure}
		\begin{subfigure}[b]{0.8\textwidth}
			\centering  
			\begin{tabular}{m{0.02\textwidth} m{0.2\textwidth} m{0.2\textwidth} m{0.2\textwidth} m{0.2\textwidth}}
				
				\multicolumn{1}{>{\centering\arraybackslash}m{0.02\textwidth}}{} 
				& \multicolumn{1}{>{\centering\arraybackslash}m{0.2\textwidth}}{\textbf{F-Score}}
				& \multicolumn{1}{>{\centering\arraybackslash}m{0.2\textwidth}}{\textbf{Area}}
				& \multicolumn{1}{>{\centering\arraybackslash}m{0.2\textwidth}}{\textbf{Combined A}}
				& \multicolumn{1}{>{\centering\arraybackslash}m{0.2\textwidth}}{\textbf{Combined B}}\\ 
				
				\textbf{SE} & \includegraphics[width=0.2\textwidth]{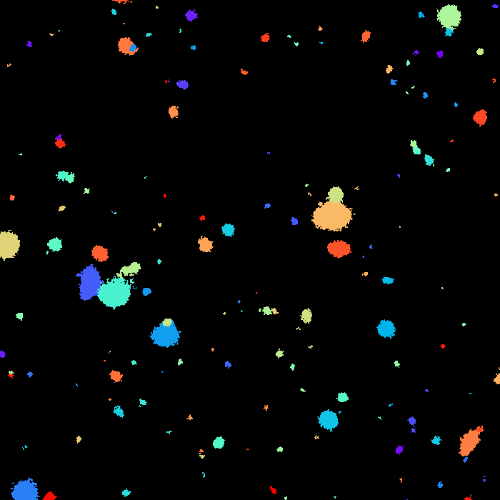} & \includegraphics[width=0.2\textwidth]{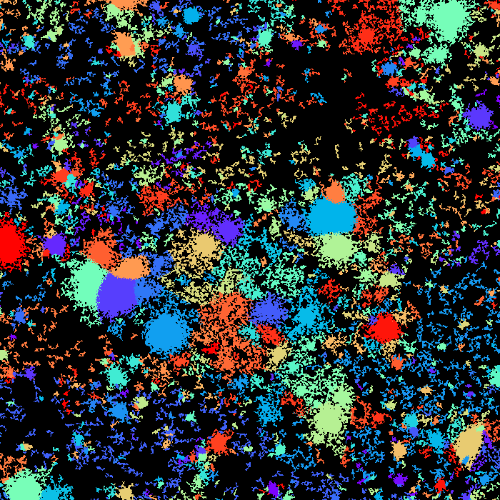} &
				\includegraphics[width=0.2\textwidth]{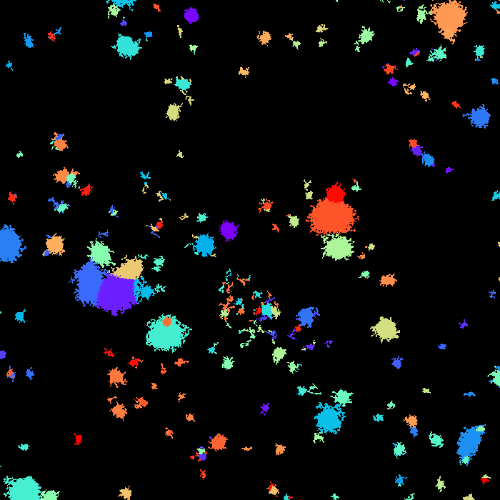}& 
				\includegraphics[width=0.2\textwidth]{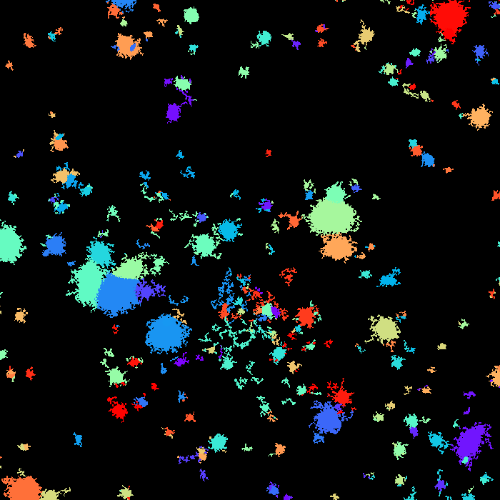}\\
				\textbf{NC} & \includegraphics[width=0.2\textwidth]{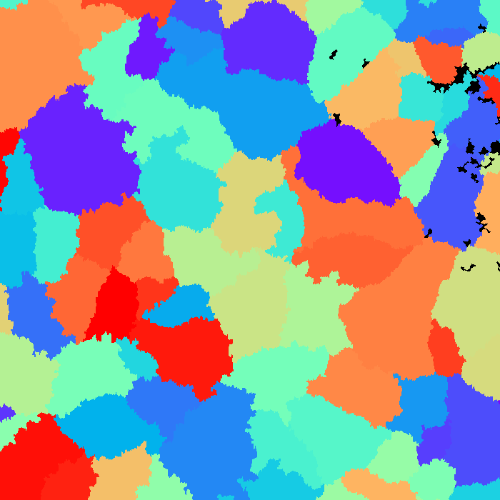} & \includegraphics[width=0.2\textwidth]{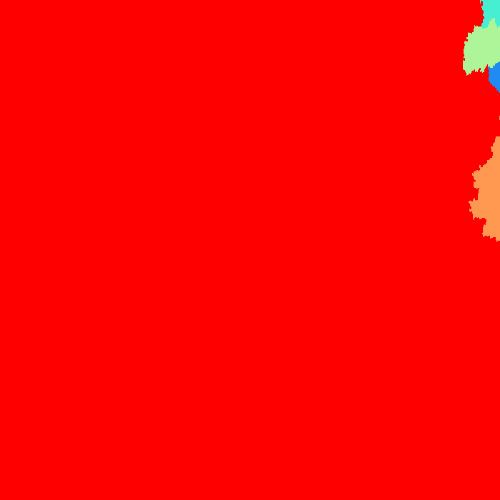} &
				\includegraphics[width=0.2\textwidth]{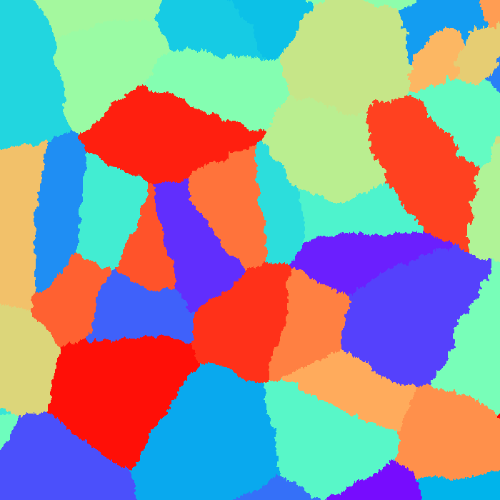}& 
				\includegraphics[width=0.2\textwidth]{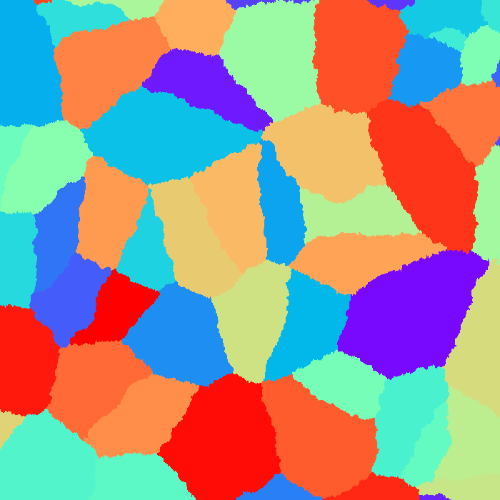}\\
				\textbf{MT}&
				\includegraphics[width=0.2\textwidth]{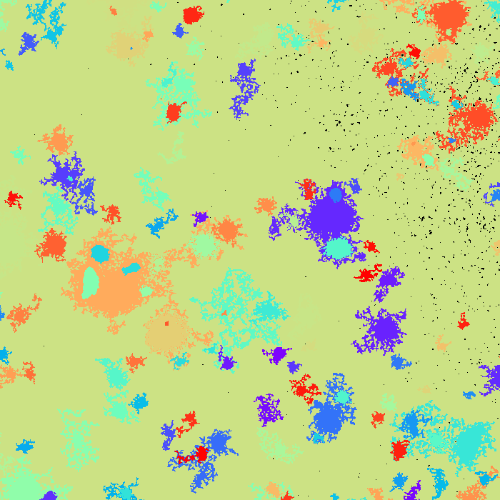} & \includegraphics[width=0.2\textwidth]{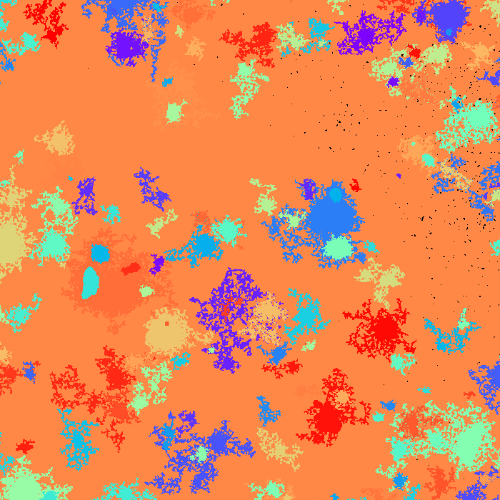} &
				\includegraphics[width=0.2\textwidth]{figs/fds/crop2/mto_1_2.png}& 
				\includegraphics[width=0.2\textwidth]{figs/fds/crop2/mto_1_2.png}\\
				\textbf{PF}&
				\includegraphics[width=0.2\textwidth]{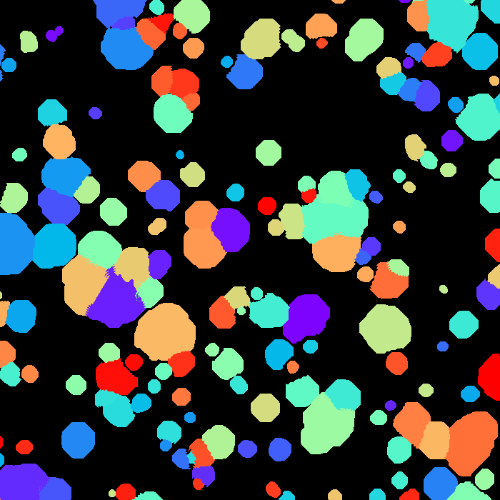} & \includegraphics[width=0.2\textwidth]{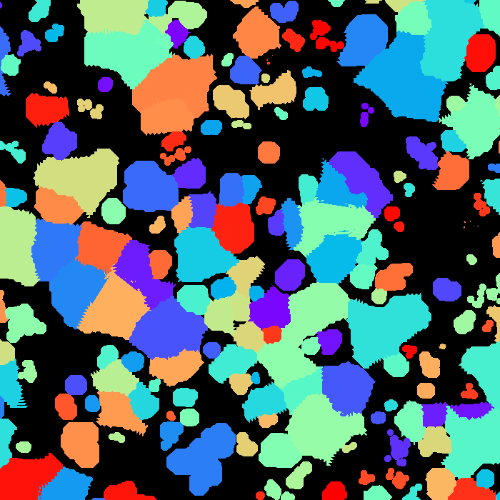} & & \\
			\end{tabular}
			\caption{Segmentations of the field, using the parameters which gave the highest median score for each combination of optimisation measure and tool.}
		\end{subfigure}  
		\caption{Segmentations of a section of a FDS field (field 11), showing a very faint structure in the lower centre.}
		\label{r11_2}
	\end{figure*}
	
	We here show a complete frame of the survey, and two smaller areas of the same frame, containing faint and challenging objects. For each combination of training image and optimisation method, the parameters with the highest median test score on the simulated dataset was used.
	
	It is clear from Fig. \ref{r11_0} that the parameters produce very similar behaviour on the real images to on the simulated images. MTObjects and NoiseChisel both capture similar areas of light, but segment them very differently; whilst ProFound and SExtractor capture only the centres of objects.
	
	Examining smaller details of the images gives more insight into behaviour on challenging sources. Fig. \ref{r11_1} shows the segmentation of a faint, elongated galaxy. SExtractor only detects a small area of the galaxy when optimised for area, and incorrectly merges it with other surrounding objects; in all other optimisations it fails to detect the galaxy at all, perhaps due to a too high detection threshold. ProFound detects small blobs covering the area of the galaxy, but does not identify an underlying structure. Similarly, NoiseChisel, whilst locating a larger area of light, breaks it into chunks appearing to correspond to smaller objects, losing the large structure. MTObjects was the only tool to capture the entire structure as one object, but incorrectly labelled it as the same object as the bright source in the bottom right corner, as well as connecting it to the outskirts of the object in the bottom left.
	
	Figure \ref{r11_2} contains another faint structure, located near a bright star, which is extremely difficult to visually detect. All four tools struggle to produce ideal results in this situation. As before, ProFound and SExtractor do not detect the faintest parts of objects, which here gives the advantage of allowing both tools to distinguish between smaller sources. SExtractor again produces a high number of false positives when optimised for area, but does begin to detect areas of structure in Combined A and B. In contrast, ProFound produces a blobby segmentation, with less visual similarity to the input image, but again covering a good deal of the smaller structures. NoiseChisel and MTObjects mark almost all of the image section as containing sources, but with a very different segmentation. NoiseChisel's area optimisation fails to detect any substructures in this part of the image, marking all objects as a single large structure. In the other optimisations, it shows very little visual similarity to the input image. MTObjects correctly detects many of the sources in the area, although it again joins the outskirts of some objects, and produces a ragged appearance.

	\subsection{IAC Stripe 82 Legacy Project}
	
	As an added layer of generalisation, we test the parameters with the highest median test score found for each combination of training image and optimisation method on deep \textit{g-}band IAC Stripe82 images.  The limiting surface brightness is $29.1 $ mag/arcsec$^2$ (3$\sigma$, 10 $\times$ 10 arcsec$^2$) \citep{roman2018s82}.
	
	These images consist of faint and diffuse structures such as galactic cirri, tidal streams, interacting galaxies as well as scattered light from point sources. \par
	
\begin{figure*}
		\centering
		\begin{subfigure}[b]{0.55\textwidth}
			\includegraphics[width=\textwidth]{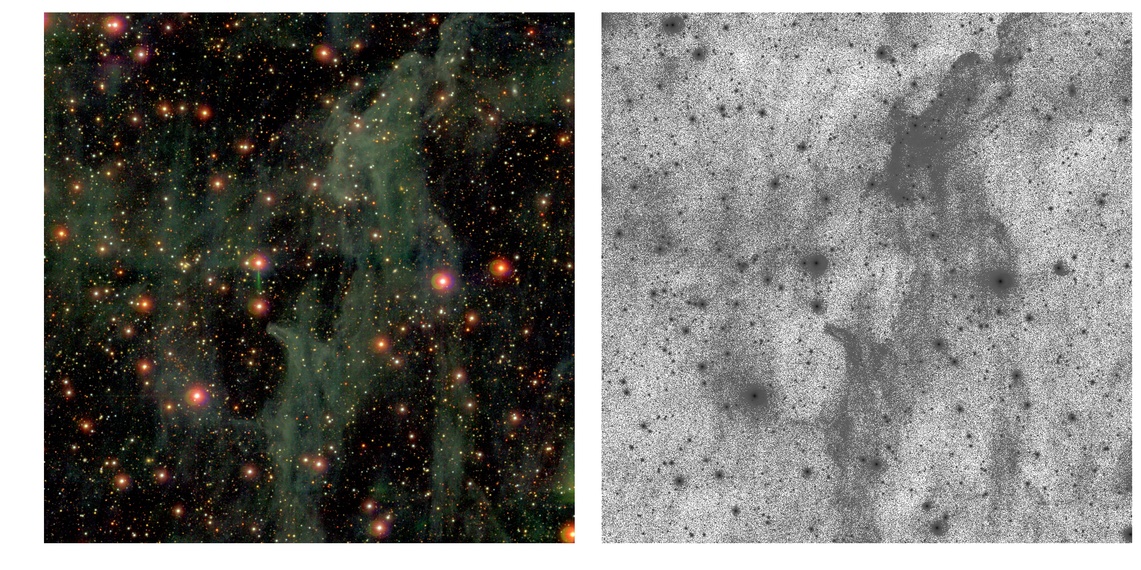}
			\caption{\textit{Left:} \textit{gri}-composite image. \textit{Right:} \textit{g}-band input image in log scale. }
		\end{subfigure}
		
		\begin{subfigure}[b]{0.8\textwidth}
			\centering  
			\begin{tabular}{m{0.02\textwidth} m{0.2\textwidth} m{0.2\textwidth} m{0.2\textwidth} m{0.2\textwidth}}
				
				\multicolumn{1}{>{\centering\arraybackslash}m{0.02\textwidth}}{} 
				& \multicolumn{1}{>{\centering\arraybackslash}m{0.2\textwidth}}{\textbf{F-Score}}
				& \multicolumn{1}{>{\centering\arraybackslash}m{0.2\textwidth}}{\textbf{Area}}
				& \multicolumn{1}{>{\centering\arraybackslash}m{0.2\textwidth}}{\textbf{Combined A}}
				& \multicolumn{1}{>{\centering\arraybackslash}m{0.2\textwidth}}{\textbf{Combined B}}\\ 
				
				\textbf{SE} & \includegraphics[width=0.2\textwidth]{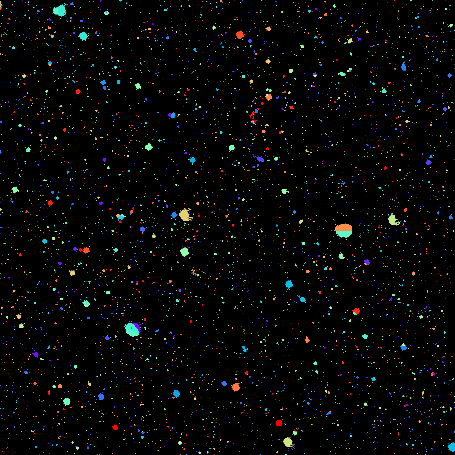} & \includegraphics[width=0.2\textwidth]{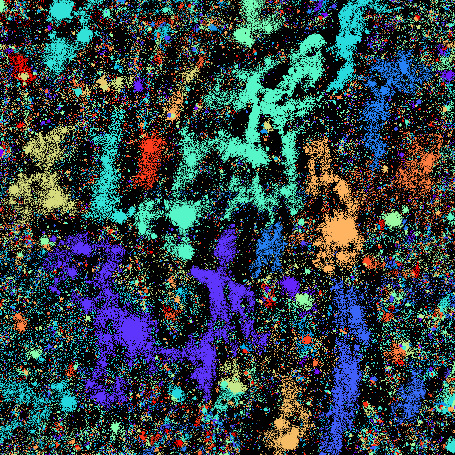} &
				\includegraphics[width=0.2\textwidth]{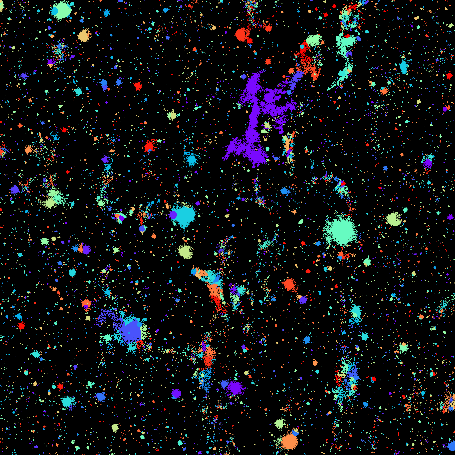}& 
				\includegraphics[width=0.2\textwidth]{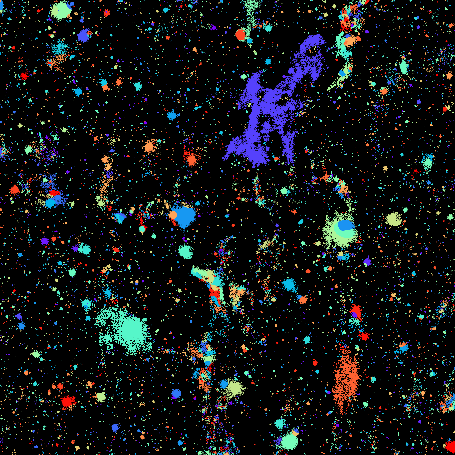}\\
				\textbf{NC} & \includegraphics[width=0.2\textwidth]{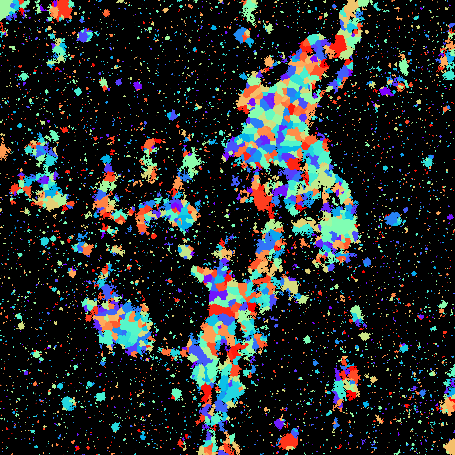} & \includegraphics[width=0.2\textwidth]{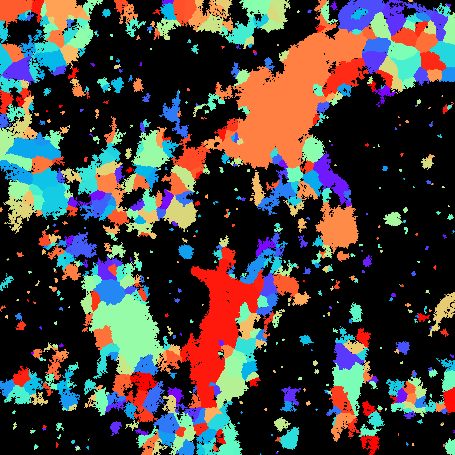} &
				\includegraphics[width=0.2\textwidth]{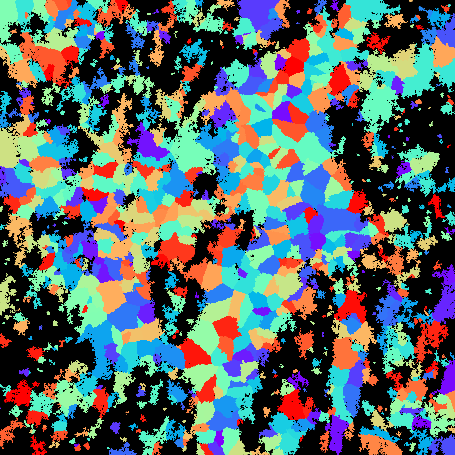}& 
				\includegraphics[width=0.2\textwidth]{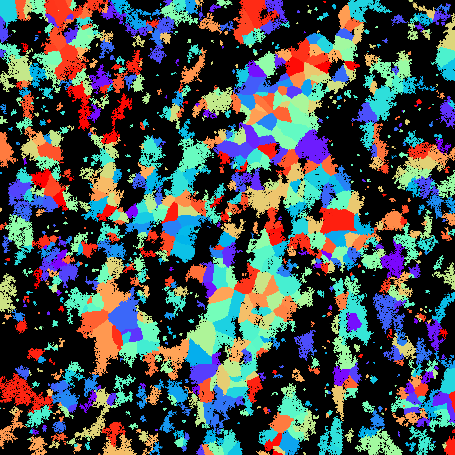}\\
				\textbf{MT}&
				\includegraphics[width=0.2\textwidth]{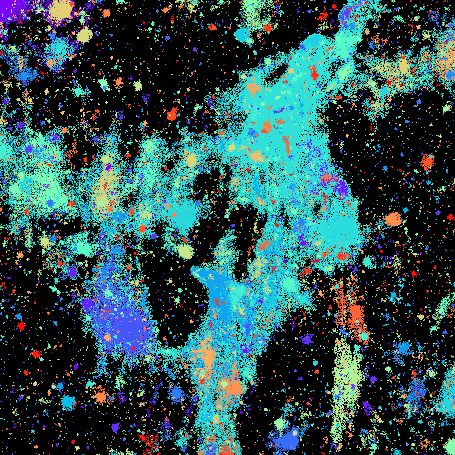} & \includegraphics[width=0.2\textwidth]{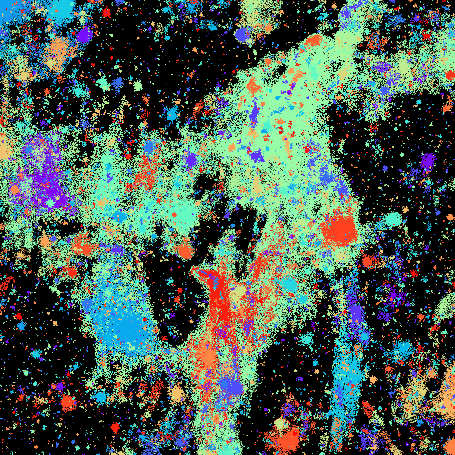} &
				\includegraphics[width=0.2\textwidth]{figs/s82/f0/mto_1_0.png}& 
				\includegraphics[width=0.2\textwidth]{figs/s82/f0/mto_1_0.png}\\
				\textbf{PF}&
				\includegraphics[width=0.2\textwidth]{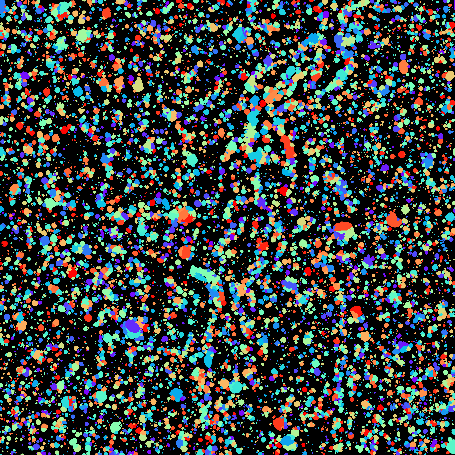} & \includegraphics[width=0.2\textwidth]{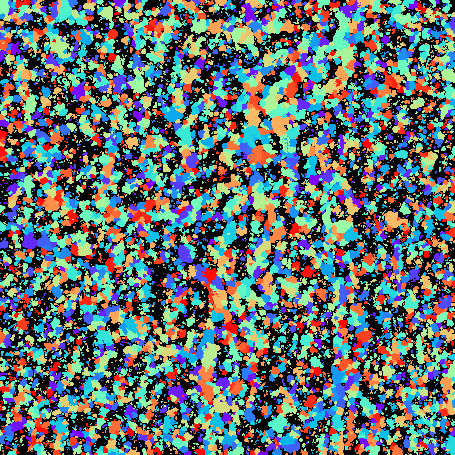} & & \\
			\end{tabular}
			\caption{Segmentation maps, using the parameters which gave the highest median score for each combination of optimisation measure and tool. For more information, see Fig. \ref{biglabels}.}
		\end{subfigure}  
		\caption{Results for IAC Stripe82 field \texttt{f0363\_g.rec.fits} showing a large structure of galactic cirri.}
		\label{s82_f0363}
	\end{figure*}

	Similarly to the segmentation of the simulated images seen Figs. \ref{biglabels} and \ref{smalllabels}, we find that SExtractor detects the least area of regions of light compared to the other tools. In particular, it misses large portions of the galactic cirrus structure in Fig. \ref{s82_f0363}, even when optimised for the area score. As in the case of the simulated images, the fact that many smaller objects (including many false positives) are detected in the background when optimised for the area score is most likely a consequence of the very low threshold used to find larger areas. However, the galactic cirrus structure is highly extended and diffuse with low and high density regions, so the tool is unable to segment the structure as a single object, and fragments it into several pieces. This `failure of detection' may, however, be taken advantage of (with some manual intervention) in studying the properties of galactic cirrus \citep{roman2019galactic}. 
	
\begin{figure*}
		\centering
		\begin{subfigure}[b]{0.3\textwidth}
			\includegraphics[width=\textwidth]{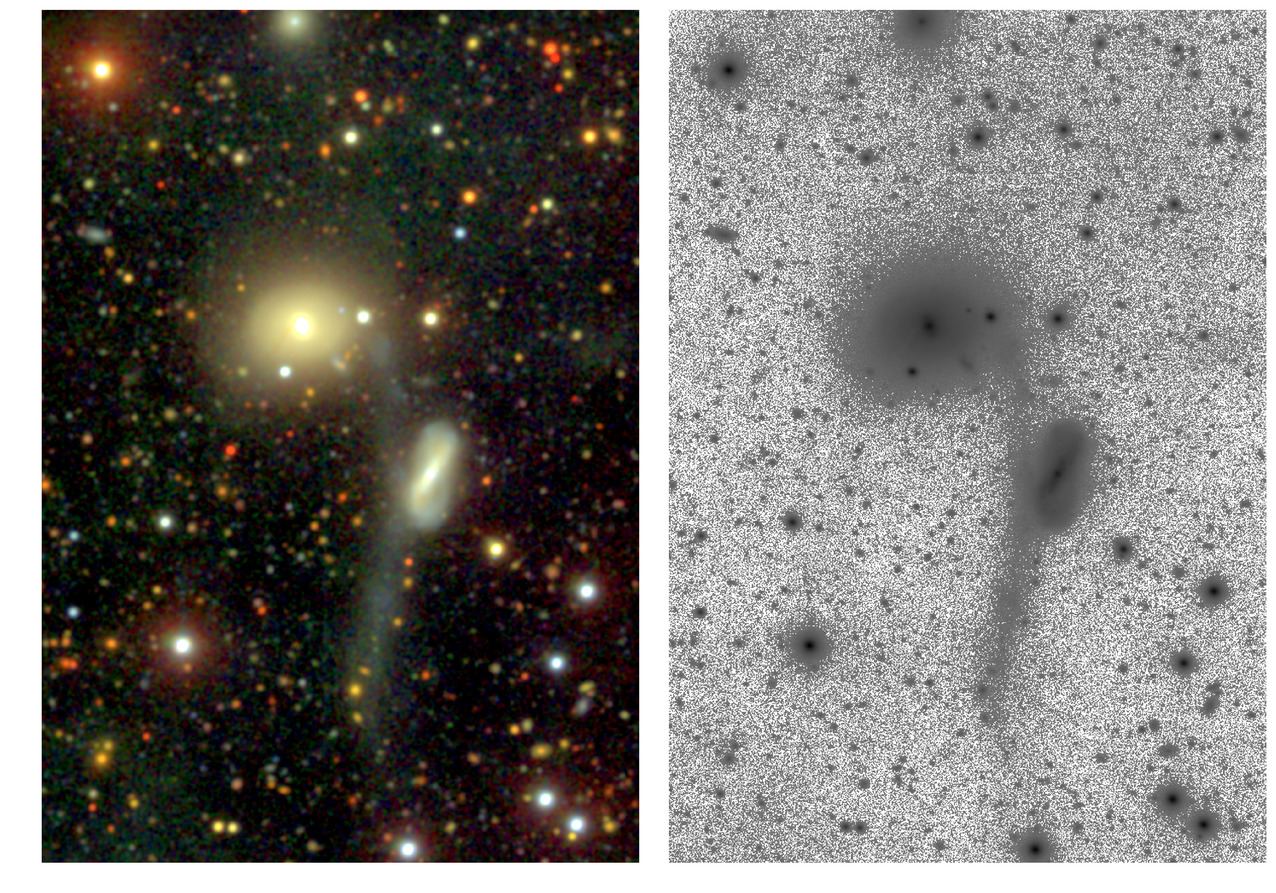}
			\caption{\textit{Left:} \textit{gri}-composite image. \textit{Right:} \textit{g}-band input image in log scale.}
		\end{subfigure}
		\begin{subfigure}[b]{0.8\textwidth}
			\centering  
			\begin{tabular}{m{0.02\textwidth} m{0.2\textwidth} m{0.2\textwidth} m{0.2\textwidth} m{0.2\textwidth}}
				
				\multicolumn{1}{>{\centering\arraybackslash}m{0.02\textwidth}}{} 
				& \multicolumn{1}{>{\centering\arraybackslash}m{0.2\textwidth}}{\textbf{F-Score}}
				& \multicolumn{1}{>{\centering\arraybackslash}m{0.2\textwidth}}{\textbf{Area}}
				& \multicolumn{1}{>{\centering\arraybackslash}m{0.2\textwidth}}{\textbf{Combined A}}
				& \multicolumn{1}{>{\centering\arraybackslash}m{0.2\textwidth}}{\textbf{Combined B}}\\ 
				
				\textbf{SE} & \includegraphics[width=0.2\textwidth]{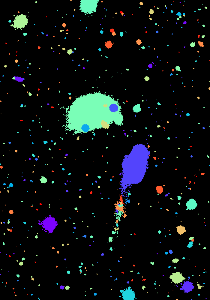} & \includegraphics[width=0.2\textwidth]{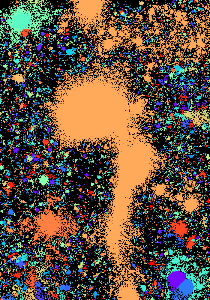} &
				\includegraphics[width=0.2\textwidth]{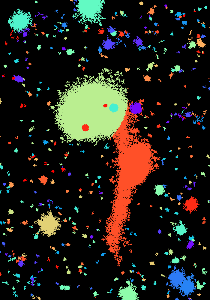}& 
				\includegraphics[width=0.2\textwidth]{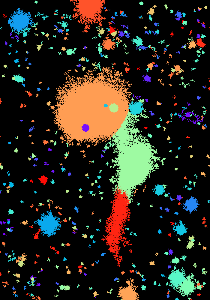}\\
				\textbf{NC} & \includegraphics[width=0.2\textwidth]{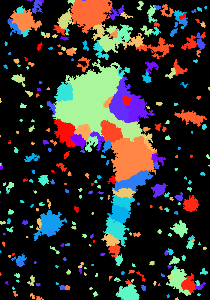} & \includegraphics[width=0.2\textwidth]{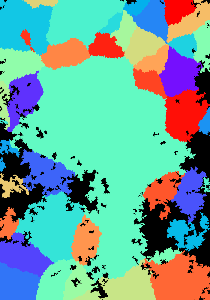} &
				\includegraphics[width=0.2\textwidth]{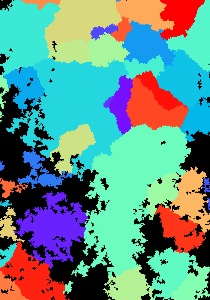}& 
				\includegraphics[width=0.2\textwidth]{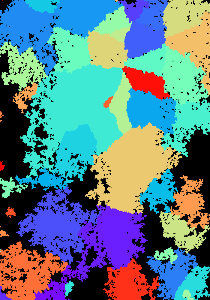}\\
				\textbf{MT}&
				\includegraphics[width=0.2\textwidth]{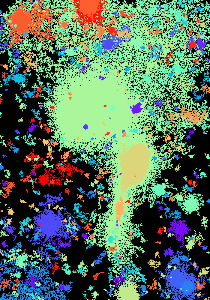} & \includegraphics[width=0.2\textwidth]{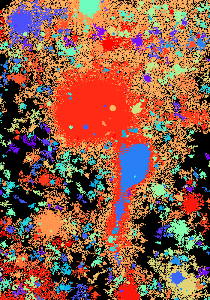} &
				\includegraphics[width=0.2\textwidth]{figs/s82/J0/mto_1_0.png}& 
				\includegraphics[width=0.2\textwidth]{figs/s82/J0/mto_1_0.png}\\
				\textbf{PF}&
				\includegraphics[width=0.2\textwidth]{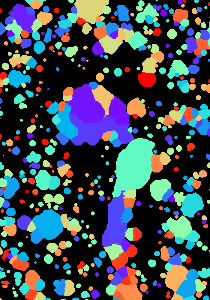} & \includegraphics[width=0.2\textwidth]{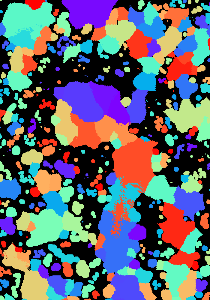} & & \\
			\end{tabular}
			\caption{Segmentation maps, using the parameters which gave the highest median score for each combination of optimisation measure and tool. For more information, see Fig. \ref{biglabels}.}
		\end{subfigure}
		\caption{Results for IAC Stripe82 field cropped to show two interacting galaxies (SDSS \texttt{J031943.04+003355.64} and SDSS \texttt{J031947.01+003504.44}).}
		\label{s82_J031943}
		
	\end{figure*}

\begin{figure*}
		\centering
		\begin{subfigure}[b]{0.6\textwidth}
			\includegraphics[width=\textwidth]{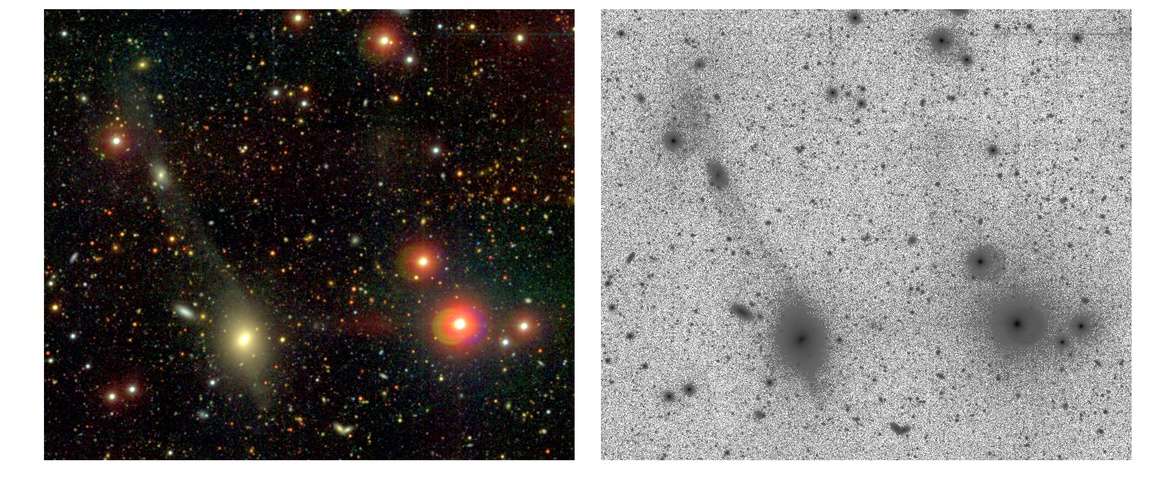}
			\caption{\textit{Left:} \textit{gri}-composite image. \textit{Right:} \textit{g}-band input image in log scale.}
		\end{subfigure}
		\begin{subfigure}[b]{0.8\textwidth}
			\centering  
			\begin{tabular}{m{0.02\textwidth} m{0.2\textwidth} m{0.2\textwidth} m{0.2\textwidth} m{0.2\textwidth}}
				
				\multicolumn{1}{>{\centering\arraybackslash}m{0.02\textwidth}}{} 
				& \multicolumn{1}{>{\centering\arraybackslash}m{0.2\textwidth}}{\textbf{F-Score}}
				& \multicolumn{1}{>{\centering\arraybackslash}m{0.2\textwidth}}{\textbf{Area}}
				& \multicolumn{1}{>{\centering\arraybackslash}m{0.2\textwidth}}{\textbf{Combined A}}
				& \multicolumn{1}{>{\centering\arraybackslash}m{0.2\textwidth}}{\textbf{Combined B}}\\ 
				
				\textbf{SE} & \includegraphics[width=0.2\textwidth]{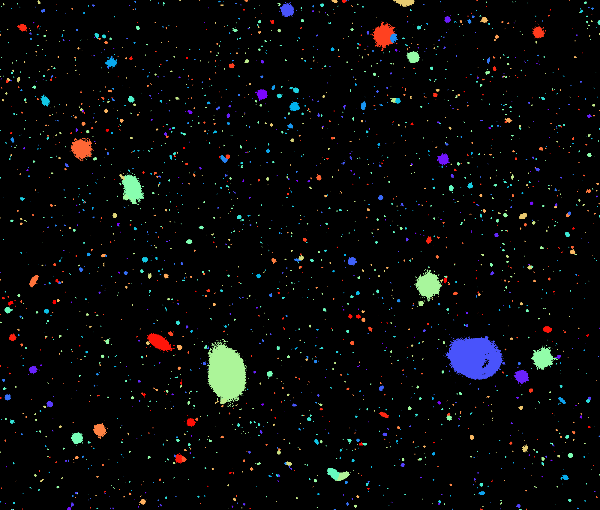} & \includegraphics[width=0.2\textwidth]{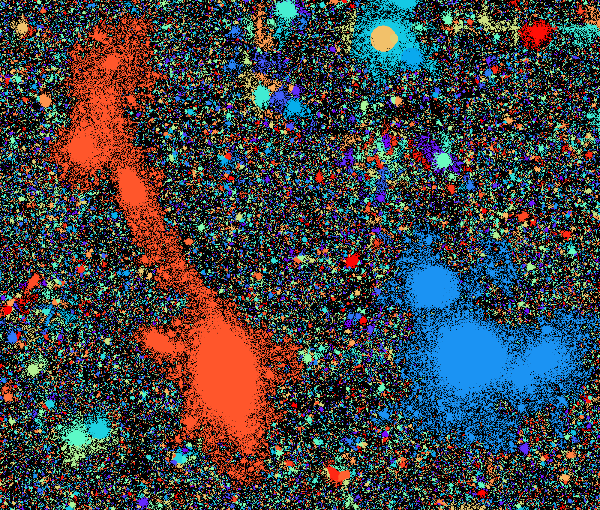} &
				\includegraphics[width=0.2\textwidth]{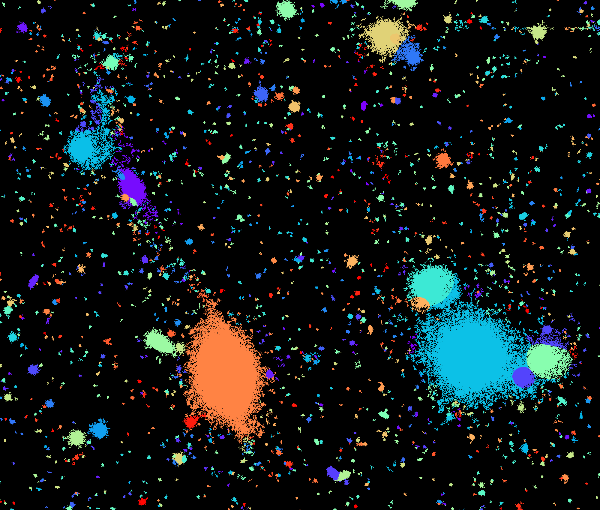}& 
				\includegraphics[width=0.2\textwidth]{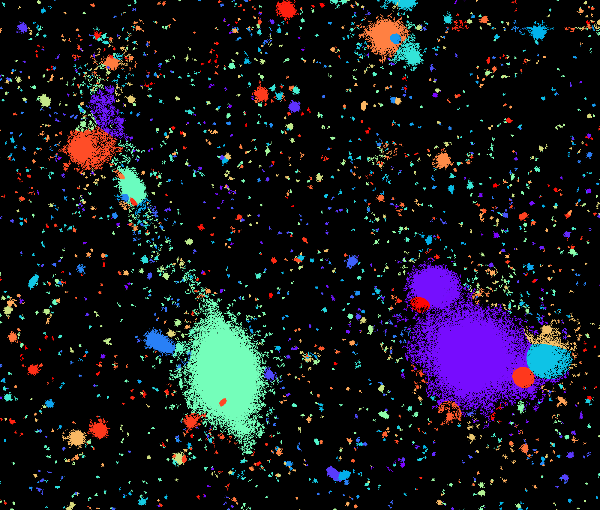}\\
				\textbf{NC} & \includegraphics[width=0.2\textwidth]{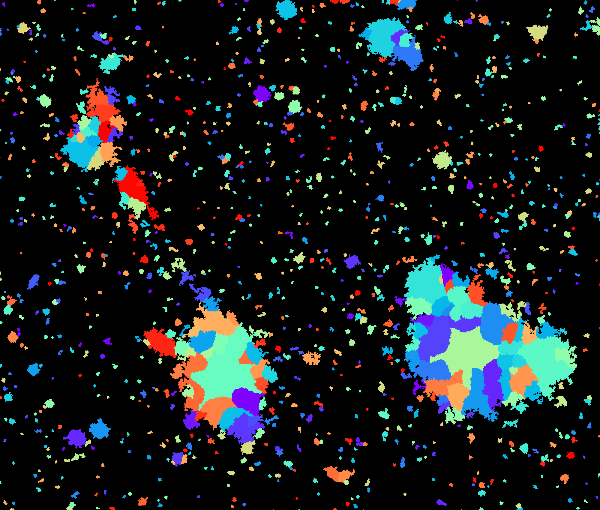} & \includegraphics[width=0.2\textwidth]{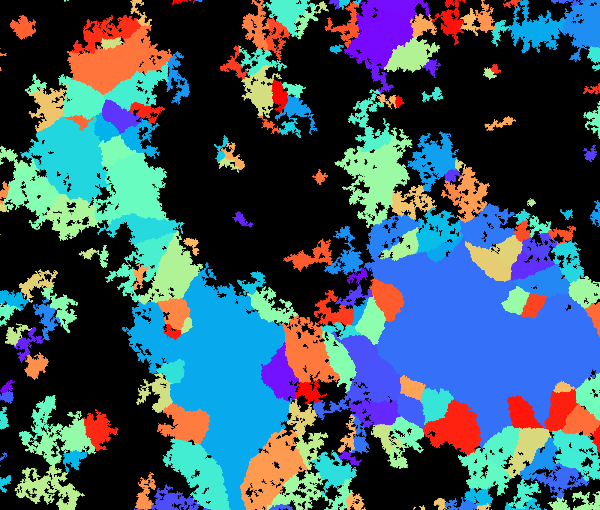} &
				\includegraphics[width=0.2\textwidth]{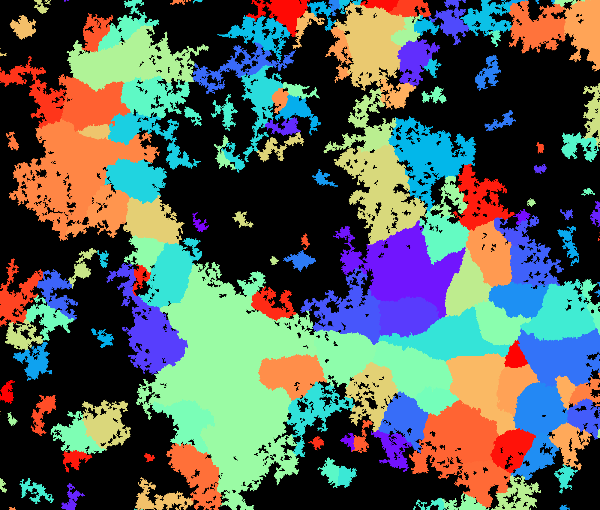}& 
				\includegraphics[width=0.2\textwidth]{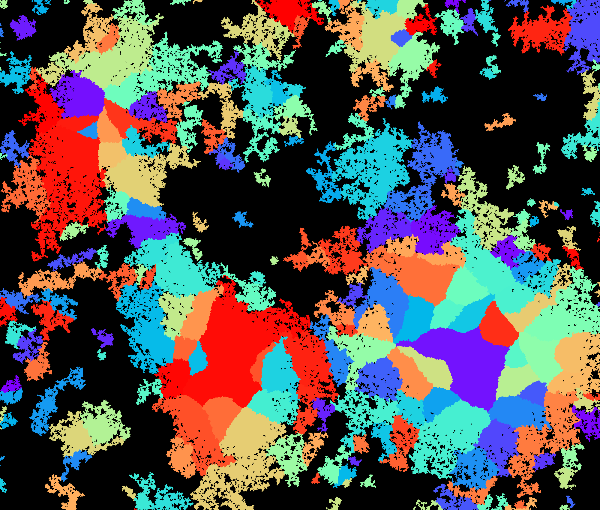}\\
				\textbf{MT}&
				\includegraphics[width=0.2\textwidth]{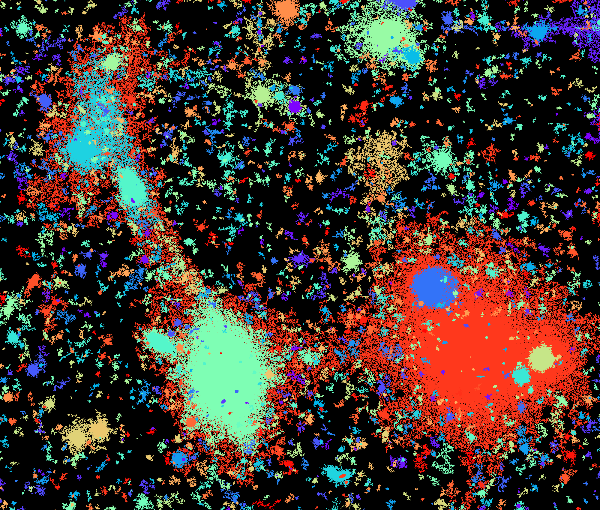} & \includegraphics[width=0.2\textwidth]{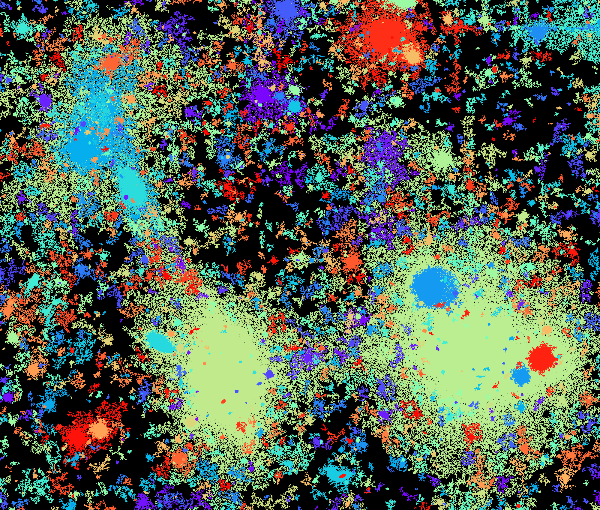} &
				\includegraphics[width=0.2\textwidth]{figs/s82/J2/mto_1_0.png}& 
				\includegraphics[width=0.2\textwidth]{figs/s82/J2/mto_1_0.png}\\
				\textbf{PF}&
				\includegraphics[width=0.2\textwidth]{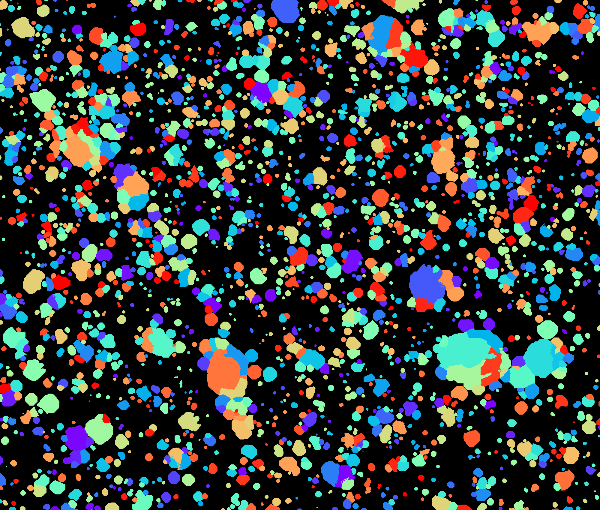} & \includegraphics[width=0.2\textwidth]{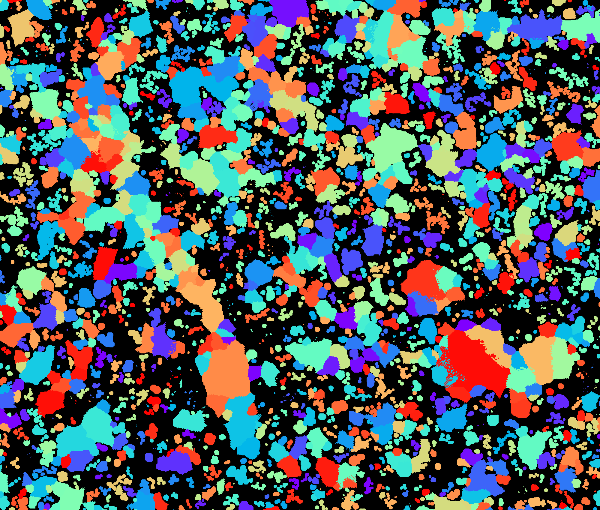} & & \\
			\end{tabular}
			\caption{Segmentation maps, using the parameters which gave the highest median score for each combination of optimisation measure and tool. For more information, see Fig. \ref{biglabels}.}
		\end{subfigure}
		\caption{Results for IAC Stripe82 field zoomed in on elliptical galaxy with an extended, very faint tidal stream (SDSS \texttt{J235618.80-001820.17}) and a bright collection of stars with a significant amount of scattered light contaminating the galaxy from the right.}
		\label{s82_J235618}
	\end{figure*}

	Remarkably, the performance of SExtractor on the image of the interacting galaxies connected with a tidal stream in Fig. \ref{s82_J031943} is much better on the main objects with parameters optimised for area score and both combined scores, whilst performing poorly in the background. Similar observations can be made in the case of the elliptical galaxy with a large stream in Fig. \ref{s82_J235618}, but this stream is much fainter than in the interacting galaxies case, and SExtractor detects the stream in fragments (similar to the galactic cirrus). \par 
	
	In contrast, for all the IAC Stripe82 images, both NoiseChisel and MTObjects detect the largest amount of light as distinct objects or diffuse regions (reflected in the highest optimisation scores).  Visually, NoiseChisel's performance seems better when optimised for F-Score compared to the other scores, but there is still diffuse light around the objects which have gone undetected. When optimised for area score or the combined scores, this missing light is recovered; but as mentioned previously, the algorithm seems to segment structures within larger objects rather arbitrarily. When comparing the outputs from each optimisation method, we can see that the substructure is segmented quite differently in all the IAC Stripe82 examples. This is probably a consequence of growing the `clumps' (as detected in the \texttt{CLUMPS} output of Segment) to cover the full detected area -- if the detected area is different, then the growth of the clumps seems to vary. This effect is visible in comparing the tool's output when optimised for all four measures in all the IAC Stripe 82 examples. The fact that the substructure over the detected regions seems visually arbitrary may not be an issue in some cases -- such as when segmentation maps are used for reducing datasets where all pixels with a significant amount of signal above the background needs to be masked for processing (see for example \citet{borlaff2019abyss}), or when the user is simply not concerned with the substructure of astronomical sources.\protect\footnote{The NoiseChisel manual ( \protect\url{https://www.gnu.org/software/gnuastro/manual/html_node/NoiseChisel.html}) states that the user may choose to run Segment after NoiseChisel depending on whether they want to analyse the substructure of sources.} 
	
	However, for studies where more accurate segmentation of tidal streams and nested objects (or substructure) is required for photometric calculations, it is not possible to automatically allocate which of these fragmented regions belong to the host structure, and the user may need to manually select regions of interest. This is especially visible for the large galactic cirrus in Fig. \ref{s82_f0363} and the faint stream in Fig. \ref{s82_J235618} where the structures are segmented into separate objects of all kinds of shapes.  \par 
	
	For the same IAC Stripe82 examples, a similar observation can be made for MTObjects, but the partitioning better follows the visual shape of all objects (background and nested). This behaviour means that the user is able to make a visual mapping between the input image and segmentation map much more easily, if they need to manually select regions of interest. In comparing, the outputs of the tool when optimised for the different scores are very similar -- the outputs for the area and combined scores are the same, and the only visible difference with F-Score is the extent to which the edges are fractured outwards. Compared to the other tools, the existence of these highly fractured edges of the segmented regions in MTObjects may not be an appealing characteristic for the user if smoother edges are required -- for instance to make photometric calculations, such as the total magnitude of objects.\footnote{Of the tools, this effect in the segmentation maps can only be controlled in NoiseChisel without compromising the extent to which objects are detected.} 
	
	Another characteristic of MTObjects can be seen in the field contaminated by a cluster of bright stars to the right of an elliptical galaxy in Fig. \ref{s82_J235618}. MTObjects allocates the diffuse stream and faint halo around the core of the galaxy to the cluster of stars. This is clearly a problem with how the detected regions are represented -- MTObjects is finding the diffuse regions in the image (at least those that could be visually identified in this example), but allocating it to the wrong object. This means that the user will need to once again manually select which regions belong to the galaxy, and this may not always be possible to identify in advance when dealing with deep datasets. Apart from these exceptions, MTObjects performs fairly similarly and consistently across the IAC Stripe82 images tested in this work.
	
	Due to speed, at the time of writing we are only able to complete the optimisation of parameters for ProFound using the F-Score and area score. In this work, we find that when optimised for these scores, only in the merging galaxies case in Fig. \ref{s82_J031943} does the tool segment the galaxy shape and its companion (though also fragmented into several arbitrary pieces, as in NoiseChisel's segmentation). In all of the other images, the large galaxies or structures are barely visible, and only because our eye is able to connect the smaller fragments into one connected region. \par

	\subsection{The Hubble Ultra Deep Field}
	In order to examine the behaviour of the tools on space-based observations, we ran the tools on the $V_{606}$-band of the Hubble Ultra Deep Field (HUDF). As mentioned previously in Sect. \ref{sec:data}, the HUDF is the deepest data used in this work, a point source depth of 29.3\,mag \citep{beckwith2006hubble} which is equivalent to a limiting surface brightness depth of $\mu_{V_{606}} \sim 32.5$\,mag/arcsec$^2$\,(3$\sigma;10\times10$\,arcsec$^2$).
	
	As the original drizzled image contained wide, zero-valued borders, we rotated and cropped it to contain as much of the field as possible, while excluding the borders. We then ran the tools on the image, using the same optimised parameters as in the previous sections. We show here the complete image, and two smaller areas of interest containing a type of feature not common in the other surveys: face-on spiral galaxies with visible substructures.
	
	Besides these artefacts, the behaviour of all four tools on the HUDF image appears to be generally similar to behaviour on the images from other surveys, despite having a higher depth and a different telescope type.
	
\begin{figure*}
		\centering
		\begin{subfigure}[b]{0.25\textwidth}
			\includegraphics[width=\textwidth]{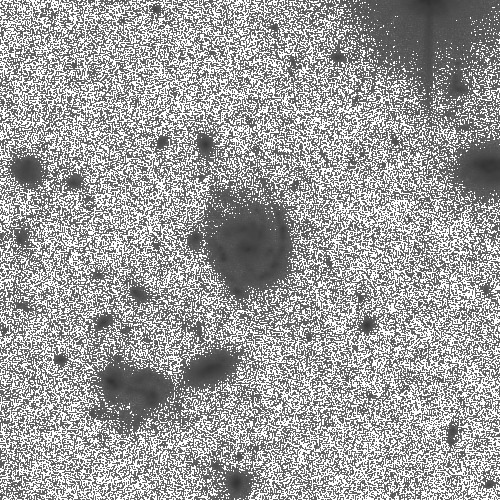}
			\caption{Input image -- the v-band of the field.}
		\end{subfigure}
		\begin{subfigure}[b]{0.8\textwidth}
			\centering  
			\begin{tabular}{m{0.02\textwidth} m{0.2\textwidth} m{0.2\textwidth} m{0.2\textwidth} m{0.2\textwidth}}
				
				\multicolumn{1}{>{\centering\arraybackslash}m{0.02\textwidth}}{} 
				& \multicolumn{1}{>{\centering\arraybackslash}m{0.2\textwidth}}{\textbf{F-Score}}
				& \multicolumn{1}{>{\centering\arraybackslash}m{0.2\textwidth}}{\textbf{Area}}
				& \multicolumn{1}{>{\centering\arraybackslash}m{0.2\textwidth}}{\textbf{Combined A}}
				& \multicolumn{1}{>{\centering\arraybackslash}m{0.2\textwidth}}{\textbf{Combined B}}\\ 
				
				\textbf{SE} & \includegraphics[width=0.2\textwidth]{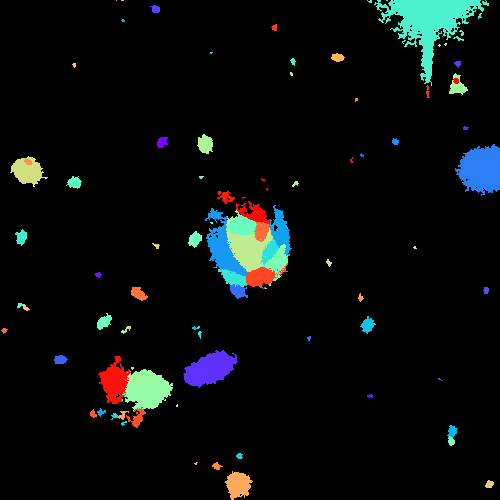} & \includegraphics[width=0.2\textwidth]{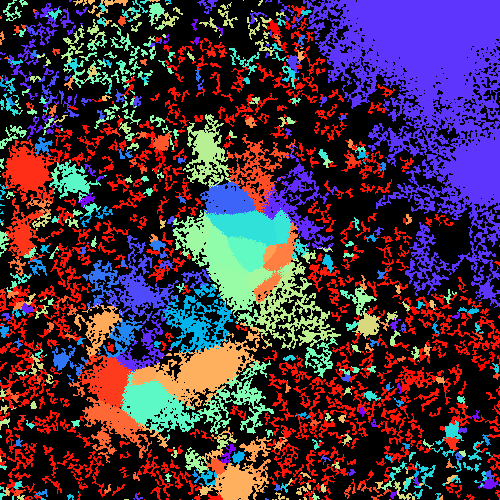} &
				\includegraphics[width=0.2\textwidth]{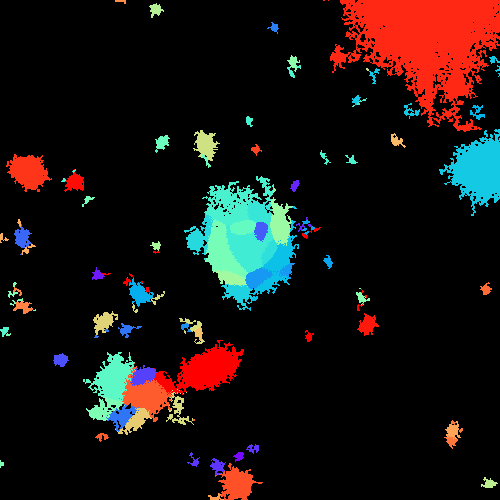}& 
				\includegraphics[width=0.2\textwidth]{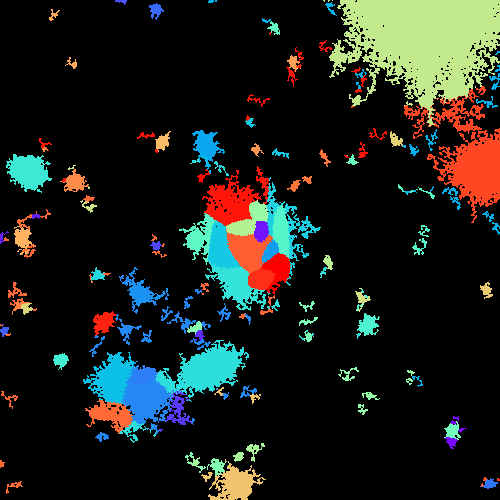}\\
				\textbf{NC} & \includegraphics[width=0.2\textwidth]{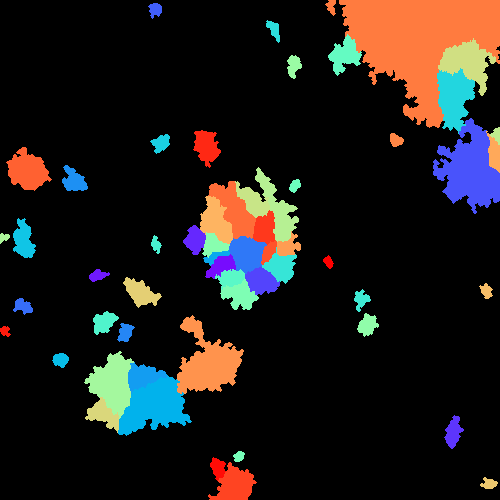} & \includegraphics[width=0.2\textwidth]{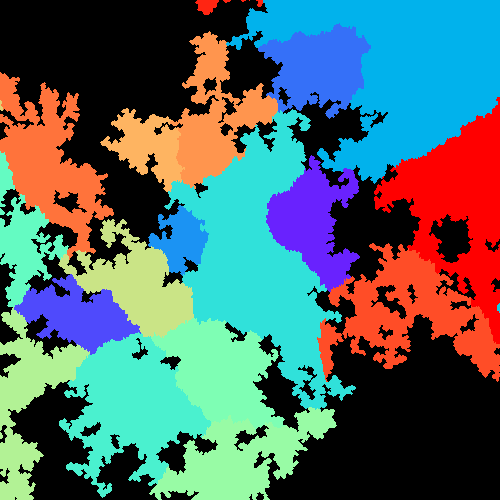} &
				\includegraphics[width=0.2\textwidth]{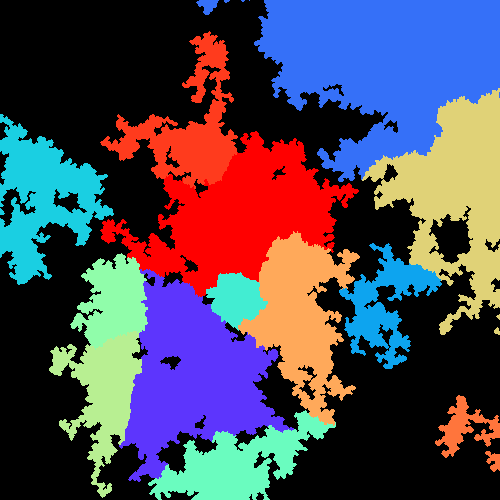}& 
				\includegraphics[width=0.2\textwidth]{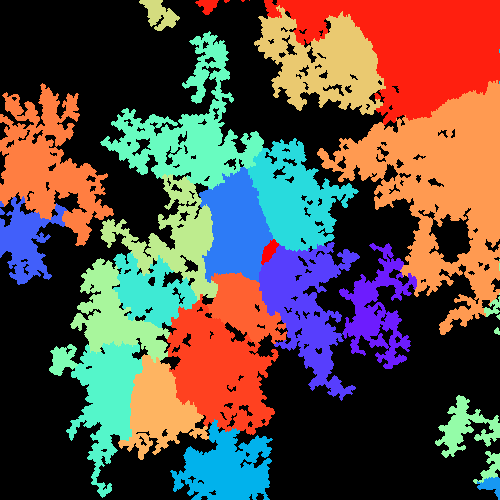}\\
				\textbf{MT}&
				\includegraphics[width=0.2\textwidth]{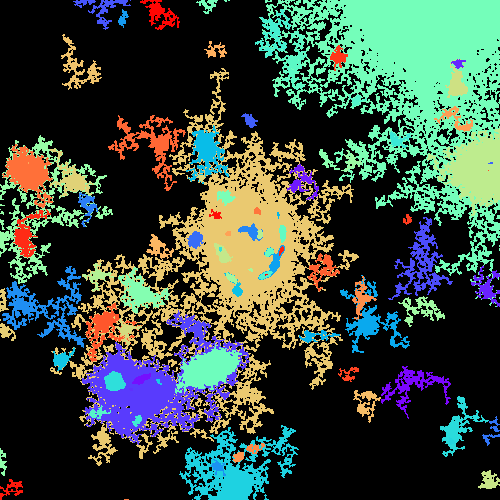} & \includegraphics[width=0.2\textwidth]{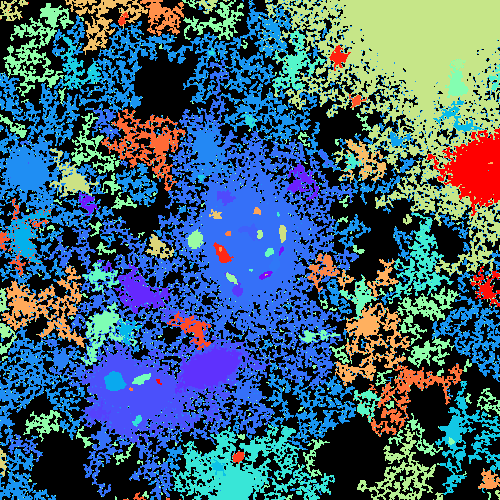} &
				\includegraphics[width=0.2\textwidth]{figs/udf/crop1/mto_1_1.png}& 
				\includegraphics[width=0.2\textwidth]{figs/udf/crop1/mto_1_1.png}\\
				\textbf{PF}&
				\includegraphics[width=0.2\textwidth]{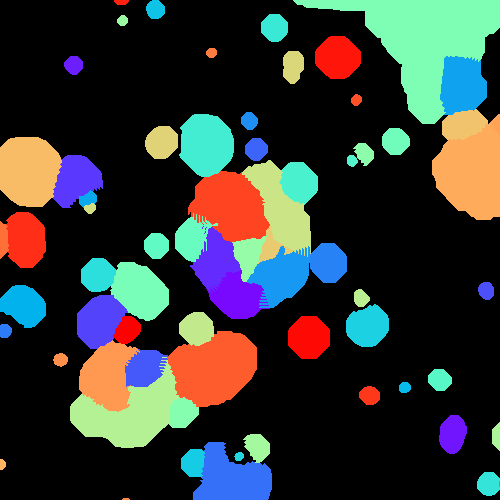} & \includegraphics[width=0.2\textwidth]{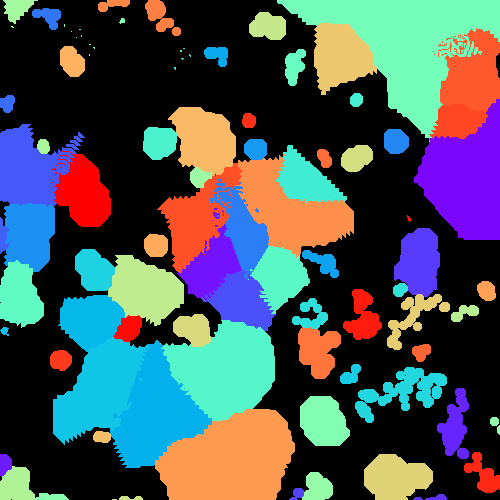} & & \\
			\end{tabular}
			
			\caption{Segmentations of the field, using the parameters which gave the highest median score for each combination of optimisation measure and tool. For more information, see Fig. \ref{biglabels}.}
		\end{subfigure}  
		\caption{Segmentations of a section of the Hubble Ultra Deep Field.}
		\label{hubble_1}
	\end{figure*}

	This is further corroborated by Fig. \ref{hubble_1}, which shows a face-on spiral galaxy, as well as several smaller elliptical galaxies. As before, SExtractor finds only the bright centres of objects, except when optimised for area; however, it noticeably divides the spiral galaxy into chunks where there is substructure. Somewhat arbitrary division of the galaxy is also visible in the results of NoiseChisel and ProFound; with NoiseChisel capturing more of the outskirts, as before. MTObjects appears to be the most successful at segmenting the spiral, with the majority of the galaxy captured as a single object, with smaller structures nested within it, although, as in previous instances, the outskirts are fractured.
	
\begin{figure*}
		\centering
		\begin{subfigure}[b]{0.25\textwidth}
			\includegraphics[width=\textwidth]{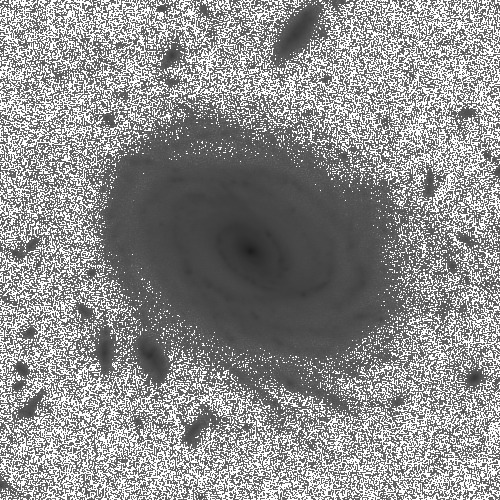}
			\caption{Input image -- the v-band of the field.}
		\end{subfigure}
		\begin{subfigure}[b]{0.8\textwidth}
			\centering  
			\begin{tabular}{m{0.02\textwidth} m{0.2\textwidth} m{0.2\textwidth} m{0.2\textwidth} m{0.2\textwidth}}
				
				\multicolumn{1}{>{\centering\arraybackslash}m{0.02\textwidth}}{} 
				& \multicolumn{1}{>{\centering\arraybackslash}m{0.2\textwidth}}{\textbf{F-Score}}
				& \multicolumn{1}{>{\centering\arraybackslash}m{0.2\textwidth}}{\textbf{Area}}
				& \multicolumn{1}{>{\centering\arraybackslash}m{0.2\textwidth}}{\textbf{Combined A}}
				& \multicolumn{1}{>{\centering\arraybackslash}m{0.2\textwidth}}{\textbf{Combined B}}\\ 
				
				\textbf{SE} & \includegraphics[width=0.2\textwidth]{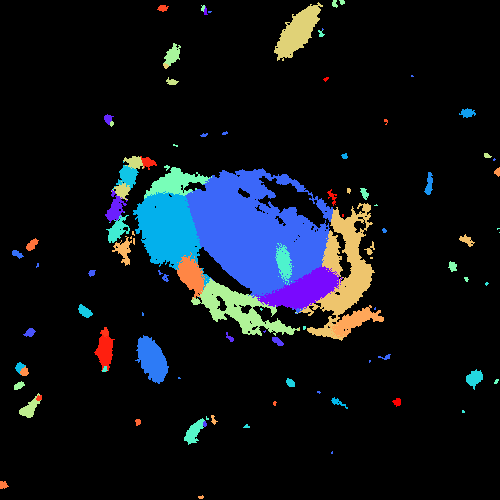} & \includegraphics[width=0.2\textwidth]{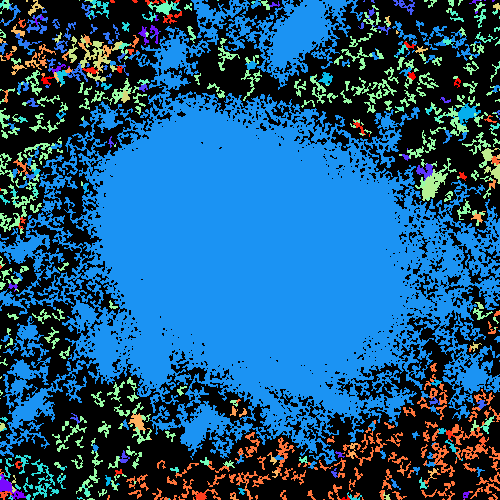} &
				\includegraphics[width=0.2\textwidth]{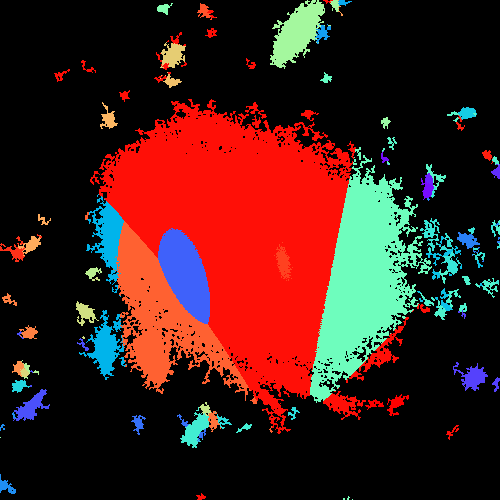}& 
				\includegraphics[width=0.2\textwidth]{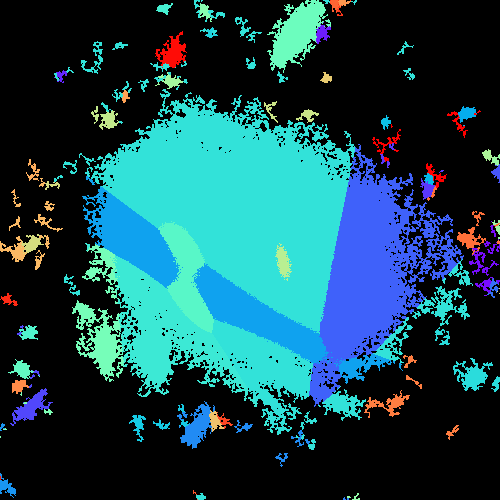}\\
				\textbf{NC} & \includegraphics[width=0.2\textwidth]{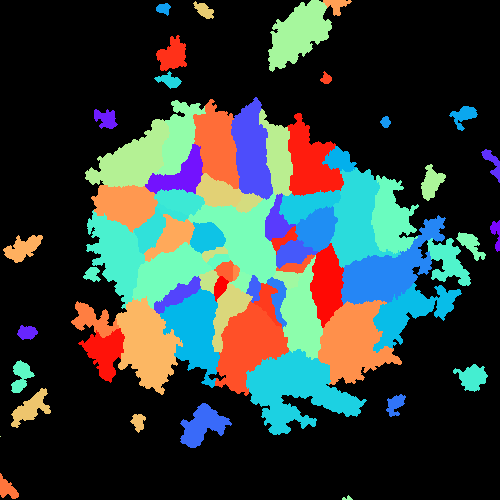} & \includegraphics[width=0.2\textwidth]{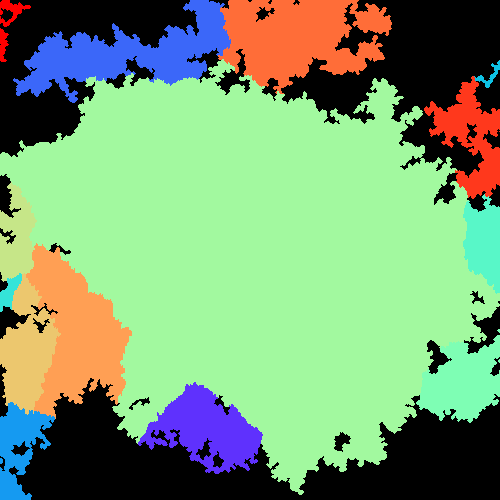} &
				\includegraphics[width=0.2\textwidth]{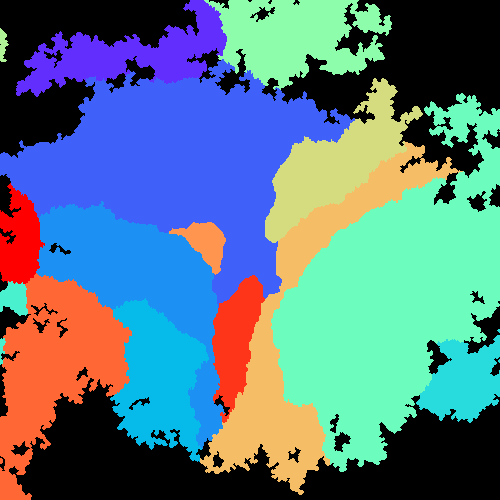}& 
				\includegraphics[width=0.2\textwidth]{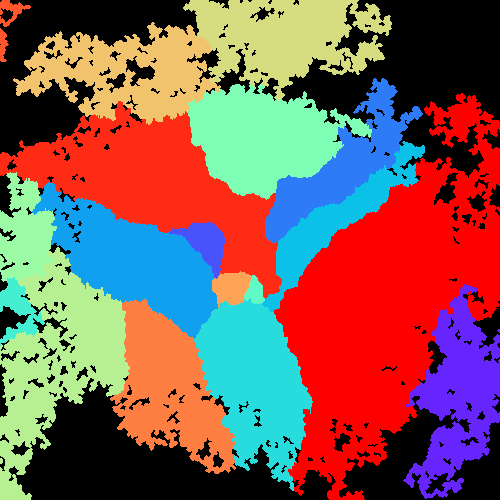}\\
				\textbf{MT}&
				\includegraphics[width=0.2\textwidth]{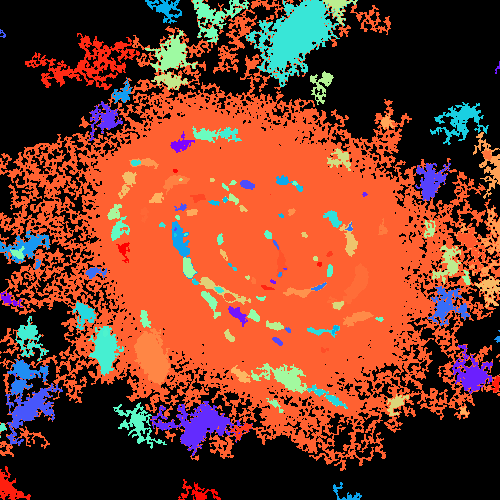} & \includegraphics[width=0.2\textwidth]{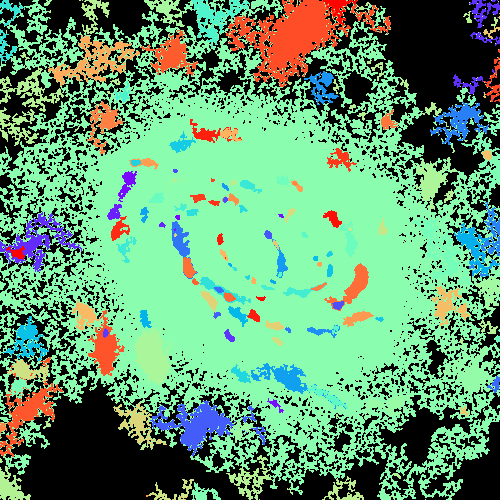} &
				\includegraphics[width=0.2\textwidth]{figs/udf/crop3/mto_1_3.png}& 
				\includegraphics[width=0.2\textwidth]{figs/udf/crop3/mto_1_3.png}\\
				\textbf{PF}&
				\includegraphics[width=0.2\textwidth]{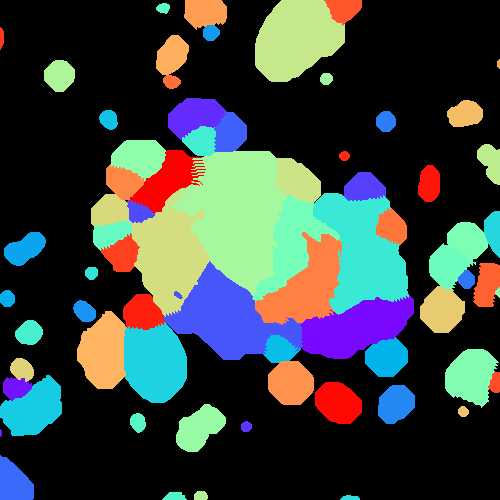} & \includegraphics[width=0.2\textwidth]{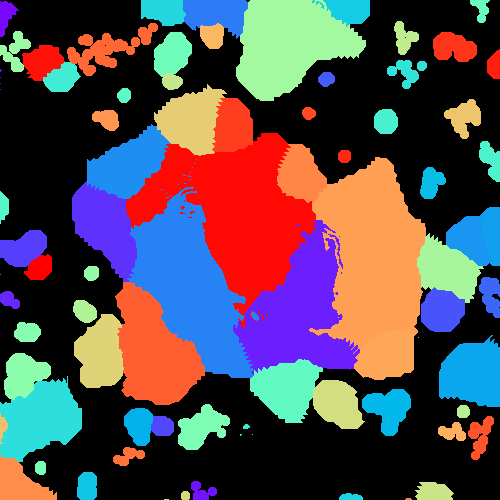} & & \\
			\end{tabular}
			
			\caption{Segmentations of the field, using the parameters which gave the highest median score for each combination of optimisation measure and tool. For more information, see Fig. \ref{biglabels}.}
		\end{subfigure}  
		\caption{Segmentations of a face-on spiral in the Hubble Ultra Deep Field.}
		\label{hubble_2}
	\end{figure*}

	In Fig. \ref{hubble_2}, which shows a larger spiral galaxy displayed at the same scale, the tools have even greater difficulty segmenting the galaxy in a meaningful way. As before, MTObjects has the most success in separating nested structures without fragmenting the overall structure of the object. NoiseChisel is also consistent with previous behaviour. In contrast, ProFound produces quite different segmentations, with a far less blobby appearance. SExtractor produces quite poor segmentations when area is included in the optimisation; with elongated ovals being found in both of the combined score images.
	
	It must be remembered that the parameters were optimised for images in quite different conditions, so it is difficult to quantify how much these inaccurate segmentations are caused by parameters ill-suited for this context. However, as the behaviour is very similar to that shown in the images from different surveys, it is reasonable to expect that it is largely caused by inherent limitations of the tools.
	
	\begin{figure*}
		\centering
		\begin{subfigure}[b]{0.25\textwidth}
			\includegraphics[width=\textwidth]{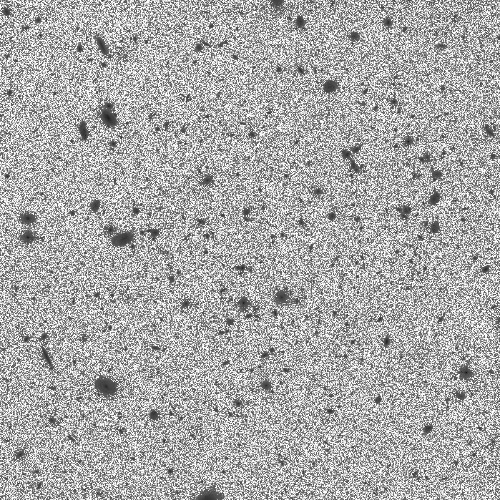}
			\caption{Input image -- the v-band of the field.}
		\end{subfigure}
		\begin{subfigure}[b]{0.8\textwidth}
			\centering  
			\begin{tabular}{m{0.02\textwidth} m{0.2\textwidth} m{0.2\textwidth} m{0.2\textwidth} m{0.2\textwidth}}
				
				\multicolumn{1}{>{\centering\arraybackslash}m{0.02\textwidth}}{} 
				& \multicolumn{1}{>{\centering\arraybackslash}m{0.2\textwidth}}{\textbf{F-Score}}
				& \multicolumn{1}{>{\centering\arraybackslash}m{0.2\textwidth}}{\textbf{Area}}
				& \multicolumn{1}{>{\centering\arraybackslash}m{0.2\textwidth}}{\textbf{Combined A}}
				& \multicolumn{1}{>{\centering\arraybackslash}m{0.2\textwidth}}{\textbf{Combined B}}\\ 
				
				\textbf{SE} & \includegraphics[width=0.2\textwidth]{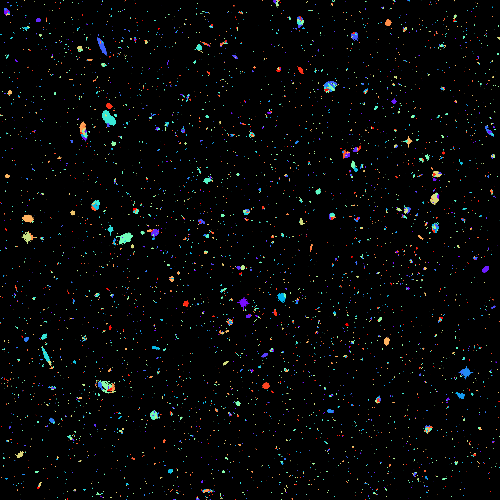} & \includegraphics[width=0.2\textwidth]{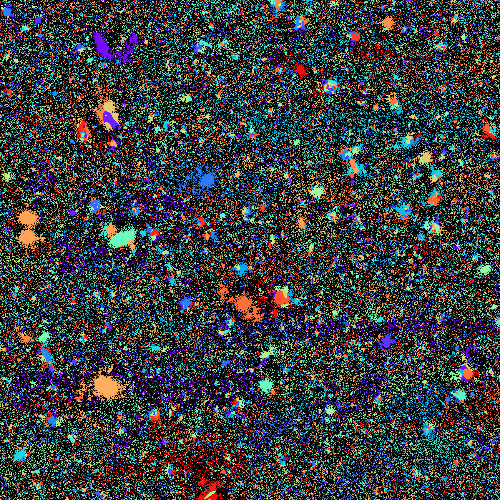} &
				\includegraphics[width=0.2\textwidth]{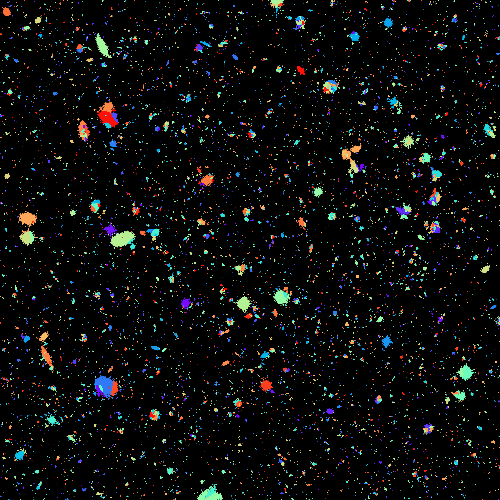}& 
				\includegraphics[width=0.2\textwidth]{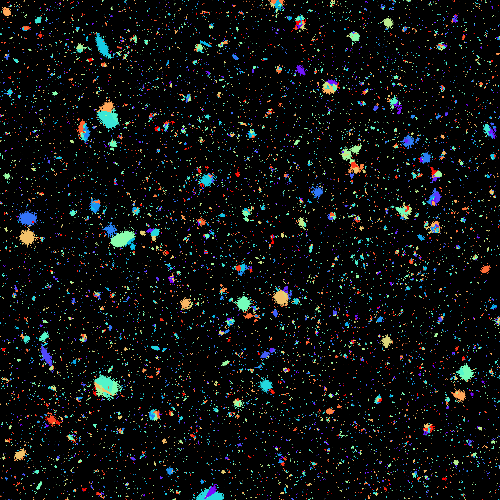}\\
				\textbf{NC} & \includegraphics[width=0.2\textwidth]{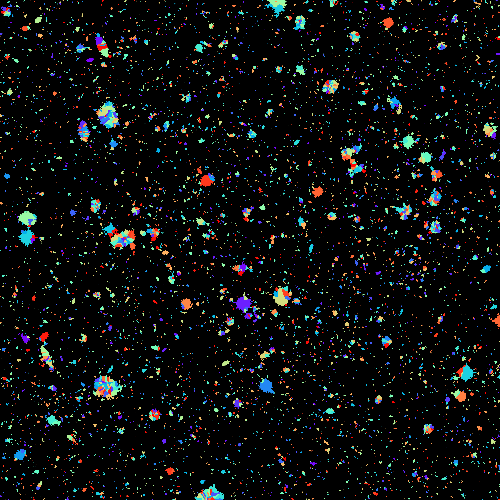} & \includegraphics[width=0.2\textwidth]{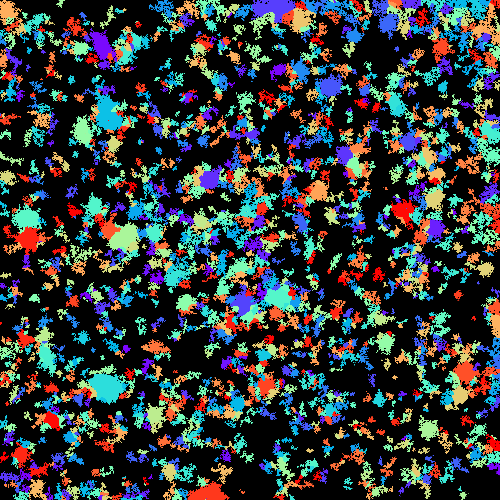} &
				\includegraphics[width=0.2\textwidth]{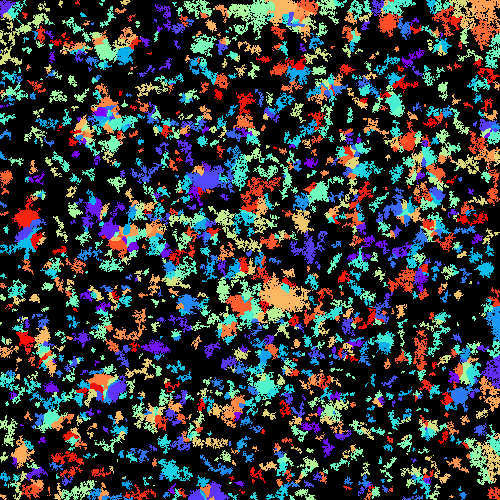}& 
				\includegraphics[width=0.2\textwidth]{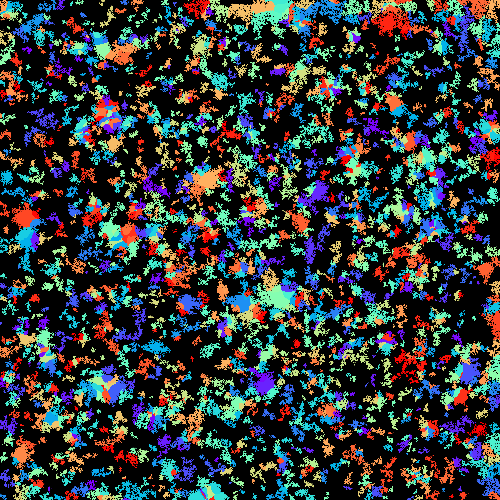}\\
				\textbf{MT}&
				\includegraphics[width=0.2\textwidth]{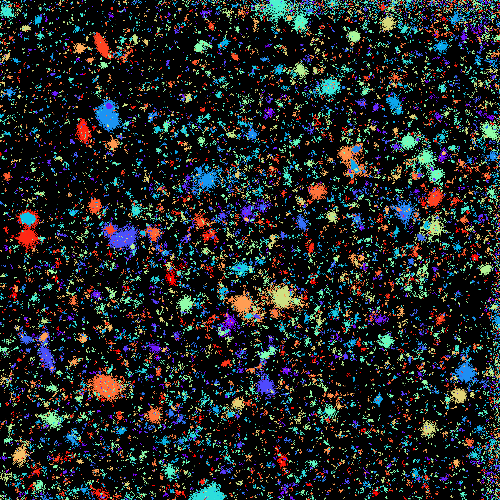} & \includegraphics[width=0.2\textwidth]{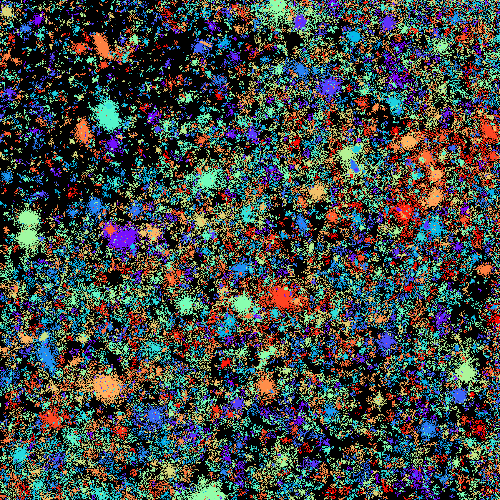} &
				\includegraphics[width=0.2\textwidth]{figs/udf/crop0/mto_1_0.png}& 
				\includegraphics[width=0.2\textwidth]{figs/udf/crop0/mto_1_0.png}\\
				\textbf{PF}&
				\includegraphics[width=0.2\textwidth]{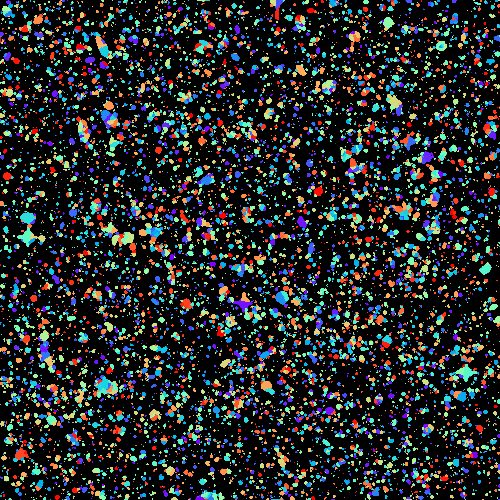} & \includegraphics[width=0.2\textwidth]{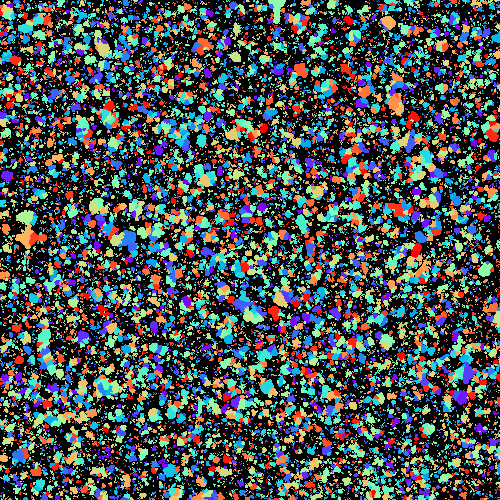} & & \\
			\end{tabular}
			
			\caption{Segmentations of the field, using the parameters which gave the highest median score for each combination of optimisation measure and tool. For more information, see Fig. \ref{biglabels}.}
		\end{subfigure}  
		\caption{Segmentations of the rotated and cropped Hubble Ultra Deep Field.}
		\label{hubble_0}
	\end{figure*}
	
	\subsection{Usability}
	As shown in the parameter tables in Appendix \ref{optables}, only MTObjects reliably found the same set of `optimal' parameters over multiple optimisations. All of the other tools appeared to have multiple locally optimum parameter combinations. This has a negative impact on ease of use -- users manually configuring a tool through trial and error may fail to find globally optimum parameters, and be unaware of this fact. The best parameters may also be dependent on the image used for optimisation -- the parameters found for one image or survey may not produce optimal results when applied to others.
	
	All four programs define parameters in terms of the individual steps of the method (e.g. use $n$ thresholds), rather than in terms of how they affect the overall detection (e.g. detect objects to a given degree of certainty). Without using an optimisation framework, users have no choice but to manually select settings that visually produce a good result, but which do not necessarily have any scientific justification for being chosen. This is further exacerbated by large parameter spaces in the cases of NoiseChisel and ProFound allowing the user to infinitely adjust the behaviour of the tools without the implications of their choices being clear. The ability to define performance in terms of the result rather than the process would greatly improve the ease of use of the tools, as well as reducing the opacity of their behaviour.
	
	These are not major problems in the case of a user processing a small number of images, but are compounded when large surveys, requiring automatic segmentation for many images, are considered. The user must select a set of parameters that produces good results for all images in their survey -- an impossible task if the tool requires manual tuning on individual images.
	
	\section{Conclusions}
	All the compared methods were capable of a reasonable level of object detection, as measured by F-score. However, ProFound and SExtractor were incapable of detecting the outskirts of objects with any degree of accuracy. NoiseChisel and MTObjects were both much more accurate at finding these fainter regions, but both had other difficulties -- the `Segment' tool used in NoiseChisel divided detected light into apparently arbitrary regions, whilst MTObjects produced extremely ragged edges and had a tendency to over-allocate faint regions to the brightest objects. NoiseChisel also produced the most accurate background values. 
	
	We found that there appears to be a trade-off between speed and accurate detection of objects' outskirts. SExtractor was capable of the highest speeds by a substantial margin, but was unable to accurately detect faint regions. MTObjects and NoiseChisel were both able to detect these regions, but at the cost of processing speed. There may potentially be improvements to be made on both tools by increased parallelisation and optimisation of the code.
	
	A common weakness in the tools was in accurately deblending nested objects. The MTObjects approach, using tree-based connected morphological filters \citep{salembier09:_connec_operat} deals relatively well when small, faint objects are nested within larger, brighter ones, but performs poorly when more similar objects overlap. In the latter case, the other methods, which are generally based on a form of watershed segmentation \citep{beucher82:_water,roerdink00} might give a better result. This is a non-trivial problem, which merits further investigation.
	
	MTObjects was the only tool to find stable parameters across multiple optimisations, suggesting that it requires the least adjustment for individual images, and may be the best-suited for use in automatic pipelines. Furthermore, in the test on simulated data, it consistently outperformed the other methods, regardless of the quality measure used. The likelihood of ranking in first place out of four, in the case of F-score and area score, in ten tests, is about $10^{-6}$ under the null hypothesis that all tools have equal performance. Despite the modest performance margin with respect to the others, the result is statistically significant.
	
	We found that the optimisation criteria must be chosen carefully in order to produce useful parameters. In particular, we found that optimising for area alone caused a substantial drop in accuracy for SExtractor and ProFound, whereas combining multiple criteria yielded more meaningful results.
	
	As discussed in the introduction, the growth of the scale of modern surveys means that there is a need for segmentation tools which are fast, automatic, and accurate. We found that of the tools tested, MTObjects was capable of the highest scores on both area and detection measures, and had the most consistent parameters, whilst SExtractor obtained the highest speeds, but with much lower accuracy. As noted earlier, a faster implementation of MTObjects already exists, and the developers of ProFound are rewriting parts of their tool to improve its speed.
	
	In addition to this analysis, we have presented a framework for automated parameter setting and evaluation of astronomical source detection tools, which is generic, and can be used with any other quality measure or model ground truth. This procedure can in the future be used to analyse improvements to existing tools, as well as evaluating the capabilities of future techniques.
	
	\section{Acknowledgements}
	We acknowledge financial support from the European Union's Horizon 2020 research and innovation programme under Marie Skłodowska-Curie grant agreement No 721463 to the SUNDIAL ITN network. 
	
	NC acknowledges support from the State Research Agency (AEI) of the Spanish Ministry of Science, Innovation and Universities (MCIU) and the European Regional Development Fund (FEDER) under the grant with reference AYA2016-76219-P, and from the Fundaci\'on BBVA under its 2017 programme of assistance to scientific research groups, for the project "Using machine-learning techniques to drag galaxies from the noise in deep imaging". The IAC project P/300724 is financed by the Ministry of Science, Innovation and Universities, through the State Budget and by the Canary Islands Department of Economy, Knowledge and Employment, through the Regional Budget of the Autonomous Community.
	
	AV acknowledges financial support from the Emil Aaltonen Foundation.
	
	The Dell R815 Opteron server was obtained through funding from the Netherlands Organisation for Scientific Research (NWO) under project number 612.001.110. 
	
	This research made use of Astropy,\footnote{http://www.astropy.org} a community-developed core Python package for Astronomy \citep{robitaille2013astropy} \citep{price2018astropy}.
	
	This work was partly done using GNU Astronomy Utilities (Gnuastro, ascl.net/1801.009) version 0.7.42-22d2. Gnuastro is a generic package for astronomical data manipulation and analysis which was initially created and developed for research funded by the Monbukagakusho (Japanese government) scholarship and European Research Council (ERC) advanced grant 339659-MUSICOS. 	
	\bibliographystyle{aa}
	\bibliography{biblio}
	
	\begin{appendix}
	    \input{filters.tex}
		\input{tables.tex}
	\end{appendix}

\end{document}

%% file: filters.tex
\section{SExtractor filters} 
\label{filterapp}

SExtractor uses a filter to pre-process the input image. A number of filters are provided with the tool, but custom filters may also be used. The SExtractor manual suggests that the symmetrical PSF of the data is an optimal filter for detecting stars \citep{bertin2006manual}, whilst documentation provided with the filters suggests that gaussian or top-hat filters are effective in detecting extended, low-surface brightness objects.

As the range of valid filters is infinite, it would not be feasible to optimise the filter in addition to the other parameters. Accordingly, we used the default filter throughout the main experiments of the paper. We subsequently tested a subset of the available filters to determine if they had a significant effect on the tool's performance:

\begin{itemize}
  \item Default -- $3\times3$ pyramidal function (approximating gaussian smoothing).
  \item Gaussian -- $9\times9$ gaussian PSF with a full width at half maximum of 5 pixels.
  \item PSF -- $9\times9$ symmetrical window of the PSF of the simulated images.
  \item Top-hat -- $5\times5$ top-hat PSF.
\end{itemize}

\begin{figure}
		\centering
		\begin{subfigure}[b]{\columnwidth}
			\includegraphics[width=\columnwidth]{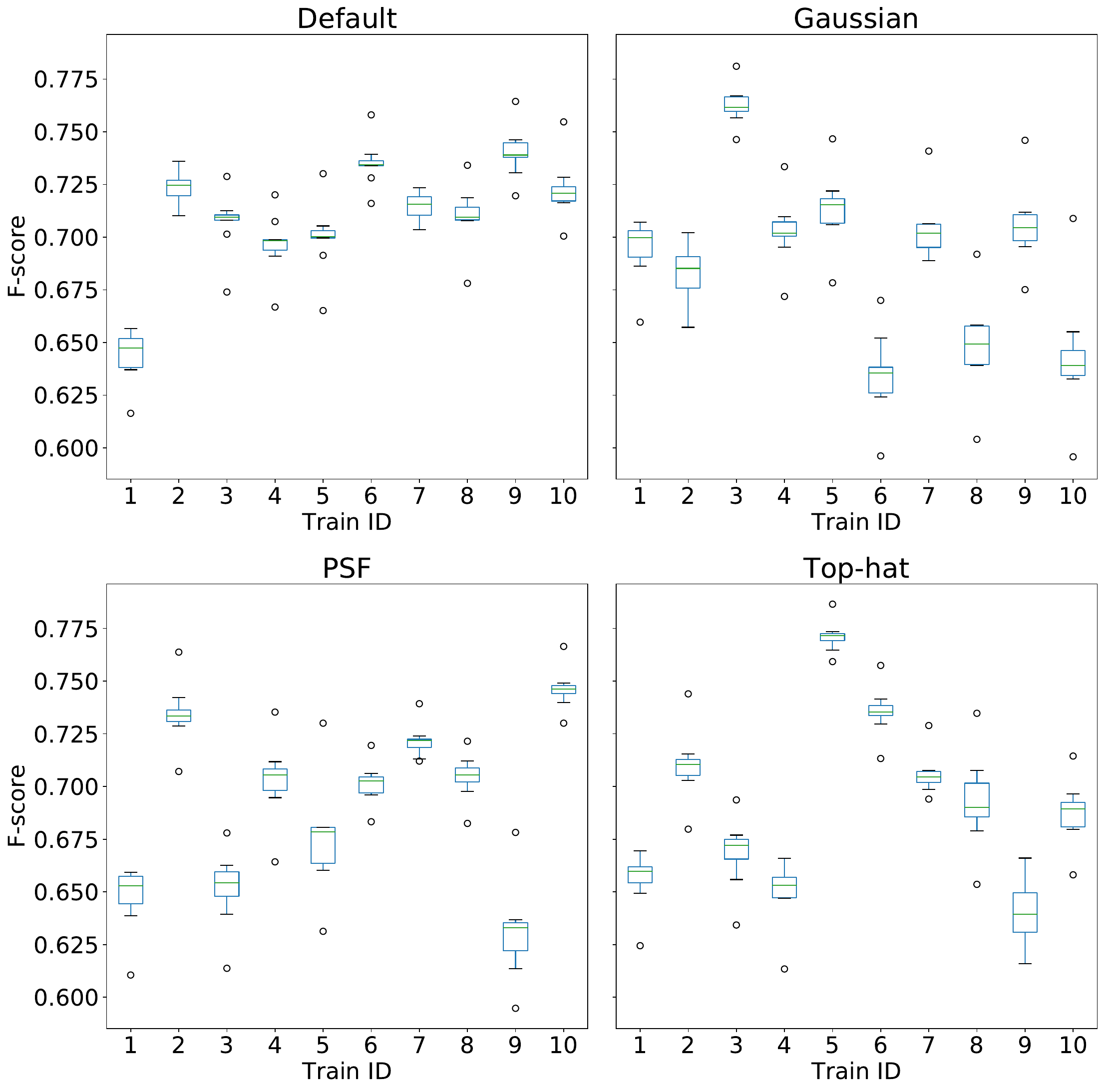}
			\caption{F-scores grouped by image used to optimise parameters}
		\end{subfigure}
		
		\begin{subfigure}[b]{\columnwidth}
			\includegraphics[width=\columnwidth]{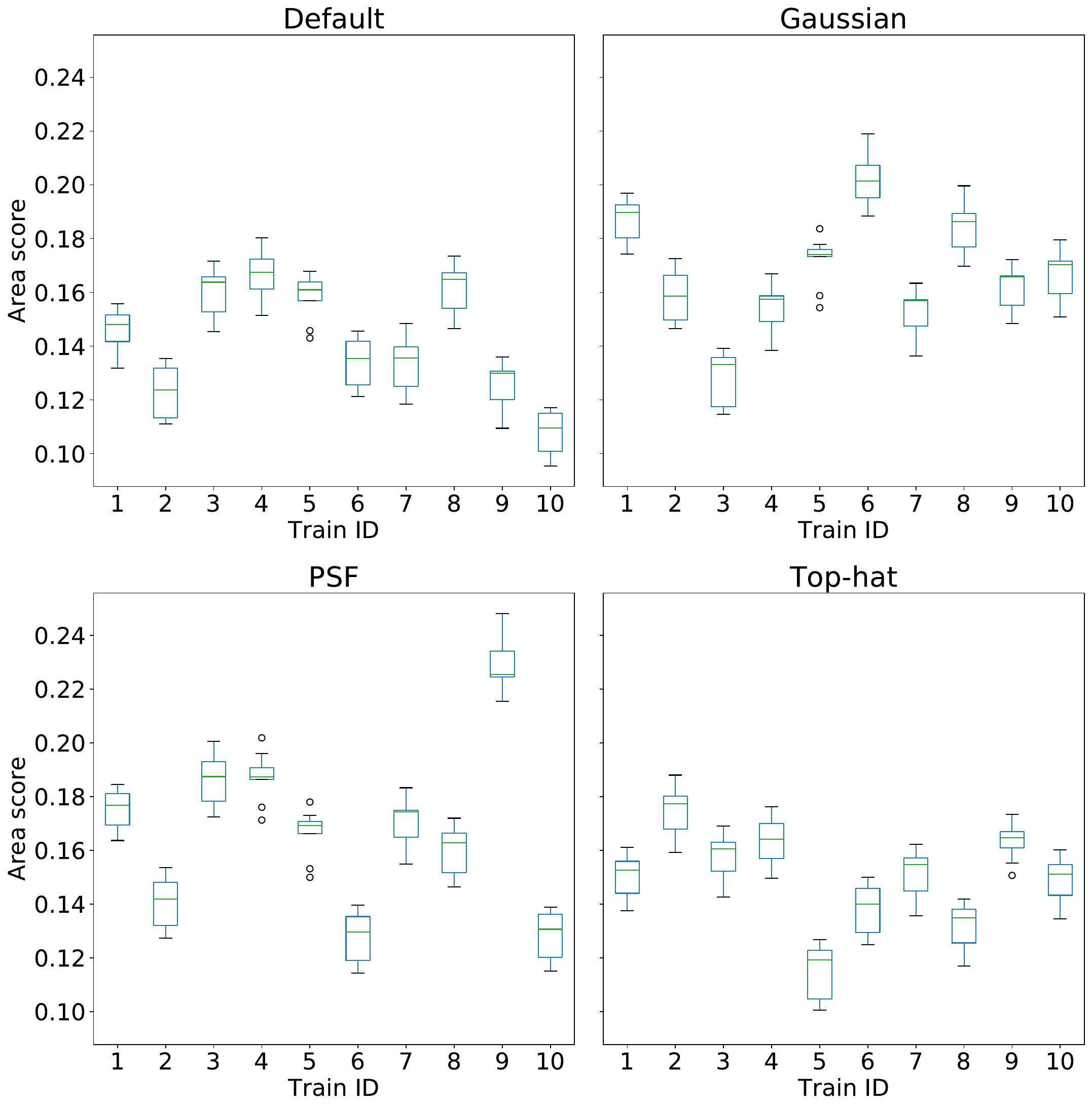}
			\caption{Area scores grouped by image used to optimise parameters}
		\end{subfigure}
		
		\caption{Optimised test distributions. Each tool's parameters were optimised for Combined A score on each of the ten images, and evaluated on the remaining nine images. Boxes extend from first (Q1) to third (Q3) quartiles of the results, with median values marked; whiskers extend to the furthest F-score less than $1.5 * (Q3-Q1)$ from each end of the box.}
		\label{filtergraph}
		
	\end{figure}

We optimised SExtractor's parameters for Combined A score for  as described in Sect. \ref{sec:opt}.
Fig. \ref{filtergraph} shows the distribution of F-scores and area scores for each of the four filters.

\begin{figure}
		\centering
			\includegraphics[width=\columnwidth]{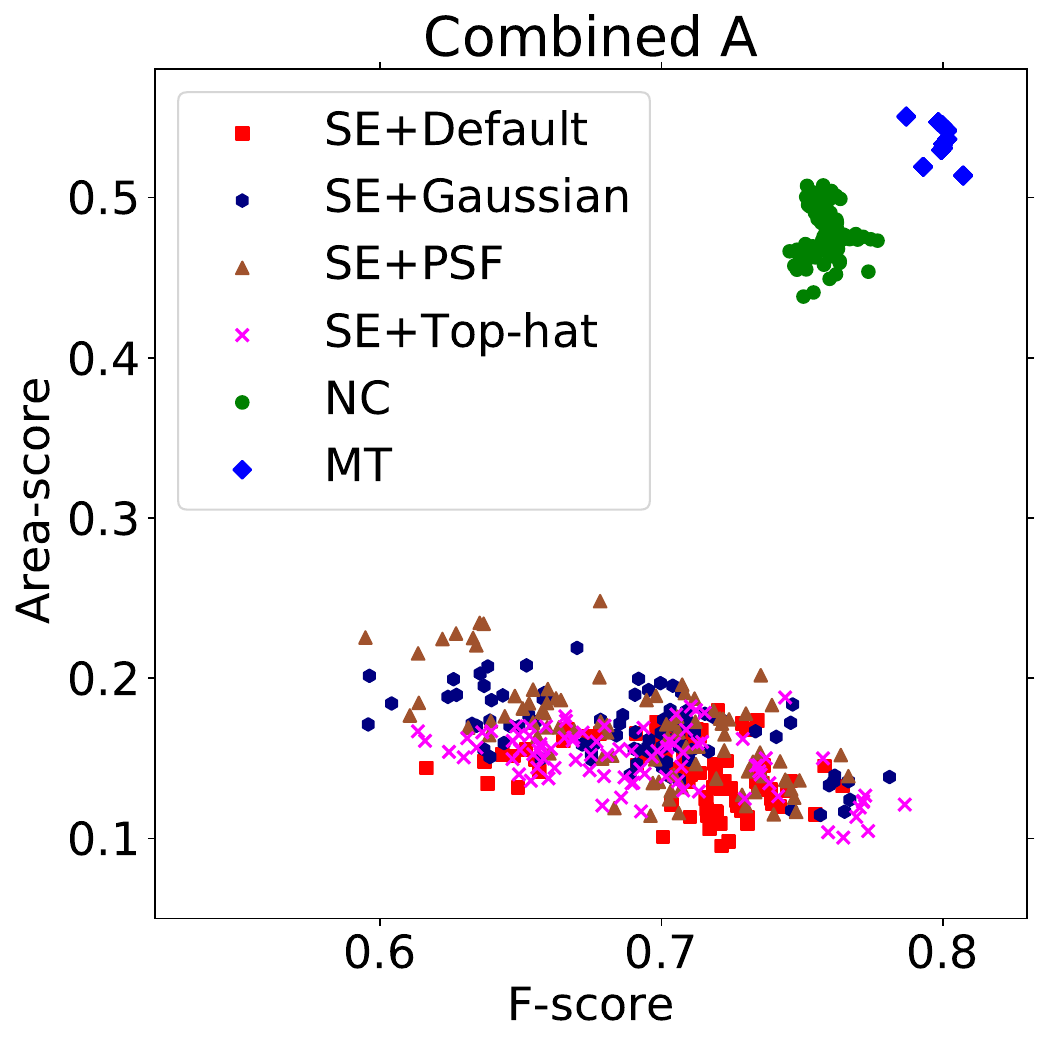}

		\caption{F-score vs area score -- each tool's parameters were optimised for Combined A score on each of the ten images, and evaluated on the remaining nine images.}
		\label{filtercomparison}
	\end{figure}

We found that the different filters had very little difference on F-score, but that there was a slightly higher area-score on average when using the gaussian filter as compared to the default. Whilst the gaussian filter could therefore be recommended in this situation, the choice of filter had no effect on the overall conclusions -- as shown in Fig. \ref{combo}, both MTObjects and NoiseChisel achieved substantially higher area scores of 0.4-0.6 compared to SExtractor's scores of 0.1-0.25. Plotting the four SExtractor filters on the same axes as Fig. \ref{combo} shows the relative similarity of the scores, as in Fig. \ref{filtercomparison}

%% file: tables.tex
\section{Optimised parameter tables} \label{optables}
Parameter sets marked with an asterisk produced the highest median test score for their optimisation metric and tool.
\label{app:a}
\begin{sidewaystable*}
\caption{Optimised parameters -- Source Extractor}
\label{app:se}
\scriptsize
\centering
\begin{tabular}{llp{1.3cm}p{1.9cm}*{4}{p{1.8cm}}}
\hline\hline 
Mode    & Image & BACK SIZE & BACK \mbox{FILTERSIZE} & DEBLEND MINCONT & DEBLEND NTHRESH & DETECT MINAREA & DETECT THRESH
 \\
\hline
0 & 1 & 65 & 2 & 1.00E-03 & 33 & 6 & 1.45E+00 \\
0 & 2 & 64 & 2 & 1.00E-03 & 34 & 9 & 1.35E+00 \\    
0 & 3 & 65 & 2 & 1.00E-03 & 35 & 5 & 1.59E+00 \\
0 & 4 & 64 & 3 & 1.00E-03 & 33 & 5 & 1.48E+00 \\
0 & 5 & 64 & 2 & 1.00E-03 & 31 & 5 & 1.51E+00 \\
0 & 6 & 37 & 5 & 1.00E-03 & 23 & 7 & 1.46E+00 \\
0 & 7 & 65 & 2 & 1.00E-03 & 33 & 5 & 1.78E+00 \\
\textbf{0*} & \textbf{8}&\textbf{63} & \textbf{3} & \textbf{1.00E-03} &\textbf{30} &\textbf{5} &\textbf{1.43E+00} \\
0 & 9 & 63 & 3 & 1.00E-03 & 32 & 6 & 1.69E+00 \\
0 & 10 & 17 & 7 & 1.73E-02 & 62 & 8 & 1.28E+00 \\
1 & 1 & 88 & 5 & 1.00E-03 & 21 & 8 & 1.00E-01 \\
1 & 2 & 36 & 9 & 1.00E-03 & 8 & 17 & 1.00E-01 \\
1 & 3 & 63 & 5 & 9.82E-02 & 32 & 4 & 1.00E-01 \\
1 & 4 & 22 & 7 & 9.81E-02 & 3 & 6 & 1.00E-01 \\
\textbf{1*} & \textbf{5} & \textbf{107} & \textbf{5} & \textbf{1.00E-03} & \textbf{41} & \textbf{30} & \textbf{1.00E-01} \\
1 & 6 & 24 & 11 & 1.00E-01 & 8 & 49 & 1.00E-01 \\
1 & 7 & 111 & 6 & 2.12E-02 & 6 & 14 & 1.05E-01 \\
1 & 8 & 108 & 3 & 1.00E-03 & 26 & 34 & 1.00E-01 \\
1 & 9 & 35 & 5 & 4.40E-02 & 48 & 28 & 1.13E-01 \\
1 & 10 & 80 & 4 & 3.67E-02 & 23 & 2 & 1.06E-01 \\
2 & 1 & 84 & 7 & 6.96E-02 & 44 & 36 & 6.56E-01 \\
2 & 2 & 29 & 7 & 1.82E-02 & 3 & 49 & 5.96E-01 \\
2 & 3 & 105 & 6 & 1.00E-03 & 63 & 37 & 6.03E-01 \\
2 & 4 & 110 & 11 & 1.00E-03 & 60 & 32 & 7.28E-01 \\
2 & 5 & 98 & 9 & 1.00E-03 & 40 & 28 & 7.07E-01 \\
2 & 6 & 33 & 8 & 1.00E-03 & 47 & 29 & 6.19E-01 \\
2 & 7 & 124 & 10 & 1.00E-03 & 50 & 29 & 1.24E+00 \\
\textbf{2*} & \textbf{8} & \textbf{98} & \textbf{7} & \textbf{1.00E-03} & \textbf{29} & \textbf{46} & \textbf{5.70E-01} \\
2 & 9 & 85 & 4 & 1.00E-03 & 44 & 22 & 8.94E-01 \\
2 & 10 & 118 & 3 & 6.87E-03 & 41 & 9 & 1.07E+00 \\
3 & 1 & 123 & 7 & 1.00E-03 & 60 & 43 & 3.93E-01 \\
3 & 2 & 71 & 5 & 2.53E-02 & 23 & 41 & 4.91E-01 \\
3 & 3 & 119 & 2 & 2.33E-02 & 23 & 44 & 4.94E-01 \\
3 & 4 & 72 & 11 & 1.00E-03 & 55 & 47 & 5.23E-01 \\
3 & 5 & 61 & 7 & 9.62E-02 & 53 & 43 & 5.35E-01 \\
3 & 6 & 101 & 3 & 1.00E-03 & 38 & 41 & 5.30E-01 \\
\textbf{3*} & \textbf{7} & \textbf{119} & \textbf{6} & \textbf{1.00E-03} & \textbf{20} & \textbf{49} & \textbf{4.37E-01} \\
3 & 8 & 104 & 7 & 1.00E-03 & 49 & 37 & 5.11E-01 \\
3 & 9 & 25 & 8 & 1.07E-02 & 39 & 25 & 6.56E-01 \\
3 & 10 & 88 & 4 & 1.00E-03 & 49 & 16 & 8.23E-01 \\
\hline
\end{tabular}
\end{sidewaystable*}
\begin{sidewaystable*}

\caption{Optimised parameters -- Noise Chisel (Noise Chisel)}
\label{app:nc}

\scriptsize
\centering

\begin{tabular}{ll*{1}{p{1.0cm}}*{4}{p{1.3cm}}*{2}{p{1.0cm}}*{2}{p{1.3cm}}*{2}{p{1.0cm}}*{2}{p{1.3cm}}}
\hline\hline 

Mode & Image & tilesize & qthresh & snquant & detgrow- quant & dthresh & erode & opening & detgrow- maxholesize & meanmed- qdiff & erode- ngb & opening- ngb & minsky- frac & noerode- quant \\

\hline

0 & 1 & 27 & 4.99E-01 & 9.99E-01 & 9.99E-01 & 0.00E+00 & 3 & 4 & 100 & 5.00E-03 & 4 & 4 & 8.00E-01 & 7.00E-01 \\
0 & 2 & 90 & 4.99E-01 & 9.71E-01 & 9.99E-01 & 6.34E-01 & 4 & 3 & 52 & 2.00E-02 & 4 & 4 & 7.54E-01 & 1.00E+00 \\
0 & 3 & 50 & 4.00E-01 & 9.27E-01 & 9.98E-01 & 8.26E-01 & 5 & 4 & 19 & 2.00E-02 & 4 & 4 & 6.62E-01 & 9.73E-01 \\
\textbf{0*} & \textbf{4} & \textbf{26} & \textbf{4.99E-01} & \textbf{9.99E-01} & \textbf{9.86E-01} & \textbf{4.03E-01} & \textbf{3} & \textbf{1} & \textbf{59} & \textbf{2.00E-02} & \textbf{4} & \textbf{8} & \textbf{8.00E-01} & \textbf{1.00E+00} \\
0 & 5 & 49 & 4.99E-01 & 6.12E-01 & 9.46E-01 & 6.87E-01 & 1 & 5 & 77 & 1.98E-02 & 8 & 8 & 8.00E-01 & 7.02E-01 \\
0 & 6 & 40 & 4.99E-01 & 9.99E-01 & 9.99E-01 & 3.19E-01 & 10 & 1 & 86 & 2.00E-02 & 4 & 4 & 8.00E-01 & 1.00E+00 \\
0 & 7 & 87 & 4.33E-01 & 9.71E-01 & 9.99E-01 & 5.20E-01 & 6 & 2 & 14 & 6.30E-03 & 8 & 8 & 8.00E-01 & 9.79E-01 \\
0 & 8 & 26 & 2.00E-01 & 9.99E-01 & 9.99E-01 & 0.00E+00 & 2 & 5 & 48 & 2.00E-02 & 4 & 4 & 8.00E-01 & 9.48E-01 \\
0 & 9 & 98 & 4.99E-01 & 6.00E-01 & 9.99E-01 & 5.39E-01 & 8 & 4 & 88 & 5.00E-03 & 8 & 8 & 4.00E-01 & 7.00E-01 \\
0 & 10 & 87 & 4.99E-01 & 9.99E-01 & 8.28E-01 & 1.67E-01 & 5 & 2 & 26 & 2.00E-02 & 4 & 8 & 6.03E-01 & 7.00E-01 \\
\textbf{1*} & \textbf{1} & \textbf{75} & \textbf{3.69E-01} & \textbf{9.99E-01} & \textbf{6.00E-01} & \textbf{1.00E+00} & \textbf{6} & \textbf{5} & \textbf{79} & \textbf{5.00E-03} & \textbf{4} & \textbf{8} & \textbf{4.00E-01} & \textbf{7.75E-01} \\
1 & 2 & 40 & 4.99E-01 & 9.99E-01 & 6.00E-01 & 1.00E+00 & 1 & 5 & 25 & 2.00E-02 & 4 & 8 & 4.00E-01 & 1.00E+00 \\
1 & 3 & 27 & 4.99E-01 & 6.00E-01 & 6.00E-01 & 0.00E+00 & 1 & 5 & 85 & 1.60E-02 & 8 & 8 & 4.56E-01 & 1.00E+00 \\
1 & 4 & 86 & 4.99E-01 & 9.99E-01 & 6.00E-01 & 1.00E+00 & 6 & 3 & 32 & 2.00E-02 & 4 & 8 & 8.00E-01 & 1.00E+00 \\
1 & 5 & 34 & 4.99E-01 & 6.00E-01 & 6.00E-01 & 0.00E+00 & 5 & 2 & 97 & 8.07E-03 & 4 & 4 & 4.00E-01 & 1.00E+00 \\
1 & 6 & 28 & 2.00E-01 & 6.00E-01 & 6.19E-01 & 6.86E-01 & 2 & 1 & 94 & 1.97E-02 & 4 & 8 & 4.00E-01 & 1.00E+00 \\
1 & 7 & 57 & 2.00E-01 & 9.99E-01 & 6.00E-01 & 0.00E+00 & 10 & 5 & 58 & 5.00E-03 & 8 & 4 & 4.00E-01 & 7.00E-01 \\
1 & 8 & 33 & 4.99E-01 & 6.15E-01 & 6.15E-01 & 1.69E-13 & 5 & 1 & 65 & 2.00E-02 & 8 & 4 & 8.00E-01 & 1.00E+00 \\
1 & 9 & 96 & 3.64E-01 & 9.28E-01 & 6.03E-01 & 7.15E-01 & 1 & 4 & 63 & 1.03E-02 & 8 & 8 & 6.87E-01 & 9.27E-01 \\
1 & 10 & 43 & 3.74E-01 & 8.28E-01 & 6.54E-01 & 8.98E-01 & 1 & 5 & 16 & 1.14E-02 & 4 & 8 & 4.98E-01 & 1.00E+00 \\
\textbf{2*} & \textbf{1} & \textbf{68} & \textbf{3.26E-01} & \textbf{6.00E-01} & \textbf{6.11E-01} & \textbf{6.85E-01} & \textbf{8} & \textbf{5} & \textbf{68} & \textbf{2.00E-02} & \textbf{4} & \textbf{8} & \textbf{5.16E-01} & \textbf{8.05E-01} \\
2 & 2 & 30 & 4.99E-01 & 6.00E-01 & 6.00E-01 & 6.36E-01 & 2 & 2 & 67 & 1.06E-02 & 8 & 8 & 8.00E-01 & 1.00E+00 \\
2 & 3 & 27 & 3.46E-01 & 9.99E-01 & 6.09E-01 & 0.00E+00 & 9 & 2 & 55 & 1.98E-02 & 4 & 4 & 4.94E-01 & 8.84E-01 \\
2 & 4 & 28 & 4.99E-01 & 6.00E-01 & 6.16E-01 & 0.00E+00 & 4 & 3 & 80 & 2.00E-02 & 8 & 4 & 4.00E-01 & 9.03E-01 \\
2 & 5 & 89 & 2.00E-01 & 6.58E-01 & 6.18E-01 & 7.90E-01 & 5 & 2 & 8 & 2.00E-02 & 8 & 8 & 8.00E-01 & 7.09E-01 \\
2 & 6 & 29 & 2.91E-01 & 9.99E-01 & 6.10E-01 & 1.19E-02 & 3 & 1 & 68 & 2.00E-02 & 4 & 4 & 5.01E-01 & 1.00E+00 \\
2 & 7 & 30 & 3.50E-01 & 6.00E-01 & 6.00E-01 & 0.00E+00 & 8 & 5 & 28 & 7.24E-03 & 8 & 8 & 6.97E-01 & 7.00E-01 \\
2 & 8 & 26 & 2.00E-01 & 6.00E-01 & 6.00E-01 & 2.73E-02 & 9 & 5 & 23 & 2.00E-02 & 8 & 8 & 5.20E-01 & 7.00E-01 \\
2 & 9 & 45 & 2.00E-01 & 8.07E-01 & 7.21E-01 & 7.51E-01 & 7 & 2 & 76 & 8.42E-03 & 8 & 8 & 7.46E-01 & 7.88E-01 \\
2 & 10 & 82 & 4.42E-01 & 6.87E-01 & 6.82E-01 & 5.27E-01 & 6 & 2 & 38 & 2.00E-02 & 4 & 8 & 7.44E-01 & 9.14E-01 \\
3 & 1 & 27 & 2.04E-01 & 9.08E-01 & 6.03E-01 & 0.00E+00 & 4 & 4 & 23 & 2.00E-02 & 8 & 4 & 7.20E-01 & 7.68E-01 \\
3 & 2 & 64 & 4.99E-01 & 6.00E-01 & 6.00E-01 & 7.87E-01 & 4 & 2 & 41 & 7.97E-03 & 8 & 8 & 4.00E-01 & 7.00E-01 \\
3 & 3 & 22 & 2.00E-01 & 9.98E-01 & 6.02E-01 & 1.57E-03 & 1 & 5 & 48 & 2.00E-02 & 8 & 8 & 4.02E-01 & 7.01E-01 \\
3 & 4 & 20 & 2.00E-01 & 9.99E-01 & 6.00E-01 & 0.00E+00 & 1 & 5 & 76 & 5.00E-03 & 8 & 8 & 4.00E-01 & 7.00E-01 \\
3 & 5 & 41 & 4.99E-01 & 9.50E-01 & 6.00E-01 & 2.16E-01 & 8 & 1 & 90 & 5.00E-03 & 4 & 8 & 4.00E-01 & 1.00E+00 \\
3 & 6 & 82 & 4.67E-01 & 9.96E-01 & 6.37E-01 & 1.07E-02 & 8 & 1 & 38 & 1.09E-02 & 8 & 4 & 8.00E-01 & 9.56E-01 \\
3 & 7 & 63 & 4.82E-01 & 6.93E-01 & 6.38E-01 & 5.90E-01 & 7 & 2 & 85 & 1.89E-02 & 8 & 8 & 4.80E-01 & 8.70E-01 \\
3 & 8 & 31 & 4.99E-01 & 7.88E-01 & 6.29E-01 & 5.51E-01 & 9 & 5 & 29 & 2.00E-02 & 4 & 4 & 5.33E-01 & 9.86E-01 \\
\textbf{3*} & \textbf{9} & \textbf{71} & \textbf{2.00E-01} & \textbf{9.99E-01} & \textbf{6.00E-01} & \textbf{0.00E+00} & \textbf{1} & \textbf{5} & \textbf{11} & \textbf{2.00E-02} & \textbf{4} & \textbf{8} & \textbf{4.00E-01} & \textbf{7.00E-01} \\
3 & 10 & 100 & 3.44E-01 & 9.99E-01 & 6.00E-01 & 0.00E+00 & 9 & 5 & 5 & 2.00E-02 & 4 & 4 & 4.00E-01 & 7.28E-01 \\

\hline
\end{tabular}
\end{sidewaystable*}

\begin{sidewaystable*}
\caption{Optimised parameters -- Noise Chisel (Segment)}
\label{app:ncs}

\scriptsize
\centering

\begin{tabular}{ll*{7}{p{1.4cm}}}
\hline\hline 
Mode & Image & tilesize & snquant & gthresh & snminarea & minriver- length & objbordersn & minskyfrac \\

\hline
0 & 1 & 72 & 9.99E-01 & 1.00E+00 & 25 & 23 & 1.28E+01 & 4.00E-01 \\
0 & 2 & 90 & 9.99E-01 & 1.00E+00 & 18 & 12 & 1.86E+01 & 5.75E-01 \\
0 & 3 & 93 & 9.94E-01 & 8.65E-01 & 23 & 20 & 3.18E+01 & 6.67E-01 \\
\textbf{0*} & \textbf{4} & \textbf{60} & \textbf{9.99E-01} & \textbf{6.69E-01} & \textbf{25} & \textbf{9} & \textbf{2.56E+01} & \textbf{7.52E-01} \\
0 & 5 & 20 & 9.99E-01 & 2.44E-01 & 20 & 40 & 2.00E+01 & 8.00E-01 \\
0 & 6 & 20 & 9.99E-01 & 1.00E+00 & 19 & 5 & 1.12E+01 & 8.00E-01 \\
0 & 7 & 29 & 9.99E-01 & 1.00E+00 & 21 & 37 & 1.49E+01 & 7.28E-01 \\
0 & 8 & 72 & 9.99E-01 & 1.00E+00 & 20 & 40 & 3.16E+01 & 8.00E-01 \\
0 & 9 & 45 & 9.99E-01 & 0.00E+00 & 20 & 14 & 2.26E+01 & 8.00E-01 \\
0 & 10 & 91 & 9.99E-01 & 3.77E-01 & 25 & 25 & 1.51E+01 & 4.00E-01 \\
\textbf{1*} & \textbf{1} & \textbf{65} & \textbf{9.99E-01} & \textbf{5.41E-01} & \textbf{10} & \textbf{34} & \textbf{5.00E-01} & \textbf{4.00E-01} \\
1 & 2 & 85 & 9.99E-01 & 1.00E+00 & 25 & 35 & 1.73E+01 & 8.00E-01 \\
1 & 3 & 35 & 9.99E-01 & 0.00E+00 & 15 & 21 & 1.88E+01 & 4.00E-01 \\
1 & 4 & 65 & 9.99E-01 & 1.00E+00 & 23 & 8 & 1.80E+01 & 7.43E-01 \\
1 & 5 & 32 & 9.99E-01 & 6.34E-01 & 15 & 14 & 1.34E+00 & 8.00E-01 \\
1 & 6 & 41 & 9.99E-01 & 4.41E-01 & 21 & 36 & 1.15E+01 & 5.99E-01 \\
1 & 7 & 52 & 9.99E-01 & 0.00E+00 & 25 & 40 & 1.81E+01 & 4.00E-01 \\
1 & 8 & 46 & 9.99E-01 & 2.85E-02 & 25 & 40 & 2.12E+01 & 8.00E-01 \\
1 & 9 & 82 & 9.97E-01 & 5.72E-01 & 20 & 23 & 3.45E+01 & 7.81E-01 \\
1 & 10 & 61 & 9.99E-01 & 2.14E-01 & 17 & 31 & 8.88E+00 & 5.64E-01 \\
\textbf{2*} & \textbf{1} & \textbf{25} & \textbf{9.99E-01} & \textbf{3.28E-01} & \textbf{25} & \textbf{24} & \textbf{3.11E+01} & \textbf{4.90E-01} \\
2 & 2 & 45 & 9.99E-01 & 4.89E-01 & 24 & 21 & 1.77E+01 & 4.00E-01 \\
2 & 3 & 80 & 9.99E-01 & 6.60E-03 & 22 & 30 & 5.88E+00 & 4.09E-01 \\
2 & 4 & 81 & 9.99E-01 & 0.00E+00 & 22 & 8 & 1.03E+01 & 8.00E-01 \\
2 & 5 & 31 & 9.99E-01 & 2.48E-01 & 22 & 16 & 7.32E+00 & 4.00E-01 \\
2 & 6 & 74 & 9.99E-01 & 9.37E-03 & 25 & 22 & 6.71E+00 & 4.10E-01 \\
2 & 7 & 38 & 9.99E-01 & 2.11E-01 & 20 & 31 & 1.29E+01 & 4.00E-01 \\
2 & 8 & 20 & 9.99E-01 & 0.00E+00 & 25 & 33 & 7.35E+00 & 4.00E-01 \\
2 & 9 & 90 & 9.99E-01 & 8.11E-01 & 14 & 27 & 3.96E+01 & 4.00E-01 \\
2 & 10 & 48 & 9.99E-01 & 6.99E-01 & 18 & 15 & 2.60E+01 & 6.55E-01 \\
3 & 1 & 72 & 9.99E-01 & 5.36E-16 & 24 & 13 & 1.29E+01 & 4.03E-01 \\
3 & 2 & 94 & 9.99E-01 & 8.28E-01 & 25 & 30 & 2.42E+01 & 4.00E-01 \\
3 & 3 & 31 & 9.97E-01 & 3.17E-02 & 25 & 5 & 2.21E+01 & 4.02E-01 \\
3 & 4 & 20 & 9.99E-01 & 1.00E+00 & 25 & 11 & 1.63E+01 & 4.00E-01 \\
3 & 5 & 32 & 9.99E-01 & 0.00E+00 & 25 & 25 & 1.81E+01 & 8.00E-01 \\
3 & 6 & 20 & 9.99E-01 & 3.93E-01 & 22 & 26 & 9.46E+00 & 6.79E-01 \\
3 & 7 & 78 & 9.99E-01 & 5.50E-02 & 24 & 21 & 2.89E+01 & 4.52E-01 \\
3 & 8 & 47 & 9.99E-01 & 6.88E-01 & 16 & 16 & 3.86E+01 & 8.00E-01 \\
\textbf{3*} & \textbf{9} & \textbf{74} & \textbf{9.99E-01} & \textbf{0.00E+00} & \textbf{25} & \textbf{11} & \textbf{2.38E+01} & \textbf{8.00E-01} \\
3 & 10 & 32 & 9.99E-01 & 2.18E-01 & 23 & 18 & 1.11E+01 & 5.88E-01 \\

\hline
\end{tabular}
\end{sidewaystable*}

\begin{sidewaystable*}
\caption{Optimised parameters -- MTObjects}
\label{app:mt}

\scriptsize
\centering

\begin{tabular}{ll*{2}{p{1.6cm}}}
\hline\hline 
Mode & Image & move\_factor & min\_dist \\
\hline

0 & 1 & 4.96E-02 & 1.49E-01 \\
0 & 2 & 6.14E-02 & 1.15E-01 \\
0 & 3 & 5.73E-02 & 1.57E-01 \\
0 & 4 & 8.24E-03 & 1.32E-01 \\
0 & 5 & 1.21E-01 & 1.45E-01 \\
0 & 6 & 0 & 1.50E-01 \\
\textbf{0*} & \textbf{7} & \textbf{0} & \textbf{1.13E-01} \\
0 & 8 & 9.34E-02 & 1.55E-01 \\
0 & 9 & 1.13E-01 & 1.33E-01 \\
0 & 10 & 3.07E-02 & 1.17E-01 \\
\textbf{1*} & \textbf{1} & \textbf{0}& \textbf{0} \\
1 & 2 & 0 & 0 \\
1 & 3 & 0 & 0 \\
1 & 4 & 0 & 0 \\
1 & 5 & 0 & 0 \\
1 & 6 & 0 & 0 \\
1 & 7 & 0 & 0 \\
1 & 8 & 0 & 0 \\
1 & 9 & 0 & 0 \\
1 & 10 & 0 & 0 \\
\textbf{2*} & \textbf{1} & \textbf{0}& \textbf{0} \\
2 & 2 & 0 & 0 \\
2 & 3 & 0 & 0 \\
2 & 4 & 0 & 0 \\
2 & 5 & 0 & 0 \\
2 & 6 & 0 & 0 \\
2 & 7 & 0 & 0 \\
2 & 8 & 0 & 0 \\
2 & 9 & 0 & 0 \\
2 & 10 & 0 & 0 \\
\textbf{3*} & \textbf{1} & \textbf{0}& \textbf{0} \\
3 & 2 & 0 & 0 \\
3 & 3 & 0 & 0 \\
3 & 4 & 0 & 0 \\
3 & 5 & 0 & 0 \\
3 & 6 & 0 & 0 \\
3 & 7 & 0 & 0 \\
3 & 8 & 9.36E-03 & 0 \\
3 & 9 & 0 & 0 \\
3 & 10 & 0 & 0 \\

\hline
\end{tabular}
\end{sidewaystable*}

\begin{sidewaystable*}
\caption{Optimised parameters -- ProFound}
\label{app:pf}

\scriptsize
\centering

\begin{tabular}{ll*{4}{p{1.6cm}}*{3}{p{1.2cm}}*{1}{p{1.6cm}}}
\hline\hline 
Mode & Image & skycut & tolerance & ext & sigma & pixcut & size & iters & threshold \\
\hline

0 & 1 & 7.22E-01 & 3.71E+00 & 1.92E+00 & 2.22E+00 & 4 & 5 & 7 & 7.50E-01 \\
0 & 2 & 9.34E-01 & 2.06E+00 & 2.29E+00 & 1.35E+00 & 10 & 7 & 2 & 8.26E-01 \\
0 & 3 & 1.00E+00 & 1.00E+00 & 7.03E+00 & 1.19E+00 & 7 & 7 & 4 & 7.50E-01 \\
0 & 4 & 8.23E-01 & 1.58E+00 & 8.03E+00 & 2.23E+00 & 6 & 7 & 3 & 1.20E+00 \\
0 & 5 & 1.22E+00 & 3.52E+00 & 3.89E+00 & 9.17E-01 & 6 & 5 & 6 & 1.78E+00 \\
0 & 6 & 6.54E-01 & 1.00E+00 & 3.78E+00 & 3.00E+00 & 15 & 7 & 9 & 7.50E-01 \\
0 & 7 & 9.34E-01 & 1.00E+00 & 4.55E+00 & 1.43E+00 & 16 & 5 & 0 & 1.96E+00 \\
\textbf{0*} & \textbf{8} & \textbf{5.56E-01} & \textbf{2.03E+00} & \textbf{3.28E+00} & \textbf{2.11E+00} & \textbf{15} & \textbf{7} & \textbf{6} & \textbf{9.49E-01} \\
0 & 9 & 1.79E+00 & 3.90E+00 & 3.29E+00 & 8.76E-01 & 4 & 5 & 7 & 1.35E+00 \\
0 & 10 & 6.25E-01 & 2.93E+00 & 4.33E+00 & 1.36E+00 & 14 & 9 & 8 & 2.00E+00 \\
1 & 1 & 5.26E-01 & 5.29E+00 & 2.22E+00 & 1.98E+00 & 7 & 9 & 7 & 8.68E-01 \\
1 & 2 & 2.38E-01 & 2.62E+00 & 9.01E+00 & 2.21E+00 & 11 & 7 & 6 & 1.12E+00 \\
1 & 3 & 7.65E-01 & 5.71E+00 & 2.70E+00 & 1.45E+00 & 5 & 5 & 8 & 8.42E-01 \\
1 & 4 & 6.13E-01 & 4.44E+00 & 4.77E+00 & 2.83E+00 & 5 & 7 & 7 & 8.00E-01 \\
1 & 5 & 1.57E-01 & 3.72E+00 & 4.23E+00 & 2.74E+00 & 10 & 9 & 6 & 1.24E+00 \\
1 & 6 & 2.77E-01 & 2.49E+00 & 8.57E+00 & 2.26E+00 & 15 & 7 & 9 & 8.31E-01 \\
1 & 7 & 6.23E-01 & 3.68E+00 & 5.18E+00 & 1.01E+00 & 1 & 5 & 4 & 1.04E+00 \\
\textbf{1*} & \textbf{8} & \textbf{6.45E-01} & \textbf{6.00E+00} & \textbf{8.97E+00} & \textbf{1.13E+00} & \textbf{14} & \textbf{9} & \textbf{9} & \textbf{9.10E-01} \\
1 & 9 & 2.82E-01 & 4.48E+00 & 6.28E+00 & 2.07E+00 & 3 & 7 & 4 & 7.50E-01 \\
1 & 10 & 1.33E+00 & 3.23E+00 & 3.63E+00 & 1.82E+00 & 9 & 7 & 9 & 8.23E-01 \\

\hline
\end{tabular}
\end{sidewaystable*}